# Oscillations in a flexible channel flow of a generalized Newtonian fluid

Prakash Goswami[1], Aditya Bandopadhyay[2] and Suman Chakraborty[1,2*]

[1]Department of Mechanical Engineering, Indian Institute of Technology Kharagpur, Kharagpur, West Bengal, India - 721302
[2]Advanced Technology Development Center, Indian Institute of Technology Kharagpur, Kharagpur, West Bengal, India - 721302

We study the flow of a generalized Newtonian fluid, characterized by a power-law model, through a channel consisting of a wall with a flexible membrane under longitudinal tension. It is assumed that at steady state the flow through the channel admits a constant flux unidirectional flow profile, while for the unsteady case, we employ the long wave approximation and use a set of reduced equations to describe the variation of the shape of the membrane (assumed to be massless and elastic) and the variation of the fluid-flux. By means of asymptotic expansion, multiscale analysis and full numerical solutions of the pertinent governing equations, we show that depending upon the Reynolds number and the membrane stress, the flow behaviour for a shear-thinning, shear-thickening and Newtonian fluid may be markedly different, being oscillatory for one while chaotic for the other. The results presented herein hold practical relevance for several biologically relevant processes involving transport of rheologically complex biofluids through flexible domains.

---

## 1. Introduction

Fluid flow through flexible channels or tubes involves interaction between the fluid stresses and elastic stresses, and is important to describe many biophysical transport processes (Grotberg & Jensen 2004). Owing to the coupled nature (fluid structure interaction) of the problem, the system depicts rich dynamical behaviour, a subject which has been a topic of much research for the past few decades. Motivated by physiological instances such as flows in veins, arteries, pulmonary circulation etc. and devices such as prosthetic heart, and the phenomenon of Korotkoff sounds etc., Shapiro theoretically analyzed the flow in flexible and collapsible tubes (Shapiro 1977). Building on this, there have been several attempts to physically identify the mechanism for the occurrences of self-excited oscillations of the rubber-like membrane of the flexible tubes (Cancelli & Pedley 1985; Pedley 1992; Matsuzaki & Matsumoto 1989).

In order to understand the phenomenon experimentally, the Starling resistor and its variants have been employed extensively. By keeping a variable exit pressure, several regimes of operation were identified during the action of a Starling resistor; it was shown that owing to the nonlinear nature of the problem, the route of approach towards a state dictates the entire dynamics of the system (Bertram et al. 1990). Their analysis, for the first time, highlighted a series of bifurcations leading to a, presumable, chaotic tube dynamics. Apart from laboratory experiments, there have been several attempts to highlight the complex behaviour of the systems with flexible or collapsible tubes by means of numerical analysis. Through finite element analysis for inertialess rubber-like membrane a period doubling bifurcation has been demonstrated, apart from downstream vorticity waves, in the case of self-excited oscillations (Luo & Pedley 1996). Besides this, by means of asymptotic methods as well as direct numerical simulations, attempts have been made towards identifying the mechanism of self-excited oscillations (Jensen & Heil 2003). They established the sloshing mode of instability which was further studied using one-dimensional, two-dimensional and the more general three-dimensional models (Stewart et al. 2010; Stewart et al. 2009; Whittaker et al. 2010; Whittaker, Heil, Boyle, et al. 2010; R. J. Whittaker, Heil, Jensen, et al. 2010). The one-dimensional model has been employed successfully in quite a few cases to predict, or at least, clarify some of the underlying mechanisms behind the membrane vibrations. In these models (Hughes & Lubliner 1973), it is assumed, amongst other things, that the membrane length is longer than the channel width (the long wavelength approximation), and that the flow profile is parabolic at every cross section. Thus, a depth averaged model allows us to study a wide range of parameters without having to explicitly solve the two-dimensional or three-dimensional equation. It has been shown using the one-dimensional model that the uniform modes are unstable to divergent (static) and oscillatory instabilities (Xu et al. 2013; Xu et al. 2014).

---

*email address for correspondence: suman@mech.iitkgp.ernet.in

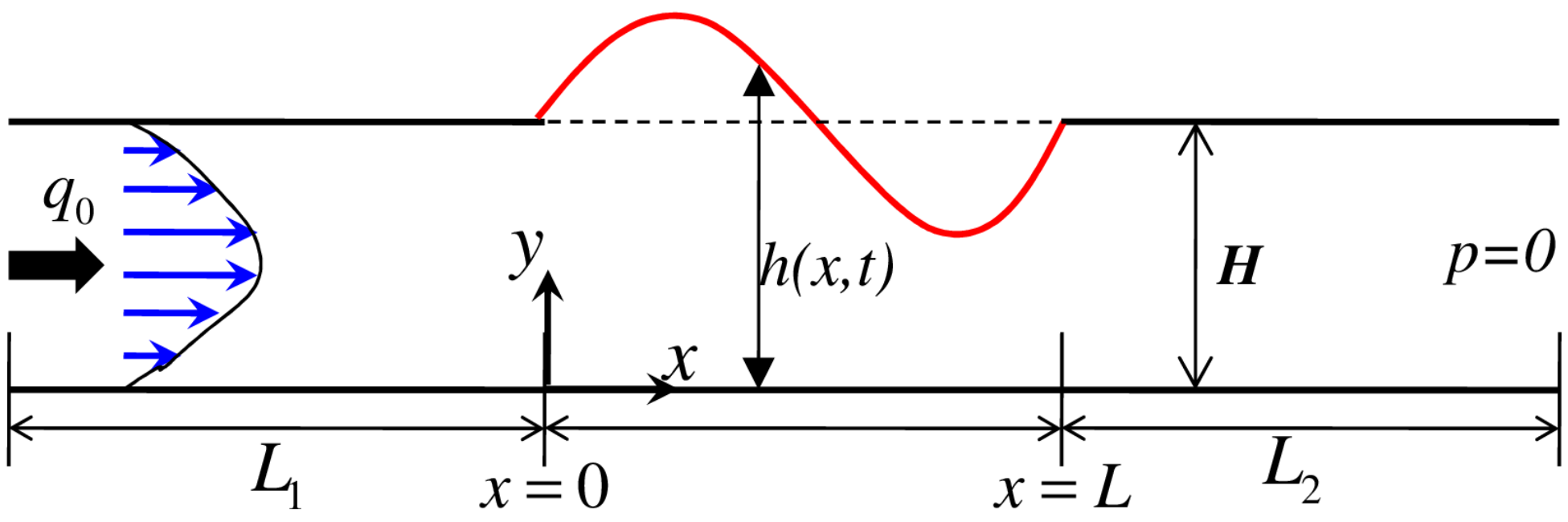

FIGURE 1: Schematic of the problem. There is a flexible membrane at the top wall in the region between $0<x<L$. The upstream flux is specified while the outlet pressure is set to zero. The lengths of the upstream and downstream sections are considered to be $L_1$ and $L_2$ respectively.

From a biophysical perspective, however, there is very little work done to identify the role of the fluid rheology towards the dynamics of collapsible tubes and channels (Grotberg 2001). Given the complex nature of the flow inherent to such dynamical systems, it is expected that the fluid rheology should play a central role towards dictating the system behaviour. Such systems are typical for many biological systems and it is quite natural for these to transport complex fluids; consequently leading to alterations in the associated observed properties and phenomena. For example, the influence of the fluid-lining viscosity on the reopening of collapsible tubes (airways) showed that shear thinning fluids lead to reduction in the opening time of a collapsed tube, while thicker non-Newtonian films led to a requirement of a larger inlet pressure for tube opening (Low 1997). Apart from such power-law models, the role of viscoelasticity of the fluid lining gel has also been considered on the airway reopening process (Hsu et al. 1996; Hsu et al. 1994). Therefore, to the best of the authors' knowledge, the influence of fluid rheology on the dynamics of a flexible channel is yet to be explored. In this work, we consider a flow domain comprising of rigid walls with a flexible-wall in between two rigid sections. The upstream fluid flux is specified while the membrane is assumed to be massless and pre-stretched. The fluid is assumed, for simplicity, to be described by the power-law (Ostwald de-Waele) constitutive relationship (Bird et al. 1977).

## 2. Problem formulation and linear solutions

### 2.1 *System description*

We follow the similar model to that of Stewart et al. (2009) and Xu et al. (2013) towards depicting the dynamics of the flexible tube in between two rigid sections (please refer to figure 1). We consider the flow of a power-law fluid in a channel of length $L_1 + L + L_2$, of which the middle portion ($0 \le x \le L$) is an elastic membrane pinned at both the ends $x = 0$ and $x = L$. The position of the membrane at a time $t$ and any axial location $x$ is dictated by the function $h(x,t)$ which takes a constant value $H$ at the rigid segments of the channel. A constant upstream flux is assumed to initiate the flow. It is also assumed that a longitudinal tension is applied to the membrane in such a way that the flow reveals a self similar parabolic nature along the channel axial position, that is at $x = -L_1$ and $x = L + L_2$. The flow field may be described by means of the continuity and momentum equations:

$$\nabla \bullet \mathbf{v} = 0,$$
$$\rho\left(\frac{\partial \mathbf{v}}{\partial t} + \mathbf{v} \bullet \nabla \mathbf{v}\right) = -\nabla p + \nabla \bullet \boldsymbol{\tau}, \tag{2.1}$$

where $\rho$ is the fluid density, $\mathbf{v}$ is flow velocity, $p$ is the pressure and $\boldsymbol{\tau}$ is the deviatoric stress tensor, which is assumed to be described by the power-law constitutive behaviour $\boldsymbol{\tau} = k_p\left(|\dot{\gamma}|^{n-1}\dot{\boldsymbol{\gamma}}\right)$, where, $k_p$ is the flow consistency index, $n$ is the behavioural index, $\dot{\boldsymbol{\gamma}}$ indicates the shear rate and is given by $\dot{\boldsymbol{\gamma}} = \left(\nabla \mathbf{v} + \nabla \mathbf{v}^T\right)$ while $|\dot{\gamma}| = \sqrt{\frac{1}{2}\dot{\boldsymbol{\gamma}} : \dot{\boldsymbol{\gamma}}}$ is the second invariant of the strain rate tensor.

Let us consider a 2-D flow with the non-dimensional variables $\breve{x} = x/L$, $\breve{y} = y/H$, $\delta = H/L(<<1)$, $\breve{u} = u/U$, $\breve{v} = v/\delta U$, $\breve{t} = Ut/L$, where $U$ is the characteristic velocity based upon the imposed volume flux and undisturbed channel height, the momentum equations can be written in the long wavelength approximation upto $O(\delta^2)$ as

$$\frac{\partial \tilde{u}}{\partial \tilde{x}}+\frac{\partial \tilde{v}}{\partial \tilde{y}}=0,$$

$$\frac{\partial \tilde{u}}{\partial \tilde{t}}+\tilde{u}\frac{\partial \tilde{u}}{\partial \tilde{x}}+\tilde{v}\frac{\partial \tilde{u}}{\partial \tilde{y}}=-\frac{\partial \tilde{p}}{\partial \tilde{x}}+\frac{1}{\mathrm{Re}}\frac{\partial}{\partial \tilde{y}}\left[\left|\frac{\partial \tilde{u}}{\partial \tilde{y}}\right|^{n-1}\frac{\partial \tilde{u}}{\partial \tilde{y}}\right], \qquad (2.2)$$

$$0=-\frac{\partial \tilde{p}}{\partial \tilde{y}}.$$

Here the variables $\tilde{p}$ and $\mathrm{Re}$ denote the non-dimensional pressure and Reynolds number respectively, and are given by $\tilde{p}=p/\rho U^2$ and $\mathrm{Re}=\dfrac{H}{L}\dfrac{\rho UH}{k_p\left(U/H\right)^{n-1}}$. The boundary conditions associated with the flow are no-slip and kinematic conditions at the channel walls, which can be given in the rigid segment of the channel, (for $-\tilde{L}_1\le \tilde{x}\le 0$ and $1\le \tilde{x}\le 1+\tilde{L}_2$ where $\tilde{L}_{1,2}=L_{1,2}/L$ ),

$$\tilde{u}=0 \text{ and } \tilde{v}=0 \text{ at } \tilde{y}=0,1 \qquad (2.3)$$

while for the flexible region of the channel, we have

$$\tilde{u}=0,\tilde{v}=0 \text{ at } \tilde{y}=0 \text{ and } \tilde{u}=0,\tilde{v}=\frac{\partial \tilde{h}}{\partial \tilde{t}} \text{ at } \tilde{y}=\tilde{h}. \qquad (2.4)$$

Integrating the set of equations (2.2) over the depth of the channel and employing (2.4), we obtained the integrated set of equations

$$\frac{\partial \tilde{q}}{\partial \tilde{x}}+\frac{\partial \tilde{h}}{\partial \tilde{t}}=0,$$

$$\frac{\partial \tilde{q}}{\partial \tilde{t}}+\frac{\partial}{\partial \tilde{x}}\left(\int_0^{\tilde{h}}\tilde{u}^2 d\tilde{y}\right)=-\frac{\partial \tilde{p}}{\partial \tilde{x}}\tilde{h}+\frac{1}{\mathrm{Re}}\left(\left|\frac{\partial \tilde{u}}{\partial \tilde{y}}\right|^{n-1}\frac{\partial \tilde{u}}{\partial \tilde{y}}\right)_{\tilde{y}=0}^{\tilde{y}=\tilde{h}}, \qquad (2.5)$$

where $\tilde{q}=\int_0^{\tilde{h}}\tilde{u}\,d\tilde{y}$ is the non-dimensional axial volume flow rate.

## 2.2 *Base velocity and nonlinear governing equations*

In what follows, the flow is initiated by a fixed upstream volume flux of strength $q_0$ and it is assumed that the flow maintains a self-similar parabolic nature throughout the channel axial position. Accordingly, we may choose the velocity scale in terms of the specified volume flux as $U \sim q_0/H$. This allows us to represent the base state velocity profile within the channel in terms of the dimensionless volume flux and channel height as (please refer to Appendix A):

$$\tilde{u}=\frac{2n+1}{n+1}\frac{\tilde{q}}{\tilde{h}}\left[1-\left|\left(\frac{2\tilde{y}}{\tilde{h}}-1\right)\right|^{\frac{1}{n}+1}\right]. \qquad (2.6)$$

Therefore, we may recast the reduced form of equation (2.5) as

$$\frac{\partial q}{\partial x}+\frac{\partial h}{\partial t}=0$$
$$\frac{\partial q}{\partial t}+\frac{4n+2}{3n+2}\frac{\partial}{\partial x}\left(\frac{q^2}{h}\right)=-\frac{\partial p}{\partial x}h-\frac{2}{\mathrm{Re}}\left(\frac{4n+2}{n}\right)^n\left|\frac{q}{h^2}\right|^{n-1}\frac{q}{h^2} \tag{2.7}$$

which is a nonlinear system involving the two state parameters $q(x,t)$ and $h(x,t)$ (please note that we have dropped the carat over the variables). It must be noted here that, the equation (2.7) is obtained via the approach developed by Shkadov (1970). By choosing a generalized parabolic velocity profile, the momentum equation is integrated to obtain a Saint-Venant type equation. Although this method is extensively used by others (Stewart et al. 2009, 2010; Xu et al. 2013, 2014 ; Xu & Jensen 2015), some inconsistencies has been found in the formulation (Luchini & Charru 2010; Ruyer-Quil & Manneville 1998; Ruyer-Quil & Manneville 2000) by means of quantitative measure. Luchini & Charru (2010) have shown that, the integrated kinetic energy equation gives consistent results. Despite having such inconsistency in the quantitative measure, the integrated momentum equation provides qualitatively similar behaviour of the membrane as that obtained from the kinetic energy formulation. In Appendix H we have given a comparison between these two formulations.

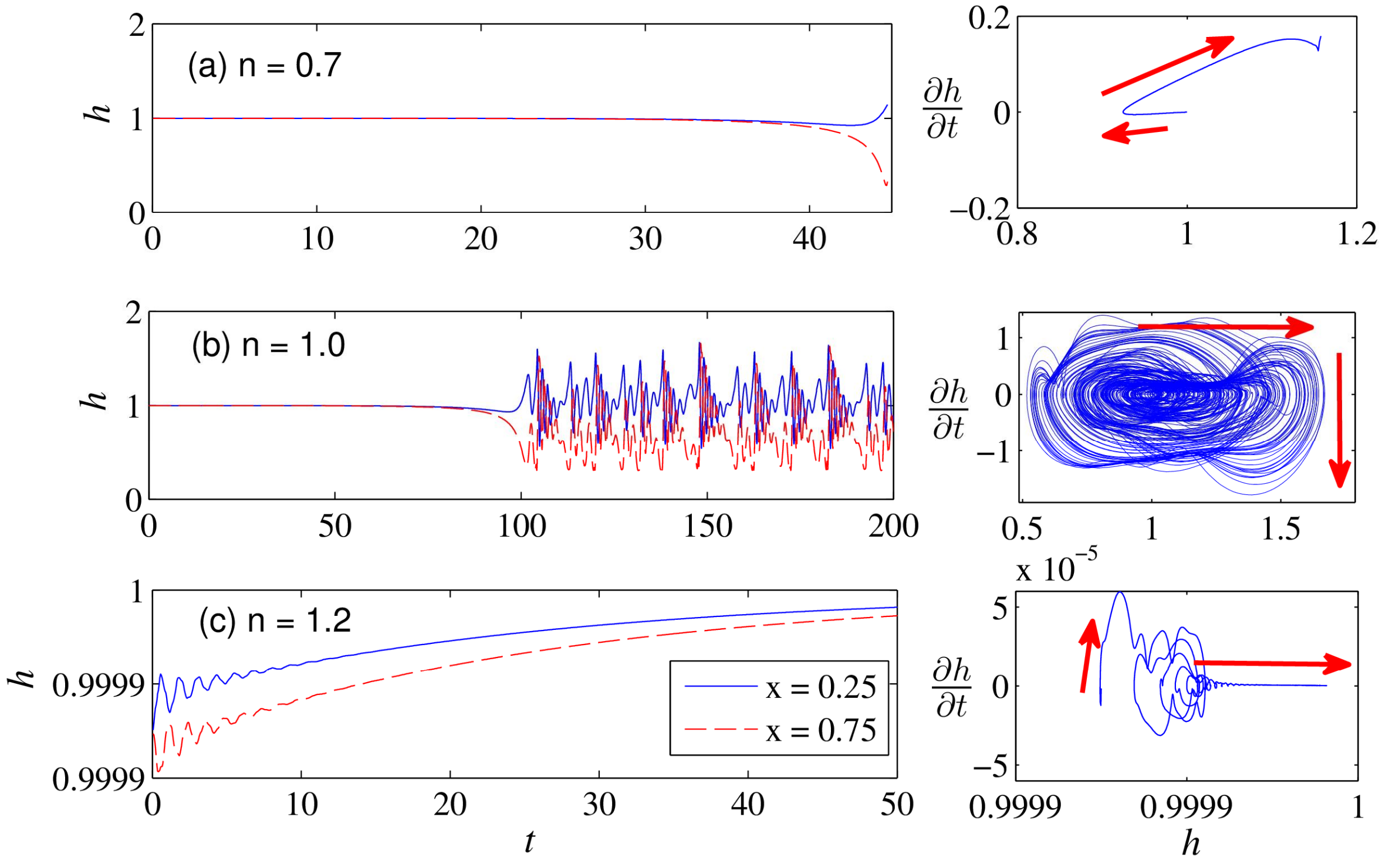

Figure 2: Temporal variation of $h(x,t)$ at two locations $x$ = 0.25 and 0.75 (marked by a solid line and dashed line respectively) for the three cases (a) $n$ = 0.7, (b) $n$ = 1.0 and (c) $n$ = 1.2. The phase portrait for the corresponding dynamics are plotted side by side for comparison for $x$ = 0.25. For brevity, we have plotted the cases for $L_2$ = 1.0, $\mathrm{Re}^{-1}$ = 0.02 and $T$ = 0.08. The red solid lines depict the direction of evolution of the phase portrait. As evident, it is divergent, oscillatory and convergent for the cases (a), (b) and (c) respectively.

In accordance with equation (2.7), a relation between the pressure and the flexible membrane may be obtained from the balance in normal forces between fluid normal stress and the elastic membrane stress which is given by ( Jensen & Heil 2003; Stewart et al. 2009)

$$p = p_e(x) - T\frac{\partial^2 h}{\partial x^2} \tag{2.8}$$

where $T\left(=T_0 H/\rho U^2 L^2\right)$ is the dimensionless tension parameter, $T_0$ is the dimensional tension acting on the membrane and $p_e(x) = \frac{2}{\mathrm{Re}}\left(\frac{4n+2}{n}\right)^n \left(1+L_2-x\right)$. Equation (2.8) is valid under the assumption that the curvature of the membrane is so smooth and the membrane wavelength is sufficiently long so that the curvature $\frac{\partial^2 h}{\partial x^2}\left(1+\delta^2\left(\frac{\partial h}{\partial x}\right)^2\right)^{-3/2} \approx \frac{\partial^2 h}{\partial x^2}$ (for detail description please refer to equation (4.13) of Appendix A). It is also assumed that the bending stiffness is negligible (thickness of the membrane is small compared to channel height) and the normal traction is dominant over the fluid load (for $\delta << 1$ , the fluid load is proportional to 1/Re). Therefore, with the aid of (2.8), we may write equation (2.7) as

$$\begin{gathered}\frac{\partial q}{\partial x} + \frac{\partial h}{\partial t} = 0,\\ \frac{\partial q}{\partial t} + \frac{4n+2}{3n+2}\frac{\partial}{\partial x}\left(\frac{q^2}{h}\right) = Th\frac{\partial^3 h}{\partial x^3} + \frac{2}{\mathrm{Re}}\left(\frac{4n+2}{n}\right)^n \left\{h - \left|\frac{q}{h^2}\right|^{n-1}\frac{q}{h^2}\right\}.\end{gathered} \tag{2.9}$$

with the boundary conditions (equation (4.7) of Appendix A)

$$\begin{aligned}&\text{at}\, x = 0, h = 1; q = 1,\\ &\text{at}\, x = 1, h = 1; T\frac{\partial^2 h}{\partial x^2} = -\left(\frac{\partial q}{\partial t} + \frac{2}{\mathrm{Re}}\left(\frac{4n+2}{n}\right)^n \left\{|q|^{n-1} q - 1\right\}\right)L_2.\end{aligned} \tag{2.10}$$

In figure 2, we depict the time series of the axial variation of the height of the flexible portion of the setup for different power law indices (a) $n$ = 0.7, (b) $n$ = 1.0 and (c) $n$ = 1.2 at two axial locations $x$ = 0.25 and 0.75 (marked as a solid line and dashed line respectively). To the right of each time series, we have depicted the phase portrait of the system at the axial location $x$ = 0.25. For the parameters considered in figure 2, i.e. $\mathrm{Re}^{-1}$ = 0.02 and $T$ = 0.08, we see that (as expected from linear stability analysis in subsection 2.3) shear thickening behaviour leads to linearly stable behaviour while the shear thinning behaviour leads to monotonic instability. For the Newtonian case, the system is expected to be linearly unstable although the full numerical solution depicts that the solution is nonlinearly stable via oscillations. The sustained nature of the oscillations was already highlighted by Xu et al. (2013). The details of the numerical scheme for solving the nonlinear set of equations (2.9) subjected to boundary conditions (2.10) are provided in Appendix B. Briefly, we have employed a spectral Chebyshev collocation method for the spatial discretization with a semi-implicit scheme for the temporal discretization.

The nature of the phase plot for the oscillatory behaviour for a Newtonian fluid (second column, second row of figure 2) is highlighted by the presence of two distinct cycles - the slow oscillation cycle and the fast oscillation cycle (two orbits like structures with one orbit being enclosed within a larger orbit). The slow and fast oscillations are observed in the time series as well. The observed multiscale behaviour in the time series for the Newtonian fluid is observed in the region where linear stability theory predicts unstable behaviour.

### 2.3 *Linear stability analysis of the base state*

We employ a perturbation on $h$ and $q$ around their base state $h=1$ and $q=1$, which is of the form, $h=1+\Re\left[H(x)e^{\sigma t}\right]$, $q=1+\Re\left[Q(x)e^{\sigma t}\right]$, where, $H(x)$, $Q(x)$ are complex functions with $|Q(x)|,|H(x)|\ll 1$ and $\sigma$ implies complex growth rate while $\Re(\bullet)$ denotes the real part of a quantity. Upon substituting the above form in equation (2.9), we obtain the linearized equations as

$$\frac{dQ}{dx}+\sigma H=0,$$
$$T\frac{d^3H}{dx^3}+\frac{4n+2}{3n+2}\left(\frac{dH}{dx}-2\frac{dQ}{dx}\right)+\frac{2}{\mathrm{Re}}\left(\frac{4n+2}{n}\right)^n\{(2n+1)H-nQ\}=\sigma Q, \tag{2.11}$$

where the boundary conditions are also suitably modified to yield

$$H=0, Q=0; \quad \text{at } x=0$$
$$H=0,\ T\frac{d^2H}{\partial x^2}=-\left(\sigma Q+\frac{2}{\mathrm{Re}}\left(\frac{4n+2}{n}\right)^n nQ\right)L_2; \quad \text{at } x=1 \tag{2.12}$$

For the inviscid limit $\mathrm{Re}^{-1}=0$, the static inviscid eigenmodes are obtained by substituting $\sigma=0$ in equation (2.11) along with the pertinent boundary conditions from equation (2.12). It must be noted that the so-called inviscid modes are the limit in which we represent the limit of $\mathrm{Re}^{-1}\to 0$. It does not represent the truly inviscid regime in which the PDE is no more second-order but only first-order. Proceeding further, we obtain $Q(x)=0$ and the equation for the membrane as

$$T\frac{d^3H}{dx^3}+\frac{4n+2}{3n+2}\frac{dH}{dx}=0, \tag{2.13}$$

subjected to $H(0)=0, H(1)=0, \left.\frac{d^2H}{dx^2}\right|_{x=1}=0$.

The solution to the above equation may be obtained as

$$H(x)=C\sin\left(\sqrt{\frac{1}{T}\frac{4n+2}{3n+2}}x\right), \tag{2.14}$$

where $C$ is an arbitrary constant the eigenvalues are obtained as

$$T_{k0}=\left(\frac{4n+2}{3n+2}\right)\frac{1}{k^2\pi^2} \quad \text{for } k=1,2,... \tag{2.15}$$

The dependence of the inviscid eigenvalues on the power-law index is apparent from the above expression. The discrete eigenvalues decay quadratically for higher modes while the deviation around the Newtonian mode ($n$ = 1) is decided by the prefactor. In particular, for a shear-thinning fluid, it is seen that the inviscid eigenvalue lies to the left (smaller value of the tension parameter) of the eigenvalue of the Newtonian one and vice versa. Physically, these eigenvalues represent the value of the tension parameter *T*, for which a stable oscillation occurs on the membrane. The *k*-th eigenmode gives *k* number of oscillations in the membrane or the function $|H(x)|$ has *k* humps. In these modes, the volume flow rate is always unity, whereas the membrane shape deforms according to $H(x)$. To maintain a fix volume flow rate at each cross-section, a pressure-gradient will be created due to the membrane deformation which normalizes the advective acceleration of the flow.

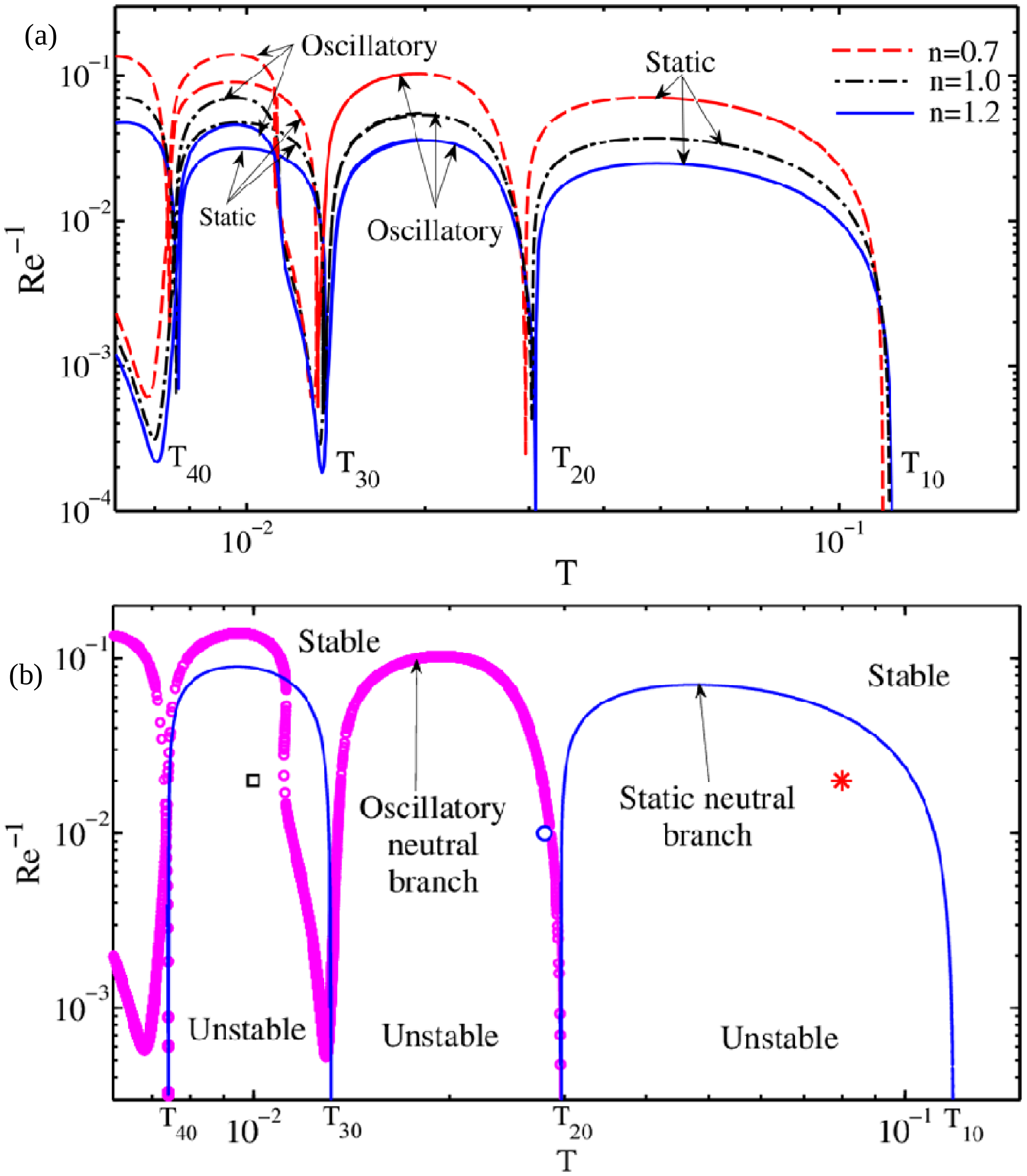

Figure 3: (a) Neutral curves obtained numerically through the solution of the eigenvalue problem (2.11) in the $\left(\mathrm{Re}^{-1}, T\right)$ space for different power-law indices and $L_2 = 1$. We have depicted three neutral curves in the figure - two static eigenmodes $\sigma = 0$ and one oscillatory eigenmode $\Re(\sigma) = 0$. The two static modes are connected by the oscillatory mode. (b) The static and oscillatory branches of the neutral curve for $n = 0.7$ are depicted for better clarity. The static branches are from $T_{10}$ to $T_{20}$ and from $T_{30}$ to $T_{40}$ (solid blue lines), whereas the oscillatory branch of the neutral curve emerges from $T_{20}$ and $T_{40}$ (magenta coloured circles). The static branch emerges from $T_{30}$ coexist with the oscillatory branch from $T_{20}$. When traversing along the neutral curve in the decreasing $T$ direction, the parametric space lies on the left of the neutral curves are unstable and on the right are stable.

In figure 3(a) we depict the stability structure of the physical problem in the $\left(\mathrm{Re}^{-1}, T\right)$ space by determining the locations where the eigenvalue obtained via numerical solution of equations (2.11) subjected to (2.12) becomes $\sigma = 0$ or $\Re(\sigma) = 0$, denoting the case of static eigenmodes or oscillatory eigenmodes respectively. It must be kept in mind that $\Re(\sigma) = 0$ need not necessarily mean a stable oscillation. It only implies that the oscillations are stable for the linearized dynamics (Hartman-Grobman theorem). The leading order eigenmodes as obtained through equation (2.15) are the inviscid limits of the eigenvalue problem. The neutral curves depicted here are in excellent agreement with the results obtained by Xu et al. (2013). The details of the numerical procedure for solving the eigenvalue problem using Chebyshev polynomials and obtaining the stability curves are described in Appendix C. The influence of the power-law index is quite dramatic in the region of the elastic parameter $T$ lying in the region midway between two leading order eigenmodes $T_{k0}$ and $T_{k+1,0}$, especially at relatively low Re - i.e. the region where the viscous effects are prominent. It is also worth noting that in the region of the inviscid eigenmodes, $T_{k0}$, the influence of the power-law index is relatively less prominent, especially at high Reynolds numbers - thereby highlighting the weak dependence of the fluid rheology near the inviscid eigenmodes. The shift in the inviscid limit of the eigenmodes (there appears to be a crossover in the neutral curves for certain value of $\mathrm{Re}^{-1}$ in the vicinity of $T_{10}$, the neutral curve for $n = 0.7$ crosses the neutral curve for $n = 1.0$, as the elastic parameter decreases), is attributed to the contribution from the momentum influx $\frac{\partial}{\partial \breve{x}}\left(\int_0^{\breve{h}} \breve{u}^2 d\breve{y}\right)$ as noted in equation (2.5). The marked change in the behaviour between cases for a fixed elasticity parameter and Reynolds number is therefore due to the power-law index. Near $T_{30}$, it is interesting to note that there are two different branches of the neutral curve - the first being the static mode while the other is the imaginary mode. We shall discuss later about the nature of bifurcations observed at the edges of these neutral curves. It is seen that for shear thickening fluid, the neutral curves uniformly shift to higher Re; in the case of shear thinning fluid the opposite trend is seen.

Figure 3(b) depicts the neutral stability curve for $n$ = 0.7. The static branch and the oscillatory branch are identified with a blue solid line and magenta circles. The static branches are from $T_{10}$ to $T_{20}$ and $T_{30}$ to $T_{40}$, whereas the oscillatory branch emerges from $T_{20}$ and coexist with the static branch from $T_{30}$. The influence of the power-law index is quite apparent and follows the same trend as mentioned above. The magnitude of the neutral eigenvalues within the region $T_{20}$ and $T_{30}$, have magnitude less than unity and beyond that region the maginudes are of greater than unity. This can be also seen from the eigenspectrum near $T_{30}$ (Figure 5). As $T \rightarrow T_{10}$ and $\mathrm{Re}^{-1} \rightarrow 0$, the membrane has the high frequency mode-1 oscillation, which supposed to induces a periodic axial sloshing motion within the flow. Jenesen & Heil (2003) pointed out that, because of this sloshing, an inflow (outflow) will occur at the upstream (downstream) of the channel if the pressure is specified at the channel outlet. This eventually disturbs the base flow structure and gives rise to oscillations in the membrane, which can be suppressed by imposing a fixed upstream flux or keeping the downstream segment sufficiently large (Liu et al. 2012; Peter S. Stewart et al. 2010; Stewart et al. 2009). With a fixed upstream pressure, the instantaneous difference between the fluxes at the downstream and upstream caused by the viscous resistance and the inertial effect, promotes the extraction of kinetic energy from the mean flow by the oscillatory membrane. With a fixed upstream flux, the mode-2 oscillation is identified as the primary cause of oscillatory instability (Liu et al. 2012). In the present formulation, for certain combination of parametric values the membrane may undergo a sustained oscillation, reminiscent of a series of slamming events from the divergent state (Xu et al. 2013), which appears from the interactive dynamics of two steady modes (figure 4). At $\left(\mathrm{Re}^{-1}, T\right) = \left(0.02, 0.08\right)$, the linear analysis predicts the membrane at mode-1 is divergently unstable (figure 3(a)) for $n = 0.7, 1.0$ (the point lies below the neutral curve) and stable for $n = 1.2$, (the point lies above the neutral curve) whereas from figure 2 one can see the membrane is divergently unstable for $n = 0.7$, and stable for $n = 1.2$, but for $n = 1.0$ it is oscillatory unstable. As $T$ decays further, there is smooth exchange from mode-1 to mode-2 oscillation, and becomes unstable as it crosses $T_{20}$. In between $T_{20}$ and $T_{30}$, the mode-2 oscillation changes to mode-3 oscillation and co-exist with a small amplitude static mode-3 oscillation.

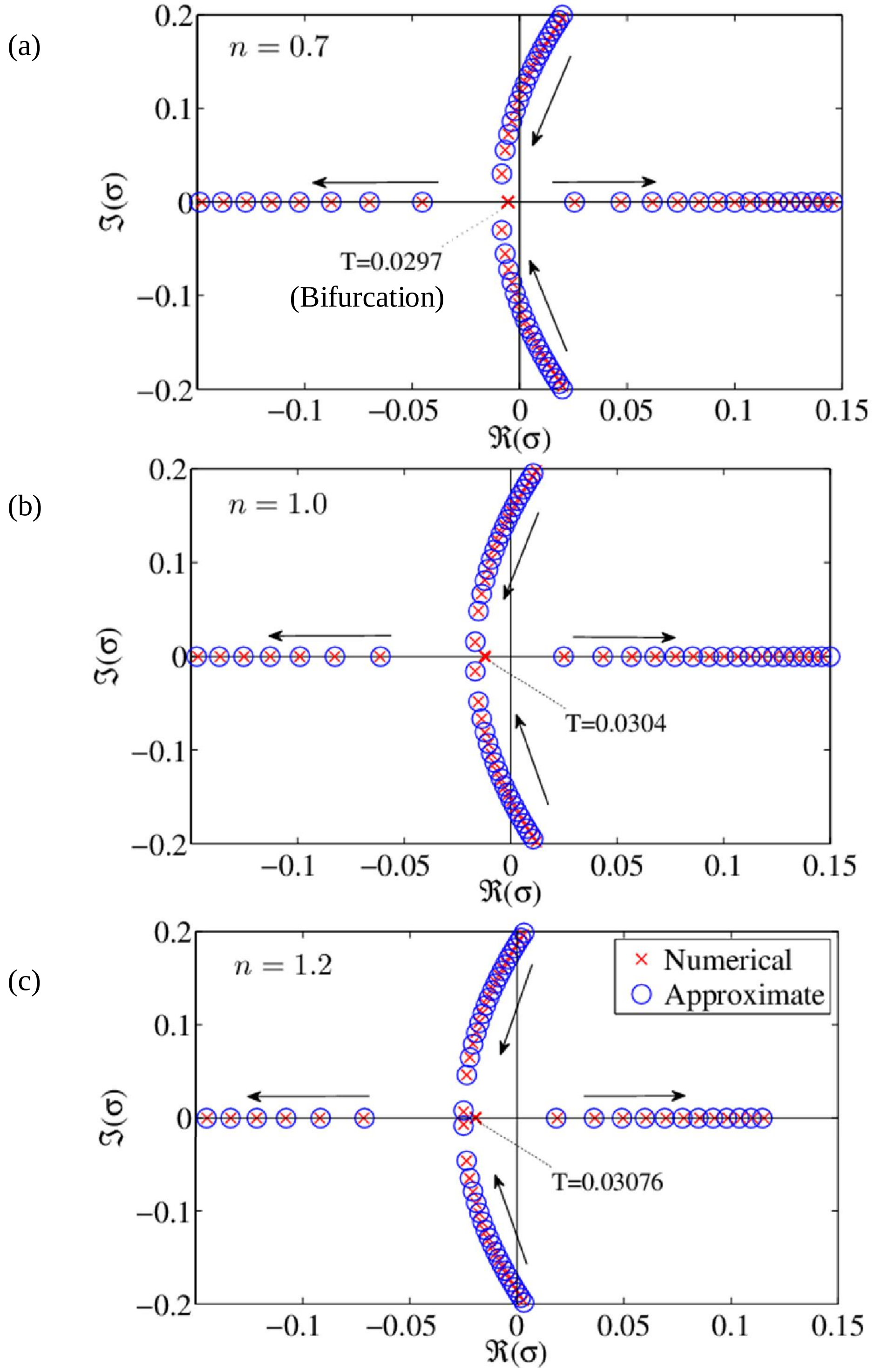

Figure 4: Evolution of eigenvalues as the elastic parameter increases, the value of $Re^{-1}$ is taken as $10^{-3}$ and $L_2 = 1$. The arrows denote the evolution of the eigenvalues as $T$ increases from 0.029 to 0.031. The asymptotic solution is denoted by the óoô symbol while the numerical solution for the linearized eigenvalue problem is denoted by ó×ô symbol. The three cases denoted here are (a) $n$ = 0.7, (b) $n$ = 1.0 and (c) $n$ = 1.2. The approximate value of $T$ at the bifurcation point for each curve is $T_{20}$.

Let us focus on the nature of the solutions observed near the eigenvalue $T_{20}$. In figure 4, we depict the evolution of the eigenvalues in the neighbourhood of $T_{20}$ as the elastic parameter $T$ varies from 0.029 to 0.031 for (a) $n$ = 0.7, (b) $n$ =1.0 and (c) $n$ = 1.2. As first demonstrated by Xu et al. (2013), as the elastic parameter $T$ is increased from a value below $T_{20}$, the pair of complex conjugate eigenvalues with positive real part gradually decreases to purely imaginary values approximately at $T_{20} - \frac{(17n+8)(9n+5)}{32(2n+1)\pi^4}\left(\frac{4n+2}{n}\right)^n \mathrm{Re}^{-1}$ (equation (2.27)) and becomes complex eigenvalues with negative real part till the value of $T$ reaches $T_{20}$, where they merges to a stable static eigenvalue. In contrast, when the parameter $T$ goes beyond the value $T_{20}$, there appears to be two negative real eigenvalues, one of which increased to zero as the value of $T$ reaches approximately to $T_{20} + \frac{15n}{64\pi^4}(3n+2)\left(\frac{4n+2}{n}\right)^{2n+1} \mathrm{Re}^{-2}$ (equation (2.24)), and becomes positive while the other remain as real and negative. From which one may identify the point $\left(\mathrm{Re}^{-1}, T\right) = \left(0, T_{20}\right)$ as a Takens-Bogdanov point. The fundamental structure of the eigenvalues remains qualitatively same as observed in figure 4. We see that as the fluid becomes more shear thickening in nature, the stability of the merged eigenvalue is enhanced (large negative values) which corroborates with the observations in figure 3 which predicts stable dynamics for shear thickening fluids in the region where the system is linearly unstable. These trends help us in asserting the fact that for shear thickening fluids, the system tends to be more stable compared to the Newtonian counterpart.

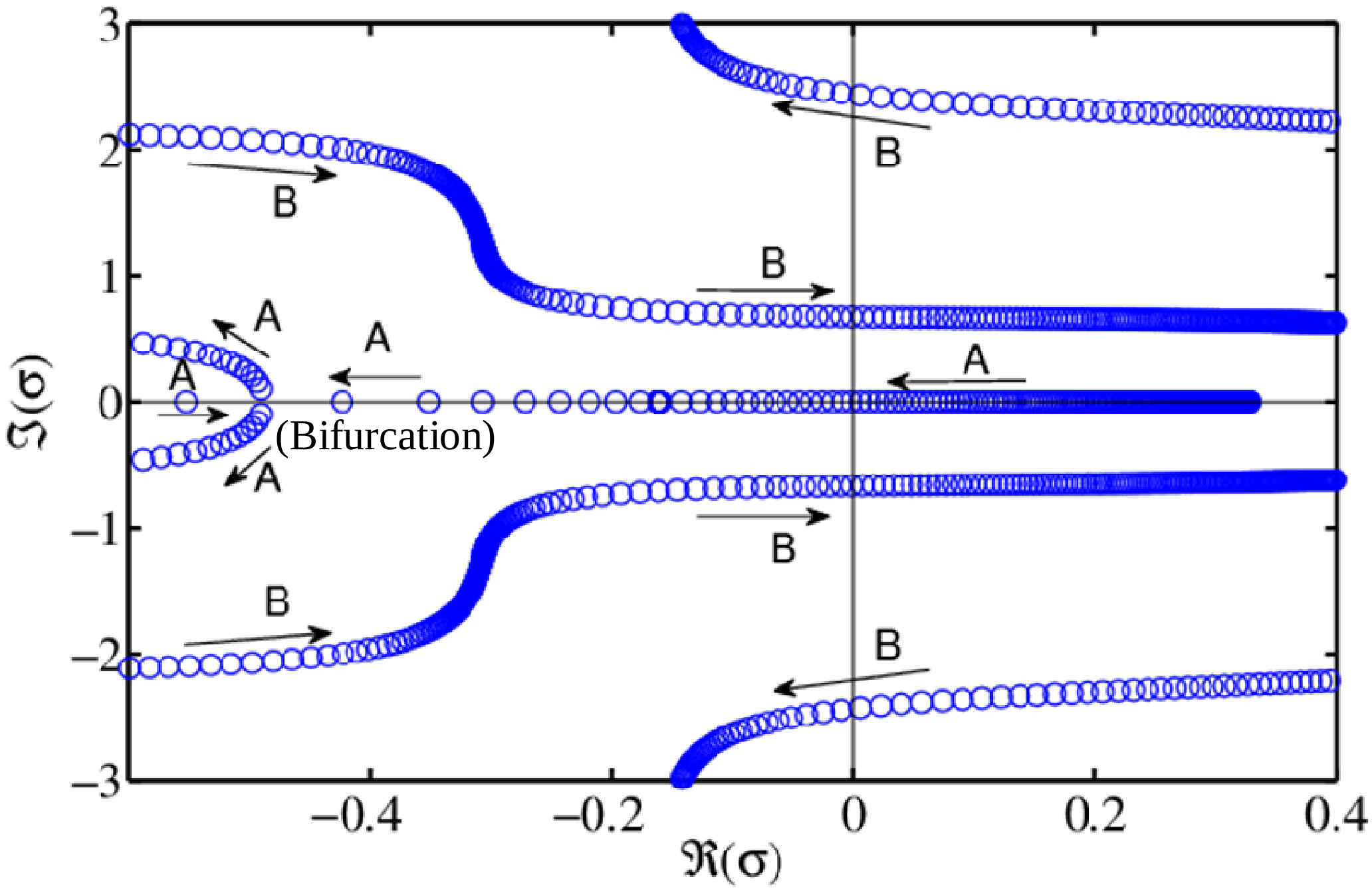

Figure 5: The eigenvalue spectrum for $Re^{-1}$=0.03, $n$ = 1, $L_2 = 1$ and $T$ varies from 0.01 to 0.02 ( near $T_{30}$ ). The arrow denotes the evolution of eigenvalues with increasing $T$, which indicates the existence of Hopf bifurcation points (where eigenpath crosses the x-axis).

In Figure 5, the eigenvalue spectrum near the point $T_{30}$ is plotted for a Newtonian fluid ( $n = 1.0$ ) and the other parametric values are considered as $\mathrm{Re}^{-1} = 0.03$ and $L_2 = 1$. Near $T_{30}$, the system has one static eigenvalue together with a conjugate pair of oscillatory or Hopf eigenvalue. With an increasing elastic parameter, $T$, the static unstable eigenvalue (denoted by óAô in figure 5) decays through the static neutral point and becomes stable. Meanwhile, the pair of oscillatory stable eigenvalue (denoted by óBô in figure 5) passes through the oscillatory neutral Hopf points and becomes unstable upto a certain extent and again it decays through the large frequency oscillatory Hopf mode and becomes stable. We have not shown the nature of the eigenvalue spectrum for the shear thickening and thinning fluids since the nature of the variations remains unchanged.

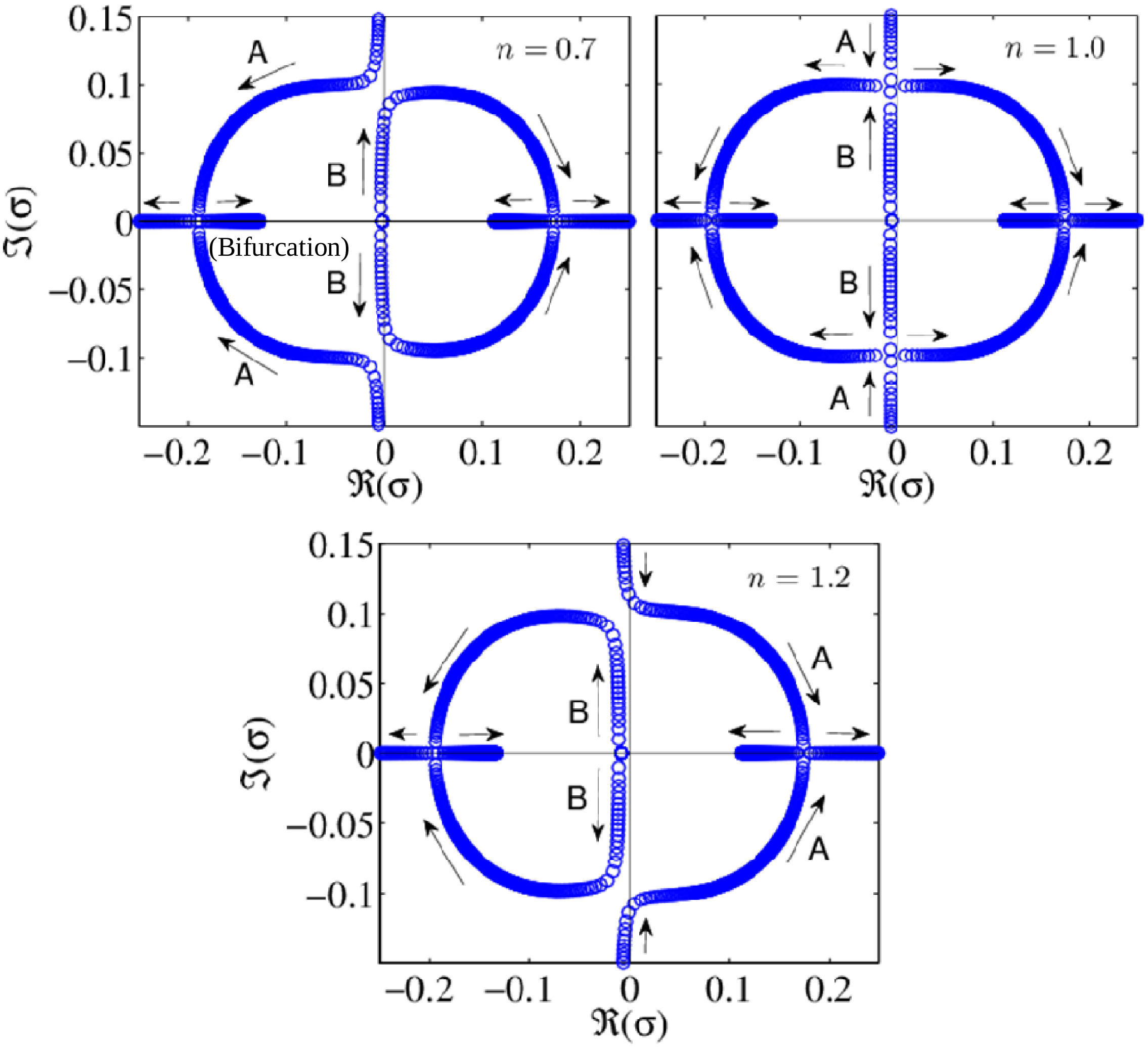

Figure 6: The eigenvalue spectrum for $L_2$=100, $\text{Re}^{-1} = 4.27\times10^{-4}$ and $(T_{20} - T)$ varies from $10^{-6}$ to $10^{-3}$. For $n = 0.7$, the eigenvalues which have almost equal negative real part, ones with the larger imaginary part (denoted by A) remains stable whereas others with the smaller imaginary part (denoted by B) becomes unstable through a Hopf point. For $n = 1.0$, those eigenvalues coalesces before parting away. In case of $n = 1.2$, the eigenvalues with larger imaginary part becomes unstable through a Hopf point whereas the eigenvalues with smaller imaginary part remains stable.

Figure 6 depicts the eigenvalue spectrum near the parameter $T_{20}$ for larger downstream segment compared to membrane length. Near the point $\left(\text{Re}^{-1}, T\right) \approx \left(0, T_{20}\right)$, in all three cases the system (2.11)-(2.12) has two pairs of conjugate eigenvalues with almost equal negative real part. As $(T_{20} - T)$ varies from $10^{-6}$ to $10^{-3}$, the eigenvalues with larger imaginary part (denoted by óAô in figure 6) decreases whereas the eigenvalues with smaller imaginary part (denoted by óBô in figure 6) increases until they are on the verge of a coalescence. Then, the type óBô eigenvalues ascends through a Hopf point (which is a point on the oscillatory neutral curve) with increasing real part and parted away with two positive real eigenvalues after coalescing at the imaginary axis. The type óAô eigenvalues descends through increasing negative real part and becomes stable with two negative real eigenvalues. For the shear thickening fluid, an opposite trend occurs. The type óAô eigenvalues becomes unstable through a Hopf point whereas the type óBô eigenvalues remains stable. In the case of a Newtonian fluid, both the type óAô and óBô coalesces at a point and then parted away, one towards the positive $\Re(\sigma)$ axis and becomes unstable and the other towards the negative $\Re(\sigma)$ axis and becomes stable. The phenomenological significance, which is given by Xu et al. (2014), is that there is a 1:1 resonance due to the self-excited oscillation of the elastic membrane. Since, the behaviour of shear thinning (or shear thickening) of a fluid is reminiscent to that of a fluid flowing with a lower (or higher) viscosity, the variation in the eigenvalue spectrum shows similar behaviour as that of the variation in the eigenvalues with respect to the Reynolds number which is elucidated by Xu et al. (2014). When the downstream rigid segment is longer than the membrane length, the disturbance at the upstream due to the mode-1 oscillation is suppressed, but the axial sloshing become important at the downstream. This sloshing motion resulting in pressure drops within the channel and provide extra mode of oscillation which promotes resonance within the flow. For a shear-thinning (shear-thickening) fluid this sloshing will be more (less) due to the less (larger) viscous resistance. So, this resonance will grow faster (slower) for a shear-thinning fluid and make the system more unstable and leads to a chaotic disturbance (sub-section 3.3).

### 2.4 *Linear stability near the inviscid limit*

We first proceed towards understanding the nature of the stability by means of asymptotic expansions about the so-called inviscid region, i.e. the infinite Reynolds number region. The base state derived in subsection 2.3 underlines the basis for the analysis in the neighbourhood of the inviscid eigenvalues which are determined by equation (2.15). Considering the parametric expansions for each of the variables $H(x)$, $Q(x)$, $\sigma$, $\mathrm{Re}^{-1}$ and $T$ near the static inviscid eigenvalues wherein the base states are now given by $H_0(x) = \sin(k\pi x)$, $Q_0(x) = 0$, $\sigma = 0$, $\mathrm{Re}^{-1} = 0$ and $T_k = T_{k0}$ (please refer to the subsection 2.3) as

$$\begin{aligned}
H(x) &= H_0(x) + \varepsilon H_1(x) + \varepsilon^2 H_2(x) + \varepsilon^3 H_3(x) + \ldots \\
Q(x) &= Q_0(x) + \varepsilon Q_1(x) + \varepsilon^2 Q_2(x) + \varepsilon^3 Q_3(x) + \ldots \\
T_k &= T_{k0} + \varepsilon T_{k1} + \varepsilon^2 T_{k2} + \varepsilon^3 T_{k3} + \ldots \\
\mathrm{Re}^{-1} &= \varepsilon R_1 + \varepsilon^2 R_2 + \varepsilon^3 R_3 + \ldots \\
\sigma &= \varepsilon \sigma_1 + \varepsilon^2 \sigma_2 + \varepsilon^3 \sigma_3 + \ldots
\end{aligned} \tag{2.16}$$

where $\varepsilon$ is a small perturbation parameter.

Substituting equation (2.16) in equation (2.11) and (2.12), we obtain various order equations for $Q(x)$ and $H(x)$. Collecting various powers in $\varepsilon$, we obtain the following differential equations (written below in a matrix form)

$$\begin{aligned}
\mathbf{L}\boldsymbol{\Phi}_i &= \mathbf{A}_i \\
\mathbf{B}_1\boldsymbol{\Phi}_i(0) = \mathbf{0},\ &\mathbf{B}_2\boldsymbol{\Phi}_i(1) = \mathbf{G}_i
\end{aligned} \tag{2.17}$$

where the linear operator $\mathbf{L}$ is denoted by $\mathbf{L} \equiv \mathbf{L}_D \frac{\partial}{\partial x} + \mathbf{L}_I$, $\mathbf{\Phi}_i = \left( Q_i, H_i, \frac{\partial H_i}{\partial x}, \frac{\partial^2 H_i}{\partial x^2} \right)^T$ denotes the vector containing the relevant variables, while $\mathbf{A}_i = \left( A_{1i}, 0, 0, A_{2i} \right)^T$ and $\mathbf{G}_i = \left( 0, 0, 0, G_i(1) \right)^T$ represent the inhomogeneous part of the differential equation and boundary condition respectively and are defined in Appendix D.

Towards finding a consistent solution, we must employ the solvability conditions on equation (2.17) (Nayfeh 2011). This is done by employing the adjoint operator of equation (2.17), which is given by (Xu et al. 2013)

$$\begin{gathered} \mathbf{L}^* \mathbf{\Psi} = \mathbf{0} \\ \mathbf{B}_1^* \mathbf{\Psi}(0) = \mathbf{0},\ \mathbf{B}_2^* \mathbf{\Psi}(0) = \mathbf{0} \end{gathered} \tag{2.18}$$

where $\mathbf{L}^* \equiv -\mathbf{L}_D^T \frac{\partial}{\partial x} + \mathbf{L}_I^T$ , $\mathbf{B}_1^*$ and $\mathbf{B}_2^*$ are given in Appendix D.

As per the definition of the adjoint operator in (2.18), the term $\left\langle \mathbf{\Phi}^T, \mathbf{L}^* \mathbf{\Psi} \right\rangle$ in the integral $\left\langle \mathbf{\Psi}^T, \mathbf{L\Phi} \right\rangle = \left\langle \mathbf{\Phi}^T, \mathbf{L}^* \mathbf{\Psi} \right\rangle + \mathbf{\Psi}^T \mathbf{L}_D \mathbf{\Phi} \Big|_{x=0}^{x=1}$ vanishes automatically and the boundary-condition operators are chosen in such a way that the integral satisfy the solvability condition

$$\left\langle \mathbf{\Psi}^T, \mathbf{A}_i \right\rangle = \mathbf{\Psi}^T \mathbf{L}_D \mathbf{\Phi}_i \Big|_{x=0}^{x=1} \tag{2.19}$$

In particular, for $k = 1$, from equations (2.17) to (2.19) we may obtain the first solvability condition as (Appendix D)

$$T_{11} = -\frac{8}{\pi^4} \left[ \sigma_1 \frac{4n+2}{3n+2} + (2n+1) \left( \frac{4n+2}{n} \right)^n R_1 \right] \tag{2.20}$$

At first order dependence on $\varepsilon$, a prediction for the eigenvalue $\sigma$ for small $\mathrm{Re}^{-1}$ and $T \to T_{10}$, can be obtained from equation (2.20) as

$$\sigma \approx -\frac{3n+2}{2} \left[ \left( \frac{4n+2}{n} \right)^n \mathrm{Re}^{-1} + \frac{\pi^4}{8(2n+1)} \left( T - T_{10} \right) \right] \tag{2.21}$$

whereas a branch of the static neutral curve (where $\sigma = 0$) may be obtained as

$$T \approx T_{10} - \left( \frac{4n+2}{n} \right)^n \frac{8(2n+1)}{\pi^4} \mathrm{Re}^{-1} \tag{2.22}$$

For $k = 2$ the first solvability condition yields $T_{21} = 0$. This gives $\left( T - T_{20} \right)$ is of $O(\varepsilon^2)$. Therefore, $\left( T - T_{20} \right)$ is either of $O(R_1^2)$ or of $O(\mathrm{R}_2)$. Choosing $\mathrm{Re}^{-1} = \varepsilon R_1$ and $\sigma = \varepsilon \sigma_1$, the second solvability condition yields

$$T_{22} = \frac{3}{16\pi^4} \left( \frac{17n+8}{3n+2} \sigma_1^2 + (18n+10) \left( \frac{4n+2}{n} \right)^n R_1 \sigma_1 + \frac{5n}{4} (3n+2) \left( \frac{4n+2}{n} \right)^{2n+1} R_1^2 \right) \tag{2.23}$$

from which we may obtain a static branch of the neutral curve near the point $\left( T_{20}, 0 \right)$ as

$$T = T_{20} + \frac{15n}{64\pi^4} (3n+2) \left( \frac{4n+2}{n} \right)^{2n+1} \mathrm{Re}^{-2} \tag{2.24}$$

Using the solvability conditions $T_{21} = 0$ and (2.23) the third solvability condition gives

$$T_{23} = \frac{3}{32(3n+2)\pi^4}\begin{bmatrix} 3(2n+1)\alpha^3 R_1^3 + 24\sigma_1^3 + 4(17n+8)\sigma_1\sigma_2 \\ +2\alpha^2 R_1(2n+1)\left\{5R_2 + 9R_1\sigma_1\right\} \\ +4\alpha\left\{(9n+5)\left(R_2\sigma_1 + R_1\sigma_2\right) + 9(2n+1)R_1\sigma_1^2\right\} \end{bmatrix} \tag{2.25}$$

where $\alpha = \left(\frac{4n+2}{n}\right)^n (3n+2)$. With the help of equation (2.25), the other branch of the neutral curve can be found by assuming $\mathrm{Re}^{-1} \approx \varepsilon^2 R_2$ and $\sigma = \varepsilon\sigma_1 + \varepsilon^2\sigma_2$, which implies an approximation for the eigenvalues as

$$\begin{aligned}\sigma \approx &-\frac{(3n+2)}{(17n+8)}\left[(9n+5)\left(\frac{4n+2}{n}\right)^n \mathrm{Re}^{-1} + 32\pi^4\frac{(2n+1)}{(17n+8)}(T - T_{20})\right] \\ &\pm 4\pi^2\sqrt{\frac{(3n+2)}{3(17n+8)}(T - T_{20})}\end{aligned} \tag{2.26}$$

This implies, the second branch of the neutral curve emerges when

$$T = T_{20} - \frac{(17n+8)(9n+5)}{32(2n+1)\pi^4}\left(\frac{4n+2}{n}\right)^n \mathrm{Re}^{-1} \tag{2.27}$$

whereas the leading order eigenvalue for the oscillatory solutions are obtained as

$$\sigma \approx \pm 4\pi^2 i\sqrt{\frac{(3n+2)}{3(17n+8)}\left(T_{20} - T\right)} \tag{2.28}$$

The approximate eigenvalues that are obtained in equation (2.21) are the points that are lies over the imaginary $\Im(\sigma) = 0$ in figure 4. The eigenvalues given by equation (2.26) gives the complex conjugate points where the Hopf points are given by (2.28). The neutral curves obtained from equation (2.24) and (2.27) are shown in figure 14(a).

The ratio of the spatially averaged kinetic energy of the fluid within the flexible membrane to a rigid membrane is given by $\kappa = \int_0^1 \left(q^2 / h\right) dx$. In figure 7 we have plotted this ratio for different values of $\mathrm{Re}^{-1}$ and $T$, keeping the downstream segment as comparable to the membrane length and the fluid as Newtonian. The values of the parameters are considered in such a way that the point $(\mathrm{Re}^{-1}, T)$ lies below the neutral curve (please refer to figure 3). The three points considered in the $(\mathrm{Re}^{-1}, T)$ domain are $(0.02, 0.08)$, $(0.01, 0.028)$ and $(0.02, 0.01)$, which are marked in figure 3(b) by the ó red asterix ($*$)ô, óblue open circle (O)ô and óblack open square (□)ô symbols, respectively . Form the neutral stability curve (linear analysis) (figure 3(b)) one can identify the parameter set $(0.02, 0.08)$ to yield a system evolution through (2.9)-(2.10) to have an unstable static mode-1 solution which may grow to a static mode-2 solution. Similarly, at $(0.01, 0.028)$ and $(0.02, 0.01)$, the system has an oscillatory unstable mode-2 and mode-3 solution, respectively.

To emphasize on these observations, in figure 7 we depict the ratio $\kappa$ with respect to time within the region where the parameter $T$ lies between $T_{10}$ and $T_{20}$. As $t$ progresses, the kinetic energy becomes periodic via a series of static mode-1 ($t=60$) to mode-2 ($t=66.6, 71.2$) oscillations (the shape of the membrane is shown in the inset of each figure for some time instants and the corresponding ratio is marked by the black dots; modes are defined according to the number of humps seen in the membrane shape $h(x,t)$). In the other two parametric situations, $(0.01, 0.028)$ and $(0.02, 0.01)$, we observe that the membrane oscillates at much higher modes which lead to an eventual divergence. The observed trends must arise from the balance of the stabilizing viscous stresses and the destabilizing influence of the membrane oscillations. Next, we shall attempt to isolate these terms so as to explicitly depict the relative nature of the aforementioned terms.

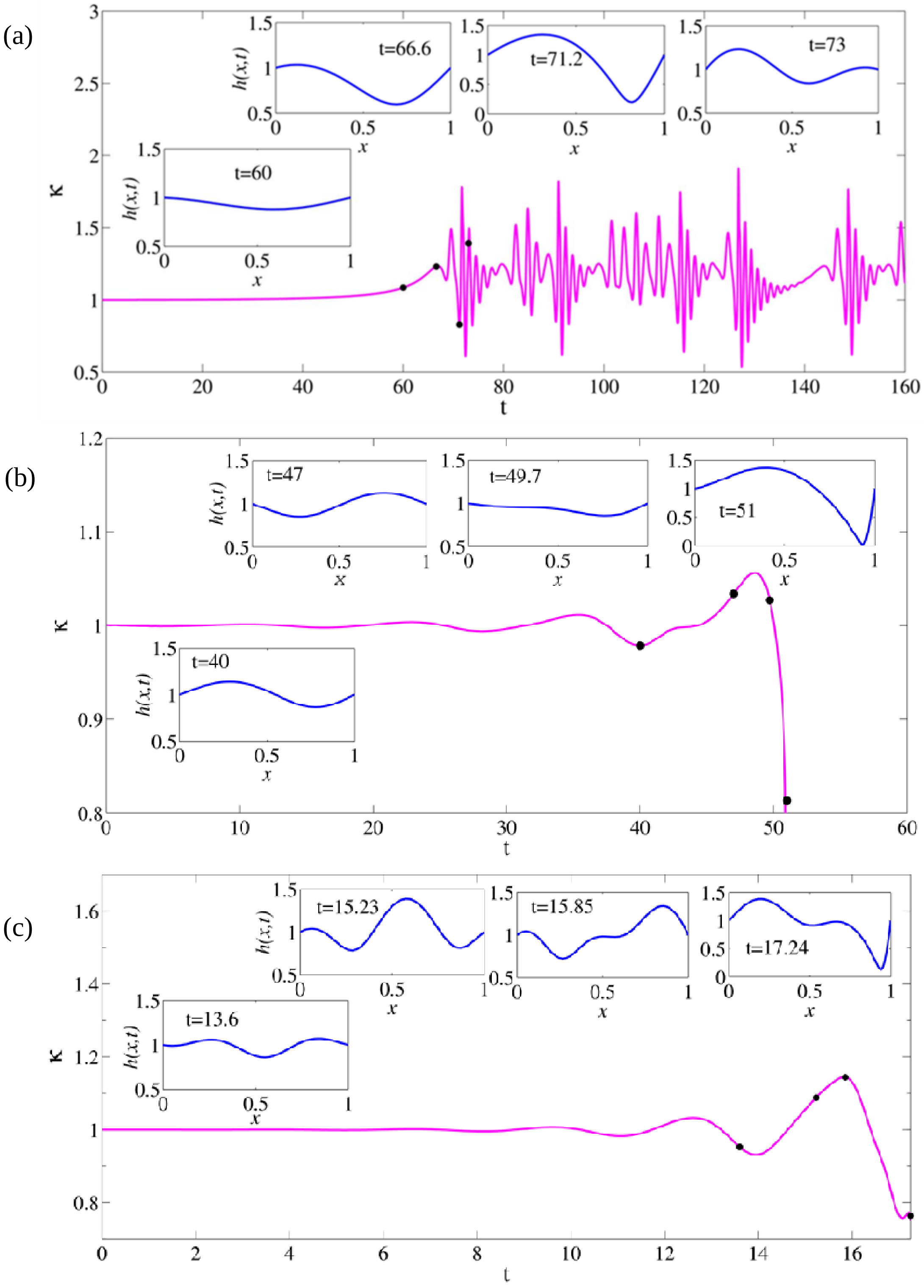

Figure 7: Temporal evolution of the ratio ($\kappa$) of the spatially averaged kinetic energy for a flexible membrane to the rigid membrane with respect for (a) $\mathrm{Re}^{-1} = 0.02$, $T = 0.08$, (b) $\mathrm{Re}^{-1} = 0.01$, $T = 0.028$, and (c) $\mathrm{Re}^{-1} = 0.02$, $T = 0.01$. The other parametric values are taken as $L_2 = 1$, $n = 1$. In the inset, the membrane profiles at different time instants are depicted and the corresponding kinetic energy ratio are shown by black dots on the $\kappa$ vs $t$ evolution. The insets from left to right represent the state for the dots in a chronological order.

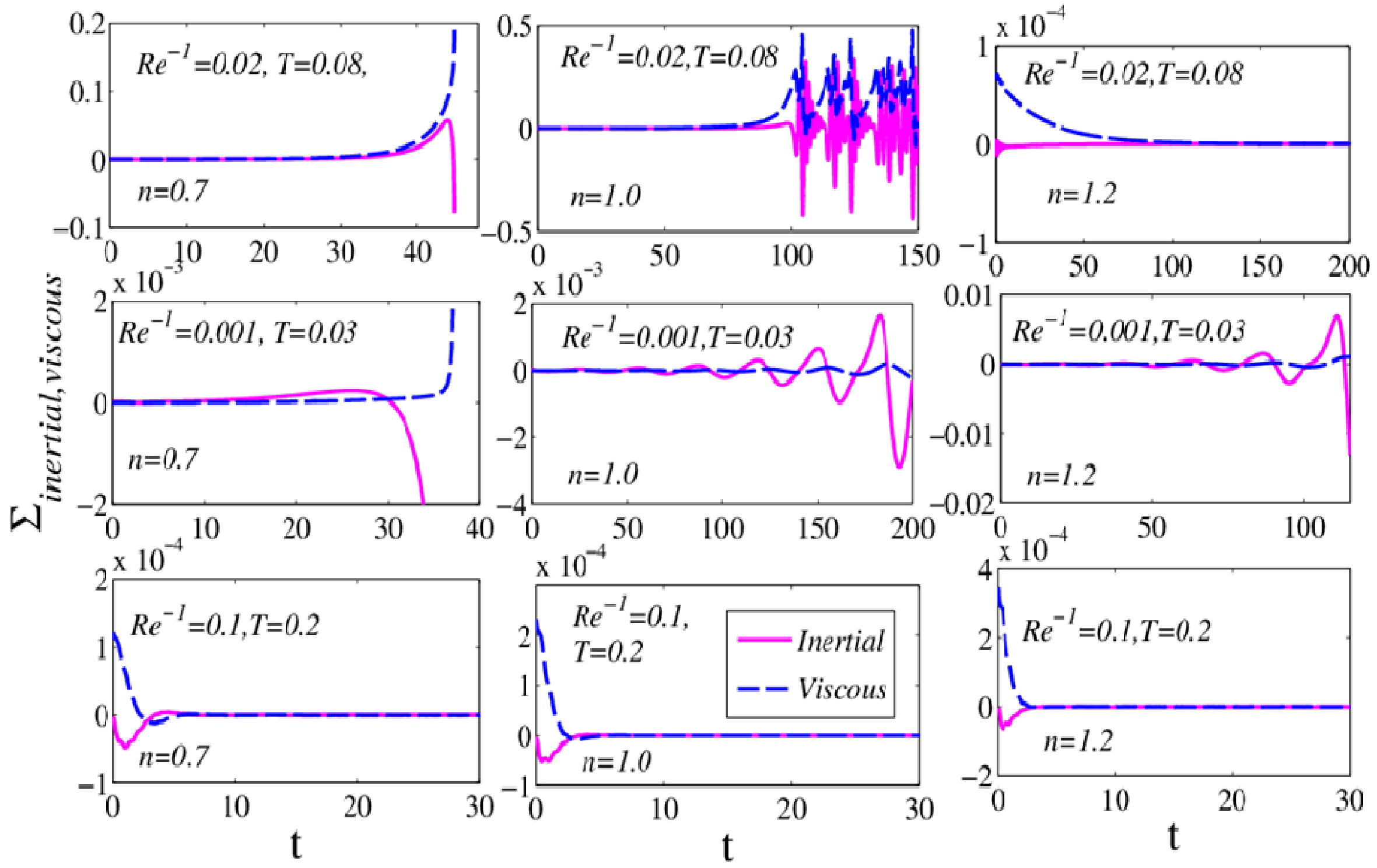

Figure 8: Spatially averaged inertial term, $\sum_{inertial}(t)$ and viscous term, $\sum_{viscous}(t)$, of equation (2.9) are plotted with respect to time for different values of the $\mathrm{Re}^{-1}$, $T$ and $n$. *1st row*: $\mathrm{Re}^{-1}=0.02$, $T=0.08$; *2nd row*: $\mathrm{Re}^{-1}=0.001$, $T=0.03$; *3rd row*: $\mathrm{Re}^{-1}=0.1$, $T=0.2$. *1st column*: $n=0.7$; *2nd column*: $n=1.0$; *3rd column*: $n=1.2$. The length of the downstream rigid segment $L_2=1$.

In this case, we track the temporal growth rate of spatially averaged inertial and viscous term in equation (2.9), i.e., $\Sigma_{inertial}(t)=\int_0^1\left[\frac{4n+2}{3n+2}\frac{\partial}{\partial x}\left(\frac{q^2}{h}\right)\right]dx$ and, $\Sigma_{viscous}(t)=\int_0^1\left[2\,\mathrm{Re}^{-1}\left(\frac{4n+2}{n}\right)^n\left(h-\left|\frac{q}{h^2}\right|^{n-1}\frac{q}{h^2}\right)\right]dx$. In figure 2 we have depicted the temporal evolution of the two terms for various parametric values (the figures in the first row correspond to $\mathrm{Re}^{-1}=0.02$, $T=0.08$; those in the second row correspond to $\mathrm{Re}^{-1}=0.001$, $T=0.03$ while the third row corresponds to $\mathrm{Re}^{-1}=0.1$, $T=0.2$. Values in different columns correspond to different values of $n$ - the first column corresponds to a shear thinning, second column corresponds to a Newtonian while the third column corresponds to a shear thickening fluid). It is clear that for the parametric values of $\mathrm{Re}^{-1}=0.02$, $T=0.08$, the shear thinning fluid depicts the scenario where the system (2.9)-(2.10) admits an unstable solution. In this situation, it is seen that the inertial and viscous growth rates ($\Sigma_{inertial}(t)$ and $\Sigma_{viscous}(t)$) diverge from each other (refer to the case $\mathrm{Re}^{-1}=0.02$, $T=0.08$, $n=0.7$ of figure 8). As the power law index is increased, i.e. the influence of the viscous stresses increases, it is observed that the behaviour changes from oscillatory to completely dissipative. This change in behaviour is quite obviously related to the fact that the viscous stresses are able to compensate the destabilizing influence of the membrane oscillation. Moreover, as we move towards higher values of the membrane tension parameter, we observe that the oscillations and instability cease to exist and yield only a decaying mode with the magnitude of the spatially averaged viscous stress larger than the inertial term. Quite obviously, the higher value of the membrane tension parameter, *T*, implies that there is a higher reluctance of the membrane to oscillate and remain horizontal. This causes the initial perturbation to die down quickly.

When *T* is chosen in the region between $T_{20}$ and $T_{30}$, the kinetic energy diverges via the rapid mode-2 oscillation of the membrane. During the course of rapid oscillation the membrane undergoes a sharp constriction before it diverges ($t \approx 51$, in the inset of figure 7(b)). A similar characteristic of the membrane (but with mode-3 oscillations) is also found in the case, when a value of $(\mathrm{Re}^{-1}, T)$ considered from the region between $T_{30}$ and $T_{40}$ (figure 7(c)). Within this parametric regime, the system (2.9)-(2.10) is oscillatory unstable. The growth in $\sum_{viscous}(t)$ is small compared to $\sum_{inertial}(t)$ but has the same frequency and is out of phase (refer to the case $\mathrm{Re}^{-1} = 0.001$, $T = 0.03$, $n = 1.0$ of figure 8). With increasing time, the magnitude of inertial term dominates over the viscous term and grows rapidly to an unstable situation.

The approximate temporal growth of the system (2.9)-(2.10) may also be characterized by the evolution of eigenvalues of the linearized system (2.11)-(2.12). At $(\mathrm{Re}^{-1}, T) = (0.02, 0.08)$, from figure 3(b), the linear analysis shows a static divergence of the system, but also the existence of a mode-2 solution, since the point lies below the neutral curve from $T_{10}$ to $T_{20}$. The numerical computation shows (figure 2) this static divergence grows with time to an oscillatory unstable solution. In case of $(\mathrm{Re}^{-1}, T) = (0.01, 0.028)$, the mode-2 oscillation grows in magnitude and diverges. At that point, the system (2.11)-(2.12) gives a pair of eigenvalue with $\Re(\sigma) > 0$, from which one can emphasize that as both the function $h(x,t)$ and $q(x,t)$ grows according to the rate $e^{\sigma t}$, which diverges with time, the ratio $\kappa$ which is represented by $(q^2/h)$ is also diverges and is seen from figure 7(b). Now, at the point $(\mathrm{Re}^{-1}, T) = (0.02, 0.01)$, a similar nature will follow, but the system exhibit an unstable mode-3 solution. At this value of $(\mathrm{Re}^{-1}, T)$ the system (2.11)-(2.12) has one unstable static mode and a pair of unstable oscillatory mode, which describes the divergence of mode-3 oscillation. For a shear-thinning fluid ($n = 0.9$) the eigenvalue corresponding to the unstable static mode have larger magnitude, hence showing faster divergence (figure 8). When we consider a shear-thickening fluid ($n = 1.05$), this static eigenvalue decreases in magnitude and becomes negative for $n = 1.2$, which indicates a stable behaviour of the membrane as well as in the kinetic energy ratio and can be seen from figure 9.

(a)

(b)

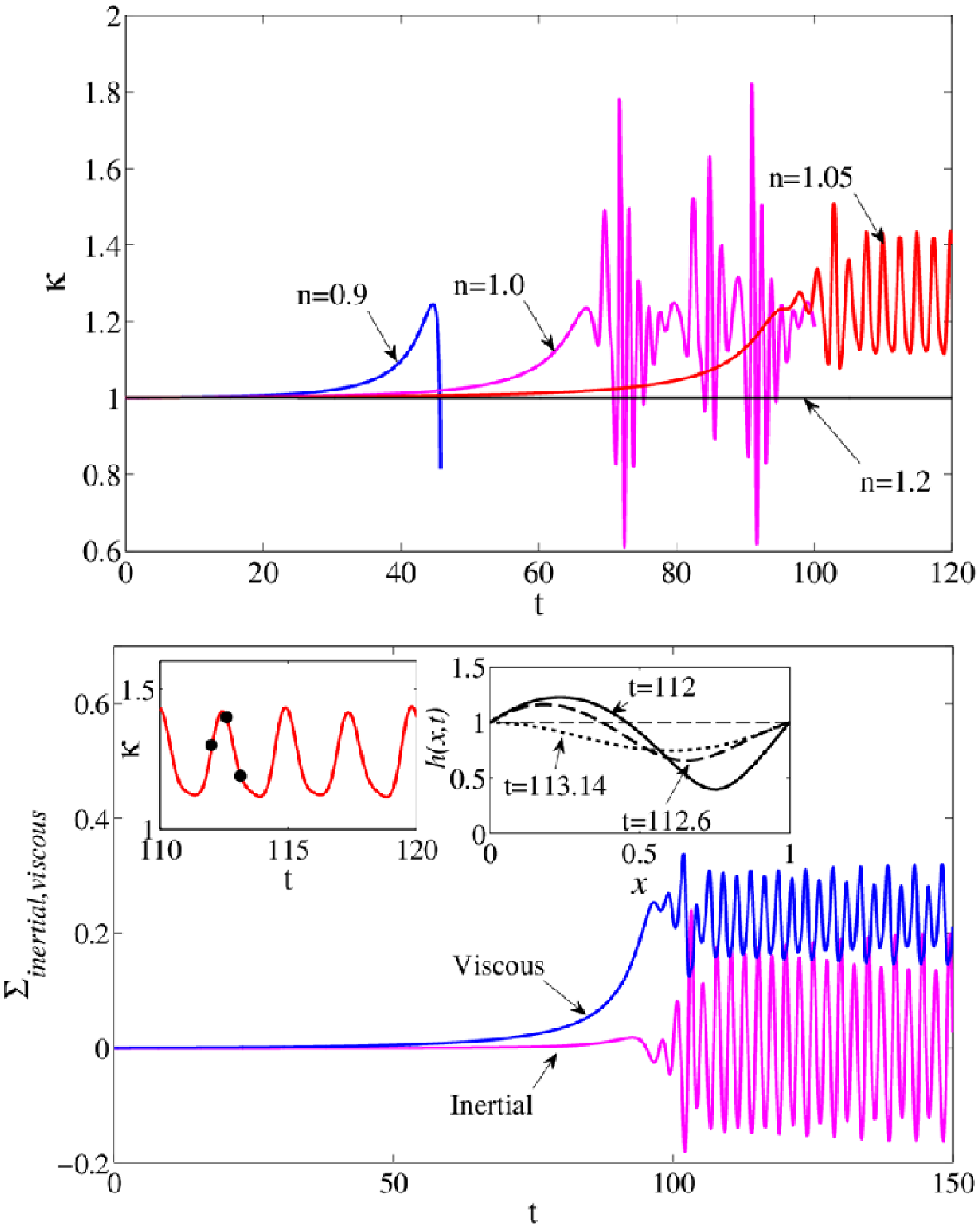

Figure 9: (a) The ratio ($\kappa$) of the spatially averaged kinetic energy of a flexible membrane to the rigid membrane for different power-law indices ($n = 0.9$, $n = 1.0$, $n = 1.05$ and $n = 1.2$). Other parametric values are chosen as $\mathrm{Re}^{-1} = 0.02$, $T = 0.08$, $L_2 = 1$. The portion of the graph from $t = 100$ to $t = 120$ for $n = 1$ is discarded for clarity of presentation. (b) Spatially averaged inertial term, $\sum_{inertial}(t)$ and viscous term, $\sum_{viscous}(t)$, of equation (2.9) are plotted with respect to time for $n = 1.05$. In the inset, the kinetic energy ratio $\kappa$ and the membrane shape $h(x,t)$ at different times are shown.

In figure 9 we have depicted the spatially averaged kinetic energy ($\kappa$) ratio for different power-law indices, $n$ (figure 9(a)) and the corresponding spatially averaged inertial and viscous term (figure 9(b)). For shear-thinning fluids ($n = 0.9$), the ratio diverges after a certain time. At that point of time, the kinetic energy shoots up because of the constriction at the membrane which almost touches the rigid wall on the other side making the gap $h(x,t) \to 0$ (a similar situation as in the inset of figure 7(b) for $t = 51$). In case of a

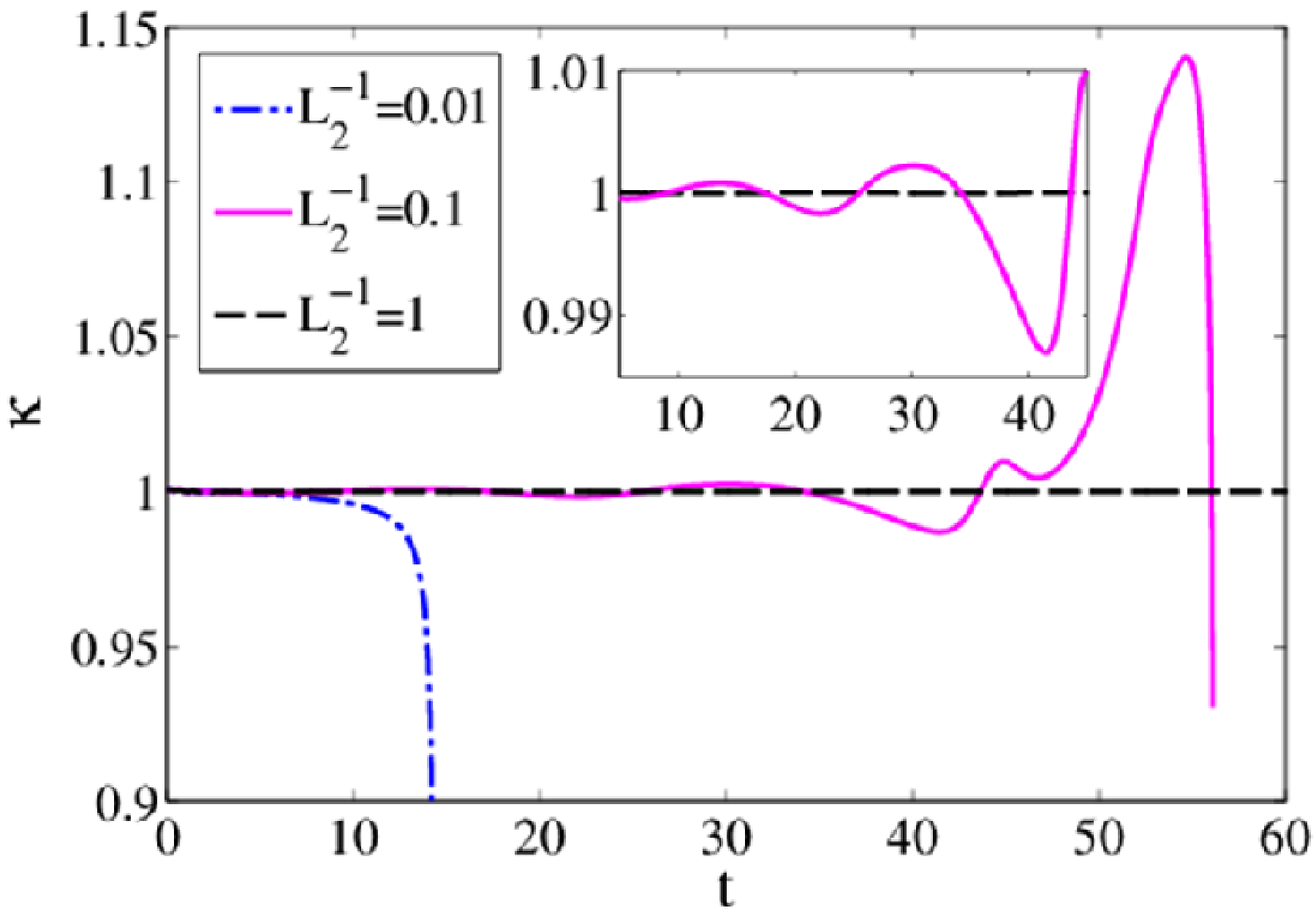

Figure 10: The ratio ($\kappa$) of the spatially averaged kinetic energy for a flexible membrane to the rigid membrane for different sizes of the downstream segment $L_2$. Other parametric values considered are $\mathrm{Re}^{-1} = 0.01$, $T = 0.03$ and $n = 1.2$. With the increasing downstream segment length $L_2$, the kinetic energy ratio diverges very fast. In the inset, the expanded view of the plot for $L_2^{-1} = 0.1$ is shown for clarity.

Newtonian fluid ($n = 1.0$), rapid oscillations are observed in the figure 9(a). These oscillations maintain a periodic nature, undergoing a shift from mode-1 to mode-2 (figure 7(a)). But in this case no sharp constriction in the shape of the membrane is observed for those parametric values.

When we consider a shear-thickening fluid (consider $n$ = 1.05 in figure 9), these rapid oscillations become uniform and stabilize to the case where the membrane undergoes a periodic change from static mode-1 to static mode-2 behaviour (refer to membrane profile $h(x,t)$ in the inset of figure 9(b)) with a relatively lower frequency as compared to $n = 1.0$. The quantities $\Sigma_{inertial}(t)$ and $\Sigma_{viscous}(t)$ also show similar behaviour (figure 9(b)) with viscous and inertial effect balancing out each other. As we further increase the power-law index, the ratio gradually reaches to $\kappa \to 1$ (for $n = 1.2$ in figure 9(a)), as if the membrane is rigid.

To elucidate the effect of larger downstream segment towards the evolution of the system dynamics, the kinetic energy ratio $\kappa$ is depicted in the figure 10 for various lengths of the downstream segments. We have considered the case of a shear-thickening fluid ($n = 1.2$) and the point on the $\left(\mathrm{Re}^{-1}, T\right)$ domain is $(0.01, 0.03)$ (marked by magenta colored ‘O’ symbol in figure 14(b)). From the figure one can see that the kinetic energy ratio is almost constant when the downstream rigid segment is comparable to the membrane length ($L_2^{-1} = 1$), whereas in case of a larger downstream segment ($L_2^{-1} = 0.1$) it diverges after going through some oscillation (corresponds to mode-2 oscillation of the membrane) which grows in magnitude with time (in the inset of figure 10). When the downstream segment is much longer than the membrane length ($L_2^{-1} = 0.01$), the value of the ratio $\kappa$ diverges rapidly. In fact, this can also be seen from figure 14(b). The point $(0.01, 0.03)$ lies above the neutral curve for $L_2 = 1$ (as denoted by the solid blue line in figure 14(b)) therefore indicating the stable behaviour of the membrane. But for $L_2 = 10$ and 100, the same point lies below the neutral stability curve (as denoted by blue colored ó×ô and ó•ô symbol, respectively, in figure 14(b)) which indicates an unstable behaviour of the membrane. The significance of the downstream segment is to allow for the pressure to equalize over the length $L_2$ relative to the length of the flexible channel, 1, since the pressure condition at the exit is prescribed.

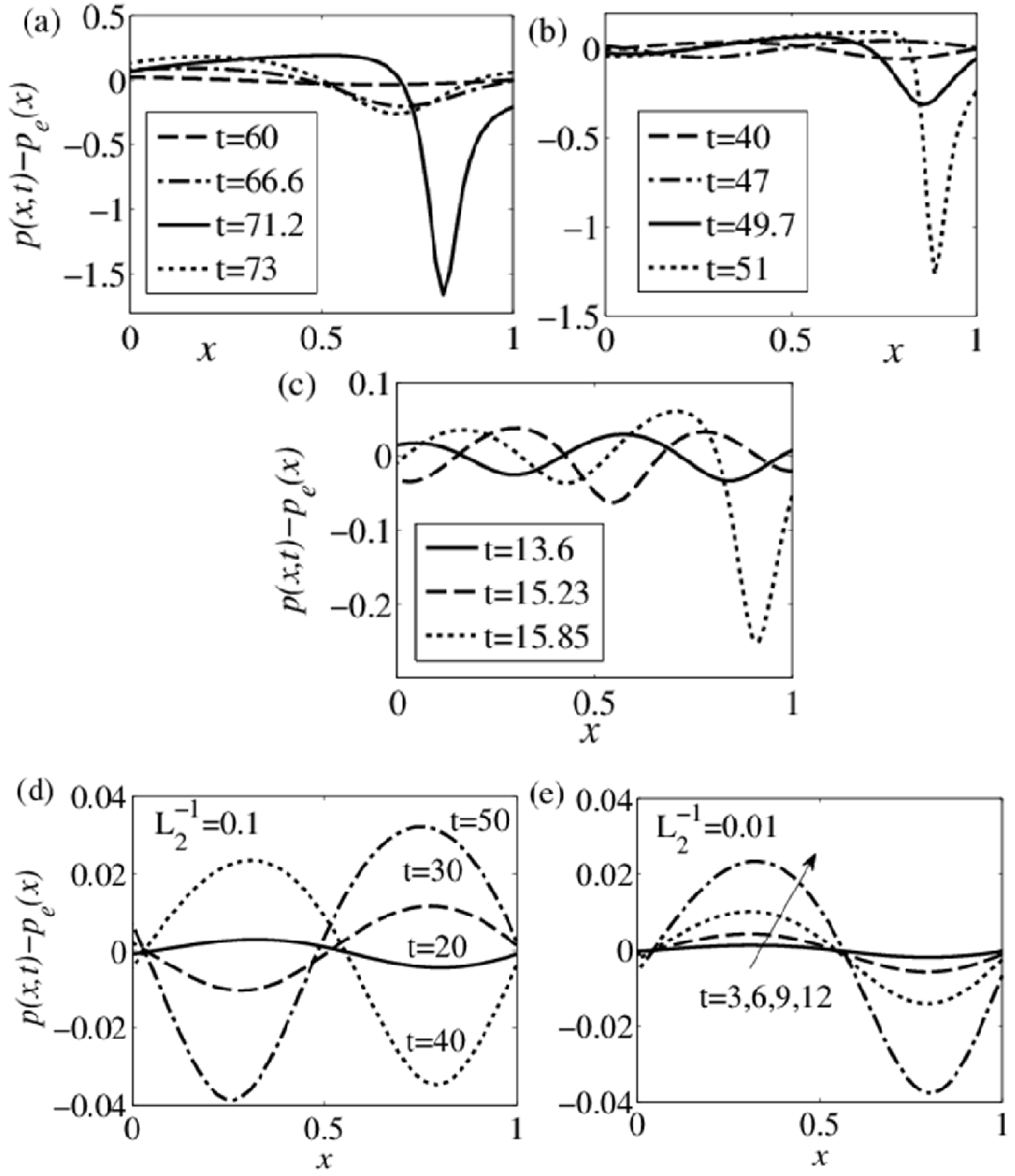

Figure 11: The pressure variation $\left(p - p_e\right)$ across the channel axial position for (a) $\mathrm{Re}^{-1} = 0.02,\ T = 0.08$; (b) $\mathrm{Re}^{-1} = 0.01,\ T = 0.028$; (c) $\mathrm{Re}^{-1} = 0.02$, $T = 0.01$. The other parametric values are taken as $L_2^{-1} = 1$, $n = 1$. The time instants are taken according to the figure 7(a-c). Pressure variation in the case of $n$ = 1.2, for (d) $L_2^{-1} = 0.1$ and (e) $L_2^{-1} = 0.01$.

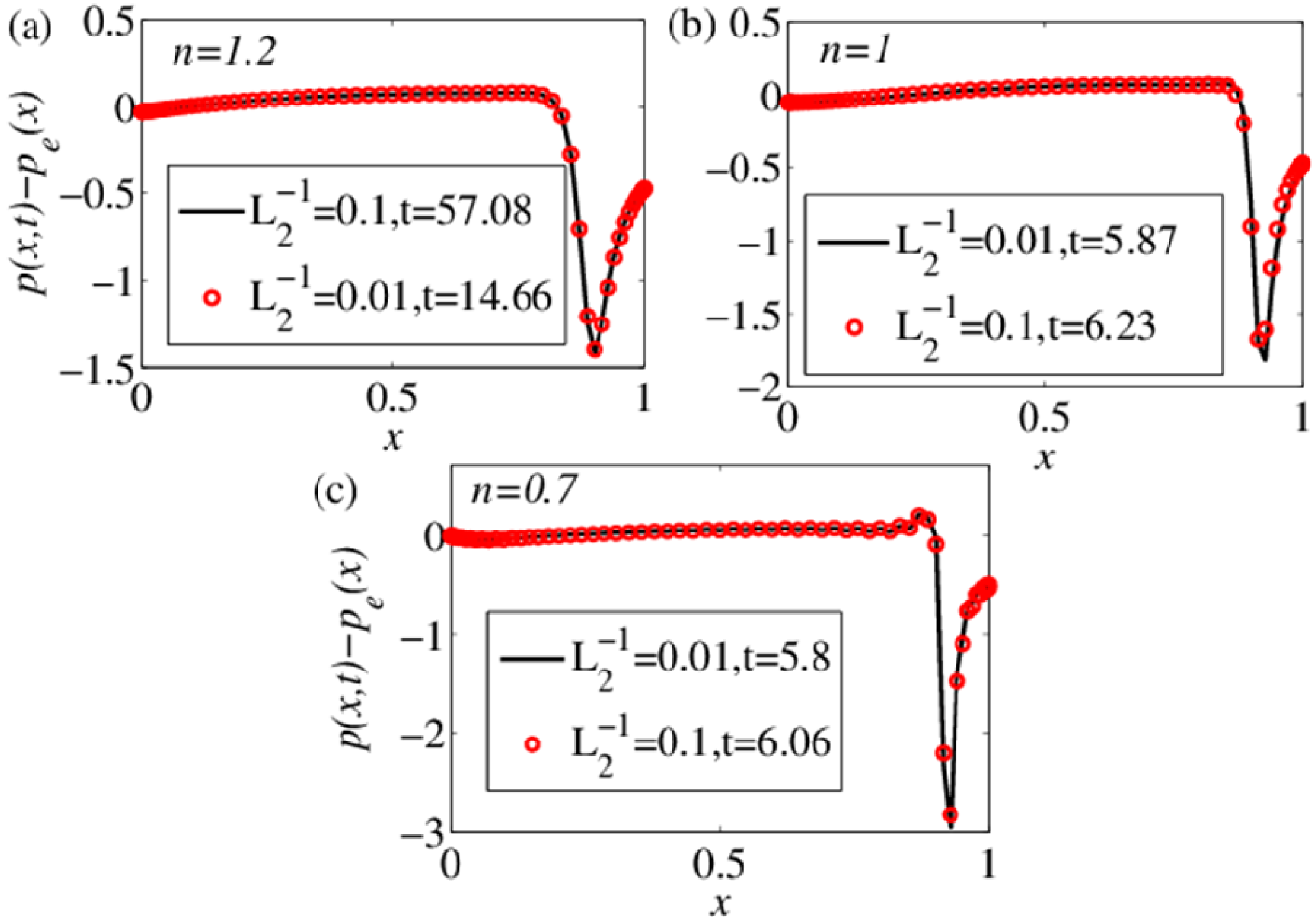

Figure 12: (a) The pressure variation $\left(p - p_e\right)$ across the channel axial position just before the collapse the pressure has similar kind of distribution for both the cases $L_2^{-1} = 0.1$ and $L_2^{-1} = 0.01$. The time taken to reach this state is more ($t = 57.08$) for $L_2^{-1} = 0.1$ compared to $t = 14.66$ for $L_2^{-1} = 0.01$. The instability arises via a growing oscillatory mode-2 amplitude for $L_2^{-1} = 0.1$, whereas it is the growth of a static mode-2 amplitude for $L_2^{-1} = 0.01$. Pressure distribution in the channel just before divergence for $L_2^{-1} = 0.1$ and $L_2^{-1} = 0.01$ for (b) $n$ = 1 and (c) $n$ = 0.7. The similarity in the mode-2 profile for the pressure distribution is observed.

In figure 11 we depict the pressure variation along the channel for the parametric value of (a) $\mathrm{Re}^{-1} = 0.02$, $T = 0.08$, (b) $\mathrm{Re}^{-1} = 0.01$, $T = 0.028$ and (c) $\mathrm{Re}^{-1} = 0.02$, $T = 0.01$; at various time instants for the case of a Newtonian fluid. Physically, it is the excess internal pressure relative to the external pressure (which is the contribution of the normal stress) is responsible for the motion of the membrane. We see that for figure 11(a), we are in the situation where the parametric range yields an oscillatory solution. This is despite the fact that there is a sharp decrease in the local pressure in the vicinity of the wall. The fact is that despite the strong pressure decay near the exit, the viscous forces are able to dampen out the influence of the membrane oscillation. In the other two cases, (b) and (c), we may readily see that the pressure falls to large values relative to the base pressures in the channel right before the system becomes unstable. The main difference in the two figures happens to be the form of the axial pressure variation; for the situation (b) $\mathrm{Re}^{-1}=0.01$, $T=0.028$, the parametric point of operation lies between the modes $T_{10}$ and $T_{20}$ while for the case (c) $\mathrm{Re}^{-1}=0.02$, $T=0.01$, we are in between the modes $T_{30}$ and $T_{40}$. This reflects in the fact that for case (b), we observe a pressure profile which resembles a mode 2 profile (with 2 extrema) while for case (c), we observe a pressure profile which resembles a mode 4 profile (with 4 extrema). Despite the change in the nature of the profile, it is observed that the event just before divergence is marked by a sharp decrease in the luminal pressure just before the exit. In order to ascertain this universality in the even right before the instability, we depict the pressure variation in the channel for various values of the power law index in figure 12 and for $L_2^{-1}=0.1$ and $L_2^{-1}=0.01$ . It is clearly seen that the event just before divergence has a similar structure for the different constitutive behaviours. It may be concluded that the sharp drop in the pressure is the main reason for the divergence to occur; this allows us to apriori make design adjustments for such kinds of flexible channel to rigid channel connections.

## 3. Nonlinear analysis

### 3.1 *Multiscale (weakly nonlinear) analysis: Hopf bifurcation*

We revisit the considerations of the stability of (2.9) considering a multiscale analysis. This is motivated by the fact that such structures are apparent in the full scale numerical solutions of the governing equations. Considering a multi-time scale expansion for $h$, and $q$ and their partial derivatives of the form

$$
\begin{gathered}
h(x;\tau_0,\tau_1,\tau_2)=1+\varepsilon H_0(x;\tau_0,\tau_1,\tau_2)+\varepsilon^2 H_1(x;\tau_0,\tau_1,\tau_2)+\varepsilon^3 H_2(x;\tau_0,\tau_1,\tau_2)+\cdots \\
q(x;\tau_0,\tau_1,\tau_2)=1+\varepsilon Q_0(x;\tau_0,\tau_1,\tau_2)+\varepsilon^2 Q_1(x;\tau_0,\tau_1,\tau_2)+\varepsilon^3 Q_2(x;\tau_0,\tau_1,\tau_2)+\cdots \\
\mathrm{Re}^{-1}=\varepsilon R_1+\varepsilon^2 R_2+\varepsilon^3 R_3+\cdots \\
T=T_{k0}+\varepsilon T_{k1}+\varepsilon^2 T_{k2}+\varepsilon^3 T_{k3}+\cdots
\end{gathered}
\tag{3.1}
$$

where $\tau_0$, $\tau_1$ and $\tau_2$ are the time scales of order $\varepsilon$, $\varepsilon^2$ and $\varepsilon^3$, respectively (i.e. $\tau_0=\varepsilon\tau$, $\tau_1=\varepsilon^2\tau$ , $\tau_2=\varepsilon^3\tau$ and the expansion for the time derivative is of the form $\frac{\partial}{\partial\tau}=\varepsilon\frac{\partial}{\partial\tau_0}+\varepsilon^2\frac{\partial}{\partial\tau_1}+\varepsilon^3\frac{\partial}{\partial\tau_2}$). Upon substituting in the equations (2.9) and boundary conditions (2.10) and collecting the terms of order $\varepsilon^j$ ($j=0,1,2,3,\cdots$), at $O(1)$, it has the unperturbed solution of the system, that is $h=1$, $q=1$ whereas the other order equations and boundary conditions take the form

$$\frac{\partial Q_j}{\partial x} = f_{1j}(x;\tau_0,\tau_1,\tau_2),$$
$$T_{k0}\frac{\partial^3 H_j}{\partial x^3} + \frac{4n+2}{3n+2}\frac{\partial H_j}{\partial x} = f_{2j}(x;\tau_0,\tau_1,\tau_2) \tag{3.2}$$
$$H_j(0) = 0, Q_j(0) = 0$$
$$H_j(1) = 0, T_{k0}\frac{\partial^2 H_j(1)}{\partial x^2} = g_j(1)$$

where the functions $f_{1j}(x;\tau_0,\tau_1,\tau_2)$, $f_{2j}(x;\tau_0,\tau_1,\tau_2)$ and $g_j(x;\tau_0,\tau_1,\tau_2)$ are defined in Appendix E. The solution of the above equations at $O(\varepsilon)$ can be obtained as

$$Q_0 = 0, H_0 = \mathrm{A}_0\sin(k\pi x) \tag{3.3}$$

where $\mathrm{A}_0 = \mathrm{A}_0(\tau_0,\tau_1,\tau_2)$ is the amplitude and $k$ is the wave number. This can be further used in the first equation of (3.2) to obtain

$$Q_1 = \frac{1}{k\pi}\frac{\partial \mathrm{A}_0}{\partial \tau_0}\left(\cos(k\pi x) - 1\right) \tag{3.4}$$

For $k=1$, the solution for $H_1$ can be obtained from the equation and boundary conditions (3.2) which requires the solvability condition

$$T_{11} = -\frac{1}{(3n+2)\pi^4 \mathrm{A}_0}\left\{8(2n+1)(3n+2)\left(\frac{4n+2}{n}\right)^n R_1\mathrm{A}_0 + \pi\mathrm{A}_0^2 + 2\frac{\partial \mathrm{A}_0}{\partial \tau_0}\right\} \tag{3.5}$$

From which a steady amplitude solution near the point can be obtained from the relation

$$(3n+2)\pi^4\mathrm{A}_0(T - T_{10}) + \mathrm{A}_0\left\{8(2n+1)(3n+2)\left(\frac{4n+2}{n}\right)^n \mathrm{Re}^{-1} + \pi\mathrm{A}_0\right\} = 0 \tag{3.6}$$

Which implies either $\mathrm{A}_0 = 0$ or

$$\mathrm{A}_0 = -\frac{(3n+2)}{8(2n+1)\pi}\left\{\pi^4(T - T_{10}) + 8(2n+1)\left(\frac{4n+2}{n}\right)^n \mathrm{Re}^{-1}\right\} \tag{3.7}$$

The solution (3.7) will collapsed into the zero root if

$$T = T_{10} - \left(\frac{4n+2}{n}\right)^n \frac{8(2n+1)}{\pi^4}\mathrm{Re}^{-1} \tag{3.8}$$

Otherwise, it will give a negative (or a positive) root according to $T$ is greater than (or less than) the value $T_{10} - \left(\frac{4n+2}{n}\right)^n \frac{8(2n+1)}{\pi^4}\mathrm{Re}^{-1}$. This will helps to an immediate identification of the transcritical bifurcation near the point $(T, \mathrm{Re}^{-1}) \approx (T_{10}, 0)$, where a stable inflated shape of the membrane gradually changes to a unstable collapsed position, when $T$ passes through the neutral curve (3.8).

For second eigenmode, that is for $k=2$, the solution for $Q_1$ and $H_1$ are obtained in Appendix E. The use of this solutions leads to the first solvability condition $T_{21} = 0$.

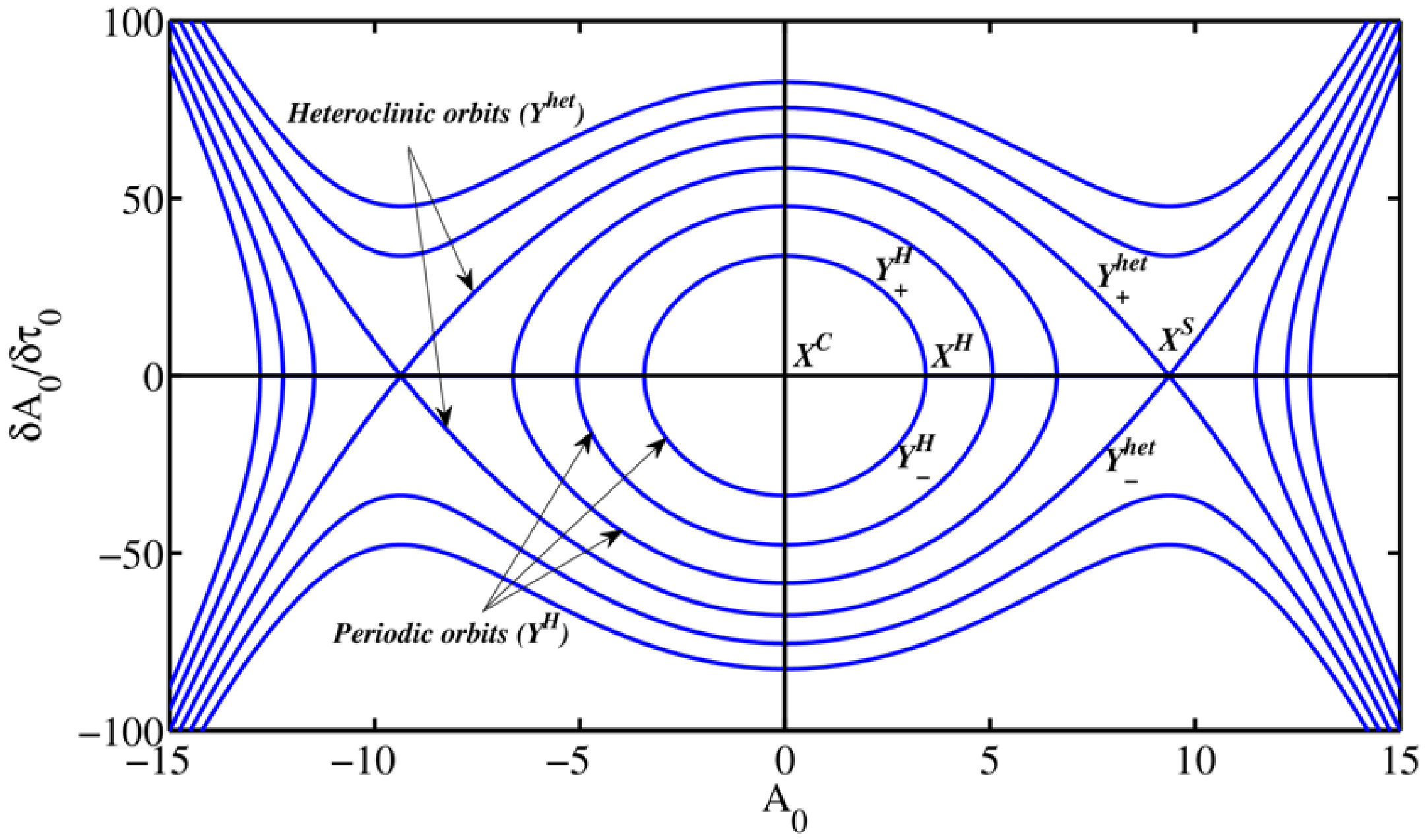

Figure 13: Phase portrait of the system (3.14) for $n=1$. The points $X^C$, $X^{\mathcal{H}}$ and $X^S$ denotes the centre, periodic and saddle points of the system. The orbit $Y^{het}_{+(-)}$ is the upper (lower) branch of heteroclinic orbit as given by (3.20) and the orbit $Y^{H}_{+(-)}$ is the upper (lower) branch of the periodic orbit as given by (3.20). The system has similar phase portrait for shear-thinning and shear-thickening fluids except a small shift in the saddle point.

Proceeding in the similar way, the solvability conditions for the problems of order $\varepsilon^3$ is obtained as

$$T_{22}=\frac{1}{32\pi^4 \mathrm{A}_0}\left(\begin{array}{l} 15\alpha(2n+1)\mathrm{A}_0 R_1^2-6\dfrac{(2n+1)}{(3n+2)}\pi^2\mathrm{A}_0^3+6\dfrac{(17n+8)}{(3n+2)}\dfrac{\partial^2 \mathrm{A}_0}{\partial\tau_0^2} \\ +4\left(\dfrac{4n+2}{n}\right)^n R_1\left\{(2n+1)(2n+7)\pi\mathrm{A}_0^2+3(9n+5)\dfrac{\partial \mathrm{A}_0}{\partial\tau_0}\right\}\end{array}\right) \tag{3.9}$$

The steady solution of $O\left(\mathrm{Re}^{-2}\right)$ near the point $\left(T,\mathrm{Re}^{-1}\right)\approx\left(T_{20},0\right)$ can be found from the amplitude equation

$$\begin{array}{c} 15(3n+2)^2(2n+1)\left(\dfrac{4n+2}{n}\right)^n \mathrm{A}_0\,\mathrm{Re}^{-2}-6(2n+1)\pi^2\mathrm{A}_0^3 \\ +4\left(\dfrac{4n+2}{n}\right)^n(3n+2)\,\mathrm{Re}^{-1}\left\{(2n+1)(2n+7)\pi\mathrm{A}_0^2+3(9n+5)\dfrac{\partial\mathrm{A}_0}{\partial\tau_0}\right\} \\ +6(17n+8)\dfrac{\partial^2\mathrm{A}_0}{\partial\tau_0^2}-32(3n+2)\pi^4\mathrm{A}_0\left(T-T_{20}\right)=0 \end{array} \tag{3.10}$$

Which yields either $\mathrm{A}_0=0$ or

$$6(2n+1)\pi^2 A_0^2 - \left(\frac{4n+2}{n}\right)^n (3n+2)(2n+1)\,\mathrm{Re}^{-1}\left(15(3n+2)\,\mathrm{Re}^{-1} + 4(2n+7)\pi A_0\right) \\ +32(3n+2)\pi^4\left(T - T_{20}\right) = 0 \quad (3.11)$$

The nature of the solutions of the above equation will be determined by its discriminant value. When $T - T_{20} > T_\alpha$, the above equation have two complex conjugate roots and for $T - T_{20} < T_\alpha$, it will have two real roots and collapsed into a double root when $T - T_{20} = T_\alpha$, where

$$T_\alpha = \frac{1}{96\pi^4}\left(\frac{4n+2}{n}\right)^{2n} (3n+2)(2n+1)\left(8n^2 + 56n + 143\right)\mathrm{Re}^{-2} \quad (3.12)$$

From which one can identify a saddle-node bifurcation at

$$T = T_{20} + \frac{1}{96\pi^4}\left(\frac{4n+2}{n}\right)^{2n} (3n+2)(2n+1)\left(8n^2 + 56n + 143\right)\mathrm{Re}^{-2} \quad (3.13)$$

So far in this weakly non-linear analysis we have obtained the steady solutions near $T_{20}$. From this analysis it is found that a saddle-node bifurcation occurs near the point $T_{20}$ in accordance with the static and Hopf modes as found from the linear analysis (please refer to equations (2.24) and (2.27)). For the oscillatory solution, in the limit $\mathrm{Re}^{-1} \to 0$, the phase portrait near the point $\left(T, \mathrm{Re}^{-1}\right) \approx \left(T_{20}, 0\right)$ is obtained from the amplitude equation

$$\frac{\partial^2 A_0}{\partial \tau_0^2} = \left(\frac{2n+1}{17n+8}\right)\pi^2 A_0^3 + \frac{16}{3}\left(\frac{3n+2}{17n+8}\right)\pi^4 A_0 T_{22} \quad (3.14)$$

The amplitude equation (3.14) resembles a Duffing oscillator. The phase portrait of equation (3.14) shows some periodic orbits and a pair of heteroclinic orbits (Figure 13). To determine the stability of these orbits a perturbation will require which comes from the solvability condition (7.13) at the order $\varepsilon^4$. Accordingly, we can have a disturbance equation which perturbs the above system, is given by equation (7.14) of Appendix E.

Defining $\tau_1 = \varepsilon\tau_0$ and $\tau_2 = \varepsilon^2\tau_0$, the amplitude equation (7.14) with the help of equation (3.14) can be re-expressed in the form

$$\frac{\partial^2 A_0}{\partial \tau_0^2} = \left(\frac{2n+1}{17n+8}\right)\pi^2 A_0^3 + \frac{16}{3}\left(\frac{3n+2}{17n+8}\right)\pi^4 A_0 T_{22} \\ + \frac{\varepsilon}{18\left(17n+8\right)^2}\begin{bmatrix} -6\alpha\left(17n+8\right)R_2\left\{(2n+1)(2n+7)\pi A_0^2 + 3(9n+5)\dfrac{\partial A_0}{\partial \tau_0}\right\} \\ -2(n-1)\pi^3 A_0^2\left\{16(3n+2)\pi^2 T_{22} + (6n+3)A_0^2\right\} \\ -36(2n+1)\pi^2\left\{16(3n+2)\pi^2 T_{22} - 3(11n+5)A_0^2\right\}\dfrac{\partial A_0}{\partial \tau_0} \\ +3\pi(17n+8)(85n+44)\left(\dfrac{\partial A_0}{\partial \tau_0}\right)^2 \end{bmatrix} \quad (3.15)$$

Equation (3.15) serves as a disturbance equation to the amplitude equation (3.14) and should be analysed to determine the stability of the orbits. Towards that we have employed the Melnikovôs approach for a Hamiltonian system as outlined by Xu et al. (2013, 2014).

### 3.3 *Hamiltonian analysis: Melnikov function*

Setting $\lambda = -T_{22}$, $\tau_0 = \tau$, $A_0 = X$ and $\frac{\partial A_0}{\partial \tau_0} = Y$, we may write equation (3.15) in the form

$$\frac{\partial X}{\partial \tau} = Y; \; \frac{\partial Y}{\partial \tau} = f(X) + \varepsilon g(X,Y) \tag{3.16}$$

where

$$f(X) = \left(\frac{2n+1}{17n+8}\right)\pi^2 X^3 - \frac{16}{3}\left(\frac{3n+2}{17n+8}\right)\pi^4 X\lambda \tag{3.17}$$

and

$$g(X,Y) = \frac{1}{18(17n+8)^2}\begin{bmatrix} -6\alpha(17n+8)R_2\left\{(2n+1)(2n+7)\pi X^2 + 3(9n+5)Y\right\} \\ -2(n-1)\pi^3 X^2\left\{-16(3n+2)\pi^2\lambda + (6n+3)X^2\right\} \\ +3\pi(17n+8)(85n+44)Y^2 \\ -36(2n+1)\pi^2\left\{-16(3n+2)\pi^2\lambda - 3(11n+5)X^2\right\}Y \end{bmatrix} \tag{3.18}$$

The Hamiltonian $\mathcal{H}$ which satisfies $\left(\frac{\partial \mathcal{H}}{\partial Y} = \frac{\partial X}{\partial \tau}; \frac{\partial \mathcal{H}}{\partial X} = -\frac{\partial Y}{\partial \tau}\right)$ for the unperturbed system $\left(\frac{\partial X}{\partial \tau} = Y, \frac{\partial Y}{\partial \tau} = f(X)\right)$ is given by

$$\mathcal{H} = \frac{Y^2}{2} - \left(\frac{2n+1}{17n+8}\right)\frac{\pi^2}{4}X^4 + \frac{16}{3}\left(\frac{3n+2}{17n+8}\right)\frac{\pi^4}{2}X^2\lambda \tag{3.19}$$

Which has a centre at $X^C = (0,0)$ and saddle points at $X^S = \left(\pm 4\pi\sqrt{\frac{\lambda}{3}\left(\frac{3n+2}{2n+1}\right)}, 0\right)$. From the phase potrait, one can see a pair of heteroclinic orbits lie along the curve

$$\frac{Y^2}{2} - \left(\frac{2n+1}{17n+8}\right)\frac{\pi^2}{4}X^4 + \frac{16}{3}\left(\frac{3n+2}{17n+8}\right)\frac{\pi^4}{2}X^2\lambda = \frac{64\pi^6(3n+2)^2\lambda^2}{9(2n+1)(17n+8)} \tag{3.20}$$

whereas the periodic orbits for the unperturbed system are lies within the limit $0 < \mathcal{H} < \frac{64\pi^6(3n+2)^2\lambda^2}{9(2n+1)(17n+8)}$.

The construction of the Melnikov function $M = \int_\tau Yg(X,Y)\,d\tau = \int_X g(X,Y)\,dX$ along the heteroclinic orbits, yield us a heteroclinic connection along the curve from (8.3) of Appendix F, as

$$T = T_{20} - \left(\frac{4n+2}{n}\right)^n \frac{15(2n+7)(17n+8)}{1376\pi^4}\mathrm{Re}^{-1} \tag{3.21}$$

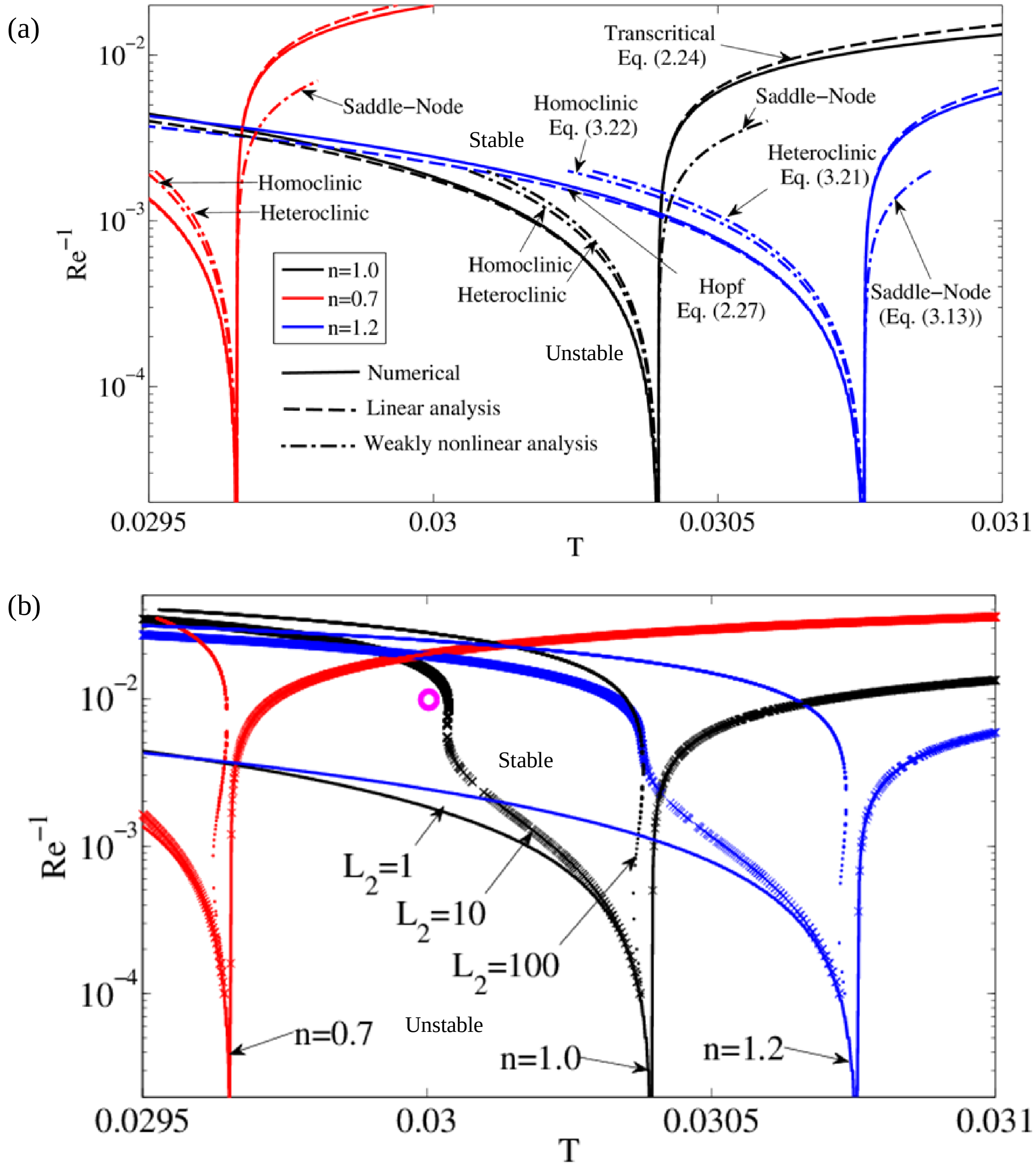

Figure 14: (a) Neutral curves in the neighbourhood of $T_{20}$ for $L_2 = 1$ and different power law indices obtained using full numerical solutions (solid line), linear analysis (– – line) and weakly nonlinear analysis (–•– line). [*Hopf: Eq.* (2.27)*; Transcritical: Eq.*(2.24)*; Saddle-Node: Eq.* (3.13)*; Heteroclinic: Eq.* (3.21)*; Homoclinic: Eq.* (3.22)] (b) Neutral curves in the neighbourhood of $T_{20}$ for different power law indices and for different lengths of the downstream segment, $L_2 = 1,10,100$, marked with a solid line, $\times$ symbol and $\bullet$ symbol respectively. Region of stability lie above the curve under consideration and is marked by stable and unstable in the plot respectively.

The function (8.6) along the periodic orbits yield us the Hopf bifurcation along (2.27), as found from the linear analysis, and a homoclinic connection along (from (8.6) and (8.7) of Appendix F)

$$T = T_{20} - \left(\frac{4n+2}{n}\right)^n \frac{15(17n+8)}{32\pi^4} \frac{4(3n+2)(2n+7)+9(9n+5)\beta}{(516n+344)+135(2n+1)\beta} \mathrm{Re}^{-1} \tag{3.22}$$

where $\beta = (3n+2)/\sqrt{(2n+1)(17n+8)}$.

The approximations for the neutral curves near the point $T_{20}$, that are obtained in equations (3.13), (3.21) and (3.22) are presented in figure14(a). In all the cases, shear-thinning ( $n = 0.7$ ), shear-thickening ( $n = 1.2$ ) and Newtonian ( $n = 1.0$ ), the neutral curve has similar qualitative behaviour as reported earlier by Xu et al. (2013). The similarity of these qualitative behaviours may be justified by the fact that, at large Reynolds number the flow becomes almost inviscid; with the inertial contribution entailing a shift in the location of the eigenvalues (please refer to sub-section 2.3). In this limit the system can be described under the framework of Bernoulli equation where the viscous effect becomes negligible (Bertram *et al.*1994) and the changes appear through the kinetic energy of the system.

Interestingly, the limits that are obtained from linear and weakly non-linear analyses (sub-sections 2.3, 2.4 and 3.1) become independent of the length of rigid downstream segment, $L_2$, and are in good agreement with the numerical solution for $L_2 = 1$. Although $L_2$ appears in the function $g_j(x;\tau_0,\tau_1,\tau_2)$ of equation (3.2), but the term including $L_2$ vanishes during the calculation of different order eigenmodes. To account the effect of $L_2$, the neutral curves near the point $T_{20}$ are plotted for $L_2 = 10$ and 100 against $L_2 = 1$ in figure 14(b). From which it can be seen that with the larger downstream segment the unstable region widens. The transcritical branch is independent of $L_2$, but there is a significant difference in the Hopf branch. The unstable region below the Hopf branch widens with a wobble in the neutral curve. The Hopf branch appears in the even modes. For sufficiently large $L_2$, for a mode-2 oscillation, a single-humped mode and a double humped mode with same frequency and wave length appears together (equation (3.25)). The interaction between these two modes may increase the unsteady behaviour of the membrane, which in turn promote the instability within the system. Xu et al. (2014) described that the two-humped mode appears from the viscous and inertial effect, whereas the single-humped mode arises from the contribution of axial sloshing motion due to the large downstream segment. It will be more interesting to see how the dynamics of system evolve along those Hopf branches with the change in the fluid rheology and the long downstream segment. Since the even modes produces the extra modes of oscillation, it is viable to analyse the system (2.9) and (2.10) for large $L_2$ in the second mode. From figure 14(b) it is observed that the qualitative behaviour of the neutral curves does not change with the rheology of the fluid. Therefore, here we have followed the weakly non-linear analysis as outlined in Xu et al. (2014) for long downstream segment.

We start from the equations (2.9) and (2.10) and reframe the second boundary condition at $x = 1$ as

$$T\ell \frac{\partial^2 h}{\partial x^2} = -\left(\frac{\partial q}{\partial t} + 2\,\mathrm{Re}^{-1}\left(\frac{4n+2}{n}\right)^n \left\{|q|^{n-1} q - 1\right\}\right) \tag{3.23}$$

where $\ell = 1/L_2$. Therefore, a small value of $\ell$ will corresponds to the long downstream segment of the channel. We consider the expansions for $h$, $q$ and the parameters $T$, $\mathrm{Re}^{-1}$ as given in equation (3.1) together with

$$\ell = \varepsilon\ell_1 + \varepsilon^2\ell_2 + \varepsilon^3\ell_3 + ... \tag{3.24}$$

At the upper Hopf branch, $(T - T_{20}) \approx \varepsilon^2 T_{22}$, $\mathrm{Re}^{-1} \approx \varepsilon R_1$ and $\ell \approx \varepsilon^2 \ell_2$. Using these expansions in equations (2.9), (2.10) and (3.23), we have system of differential equations and boundary conditions at different order of $\varepsilon$ (please refer to Appendix G). In each case we will have equations of the type (3.2), but with a modification to the boundary condition at $x = 1$. The solution of each order equations requires solvability condition which came from the next order boundary condition. The leading order system posses a solution of the form

$$Q_0 = 0, H_0 = A_0 \sin(2\pi x) + B_0 (1 - \cos(2\pi x)) \tag{3.25}$$

where $A_0 = A_0(\tau_0, \tau_1, \tau_2)$ and $B_0 = B_0(\tau_0, \tau_1, \tau_2)$ are the amplitude functions of the *sin* and (1-*cos*) modes respectively. At $O(\varepsilon^2)$, the solvability conditions are given in equation (8.10) and (8.11) of Appendix G. From the solvability condition it is revealed that the system possesses an unstable mode unless $B_0 = 0$. Setting $B_0 = 0$, we obtain the solution for the $O(\varepsilon^2)$ problem given by equation (8.12). Using these solutions in the equations of $O(\varepsilon^3)$ (please see equation (8.13)), the solvability conditions obtained are given by equations (8.14) and (8.15). Rescaling the variables by $\tilde{\ell} = \ell / \mathrm{Re}^{-2}$, $\tilde{t} = t\,\mathrm{Re}^{-1}$, $\tilde{A} = A/\mathrm{Re}^{-1}$, $\tilde{B} = B/\mathrm{Re}^{-2}$, $\tilde{T}_{22} = T_{22}/\mathrm{Re}^{-2}$, the solvability conditions can be made free from $\mathrm{Re}$, and after some algebraic manipulation we come up with a system of second order differential equation as

$$\frac{\partial^2 \tilde{A}}{\partial \tilde{t}^2} = a_{11}\tilde{A} + a_{14}\tilde{B} + a_{16}\frac{\partial \tilde{A}}{\partial \tilde{t}} + a_{18}\frac{\partial \tilde{B}}{\partial \tilde{t}} + \tilde{A}\left(a_{12}\tilde{A} + a_{13}\tilde{A}^2 + a_{15}\tilde{B} + a_{17}\frac{\partial \tilde{A}}{\partial \tilde{t}}\right) \tag{3.26}$$

$$\begin{aligned}\frac{\partial^2 \tilde{B}}{\partial \tilde{t}^2} &= \beta_{11}\tilde{A} + a_{23}\tilde{B} + a_{24}\frac{\partial \tilde{A}}{\partial \tilde{t}} + a_{27}\frac{\partial \tilde{B}}{\partial \tilde{t}} + \tilde{A}\left(\beta_{12}\tilde{B} + \beta_{13}\frac{\partial \tilde{A}}{\partial \tilde{t}} + \beta_{14}\frac{\partial \tilde{B}}{\partial \tilde{t}}\right) \\ &+ \tilde{A}^2\left(\alpha_{11} + \alpha_{12}\tilde{B} + \alpha_{13}\frac{\partial \tilde{A}}{\partial \tilde{t}}\right) + a_{21}\tilde{A}^3 + a_{22}\tilde{A}^4 + a_{25}\tilde{B}\frac{\partial \tilde{A}}{\partial \tilde{t}} + a_{26}\left(\frac{\partial \tilde{A}}{\partial \tilde{t}}\right)^2\end{aligned} \tag{3.27}$$

The coefficients involved in these equations are given in Appendix G. Equation (3.26) and (3.27) will describe the evolution of the amplitude equations for large $L_2$. To analyse these equations, they have been converted to a system of first order differential equations as given in (8.20).

To understand the dynamical behaviour of the system (3.26)-(3.27), we have solved the system (8.20) for the amplitude growth $\tilde{A}$ and $\partial\tilde{A}/\partial\tilde{t}$ for different combination of power-law index, $(T - T_{20})/\mathrm{Re}^{-2}$ and $\tilde{\ell}$. A Poincare section $\tilde{A} = 0, \partial\tilde{A}/\partial\tilde{t} < 0$ for the shear thickening fluid and $\tilde{\ell} = 100$ is shown in figure 15. As the tension parameter $(T - T_{20})/\mathrm{Re}^{-2}$ increases, the rate of change of membrane amplitude corresponding to the sinuous mode becomes chaotic in nature through period doubling. A similar kind of nature can also be observed for shear thinning and Newtonian fluids. A representative case for shear thinning fluid ($n = 0.95$) and Newtonian fluid ($n = 1$) is shown in figure G1(a-b) of Appendix G. The difference between those cases is that the period doubling occurs closer to the point $T_{20}$ for shear thinning fluid, as compared to the shear thickening fluid. Thus, in figure 16 we have marked the onset of first period-doubling for different power-law indices and $\ell/\mathrm{Re}^{-2} (= \tilde{\ell})$. As the power-law index $n$ increases, the onset of period-doubling gets delayed and occurs for more negative values of $(T - T_{20})/\mathrm{Re}^{-2}$, that is for smaller $T$ values. With the further downstream segment ($\ell/\mathrm{Re}^{-2} = 10$, $\mathrm{Re}^{-1} << 1$), the membrane amplitude requires to assume greater value of the tension parameter (larger membrane tension) to reach the period doubling. Another interesting case arises when $n = 0.9$, for which the two branches of the period doubling merge again after some values of the parameter $(T - T_{20})/\mathrm{Re}^{-2}$ for $\ell/\mathrm{Re}^{-2} = 100$, whereas same thing happens after the second period doubling for the case of $\ell/\mathrm{Re}^{-2} = 10$ (refer to figure G1(c-d) of Appendix G).

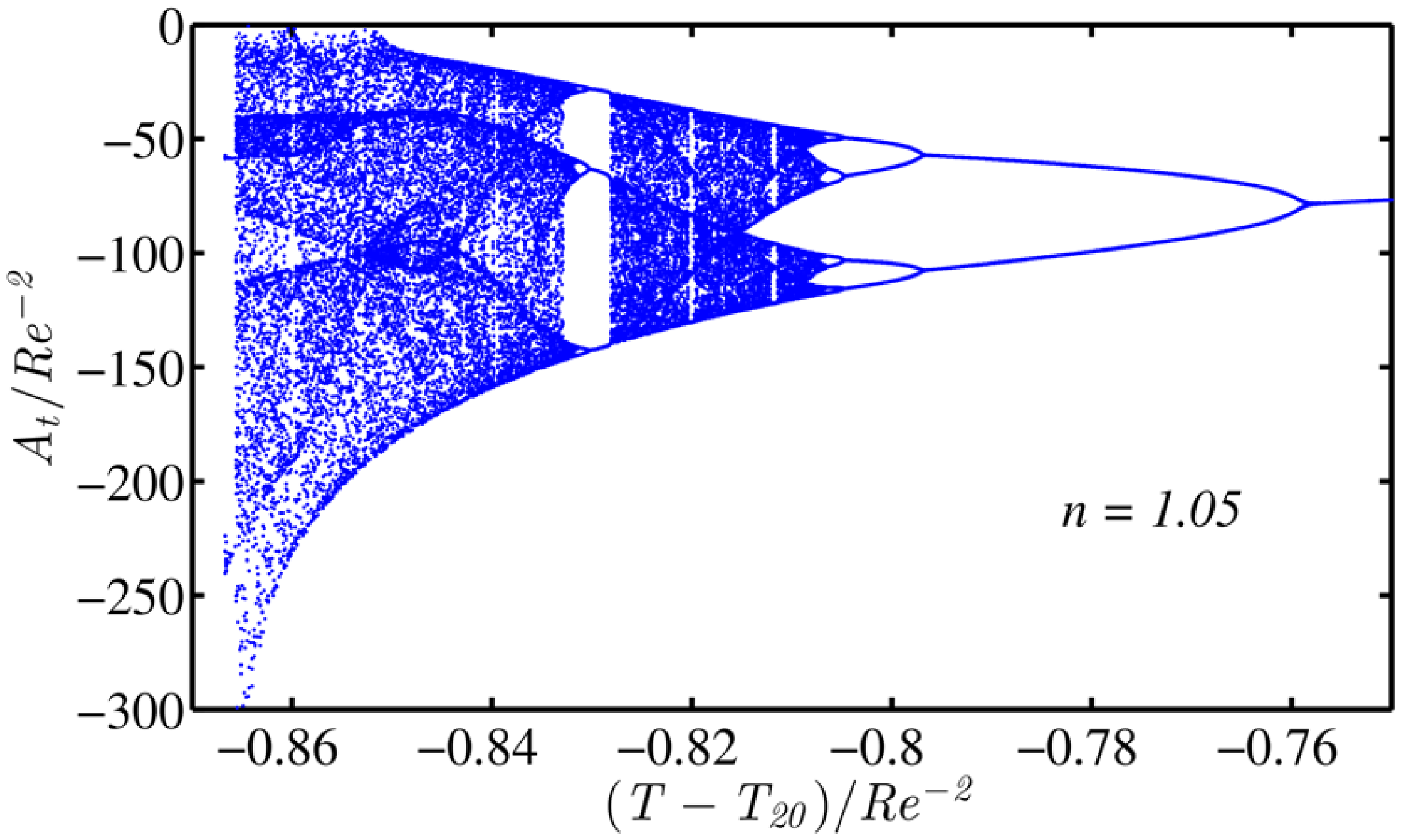

Figure 15: The bifurcation diagram for the system equation (3.26) and (3.27) for a shear thickening fluid ($n = 1.05$). As the parameter $(T - T_{20})/\mathrm{Re}^{-2}$ changes from -0.75 to -0.9, the system becomes chaotic through period doubling. The Poincare section $\tilde{A} = 0, \partial\tilde{A}/\partial\tilde{t} < 0$ is illustrated for $\tilde{\ell} = 100$. For shear thinning and Newtonian fluid, a similar kind of nature is observed, but the period-doubling occur more closer to the point $T_{20}$ (Figure G1 of Appendix G).

Thus, form the dynamical behaviour of the amplitude equations (3.26)-(3.27), one can see how the oscillating behaviour of the membrane and the flow change rapidly through the interaction behaviour of the two-humped mode ($\sin(2\pi x)$) and single-humped mode ($1 - \cos(2\pi x)$), for larger length of the downstream segment. These behaviours alter significantly with the variation in the power-law index. The self-excited oscillation grows rapidly and becomes chaotic through period-doubling of the amplitude function, which is not present in case of a channel with downstream segment comparable to membrane length.

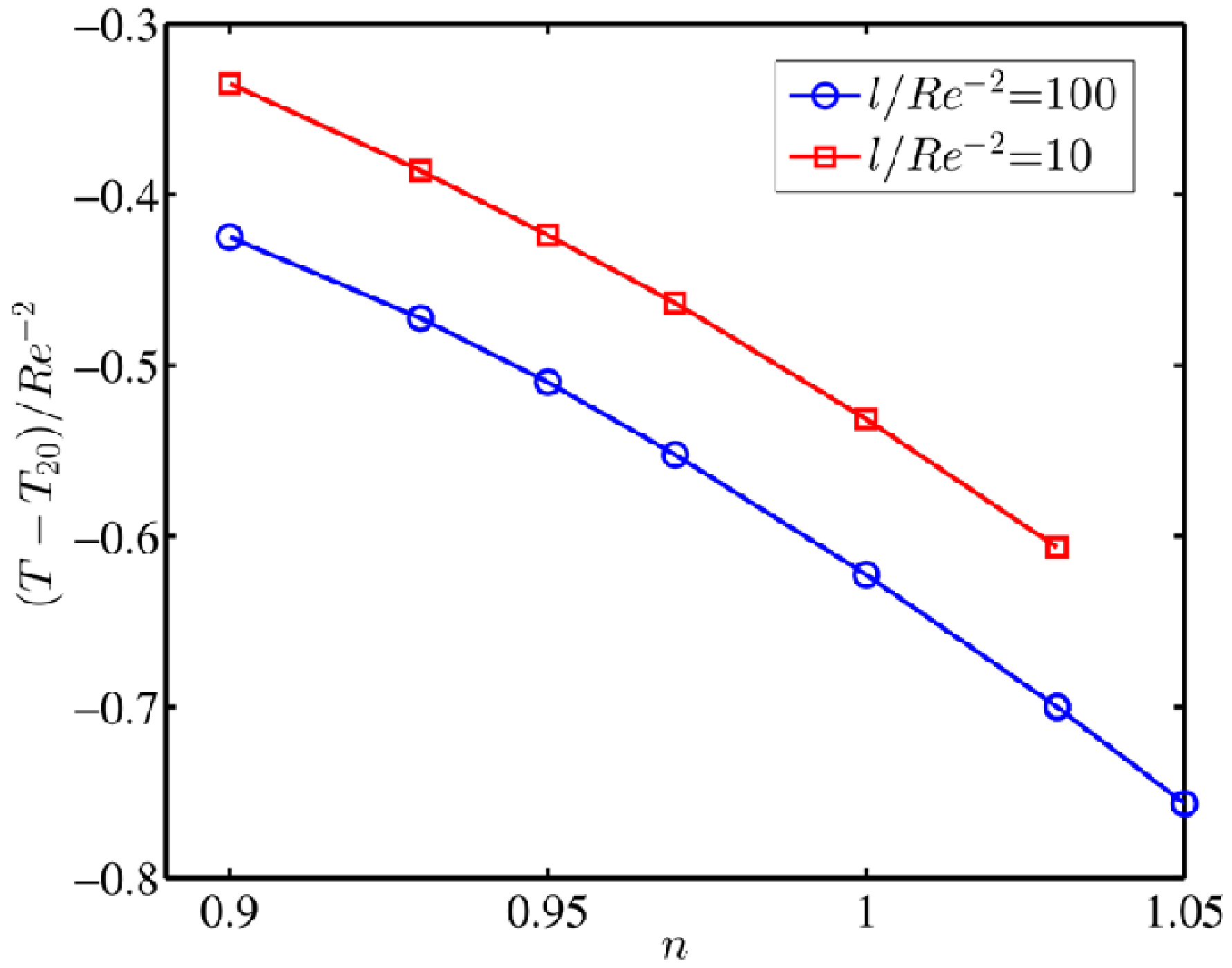

Figure 16: The onset of period doubling for different power-law index and $\ell/\mathrm{Re}^{-2}(=\tilde{\ell})$. The period-doubling occurs faster for larger downstream segment ($\ell/\mathrm{Re}^{-2}=10$ compared to $\ell/\mathrm{Re}^{-2}=100$) and it delays for increasing power-law index $n$.

## 4. Conclusions

In this paper, we have theoretically investigated the dynamical behaviour of a membrane attached between two rigid segments of a channel conveying a power-law fluid through a reduced 1-D model. Initially the flow is subjected to a fixed upstream flux, which is disturbed by the induced membrane oscillation. The equations obtained are solved numerically by Chebyshevôs spectral collocation method and it is found that the temporal behaviour of the membrane is chaotic for shear-thinning fluid whereas it is stable for shear-thickening fluids at the same parametric regime. In understanding the stability mechanism, we have employed a perturbation on the base flow and initial membrane position. At leading order of the perturbation parameter a stable oscillation occurs in the membrane, which also represents the eigenmodes at the large Reynolds number limit (inviscid limit). At this limit, the neutral stability curve has similar qualitative behaviour for shear thinning, shear thickening and the Newtonian fluid, except that the unstable region widens for a shear thinning fluid and shrinks for a shear-thickening fluid. The linear stability analysis in the inviscid limit describes the membrane behaviour by transcritical and saddle-node bifurcation points. The existence of these points is also adjudged by the weakly non-linear analysis, where a multi-time scale expansion is employed to the governing equations. The amplitude equation for the membrane exhibits a Duffing oscillator when the downstream rigid segment is comparable with the membrane length. If the downstream segment is larger than the membrane length, the amplitude equations exhibit a chaotic behaviour of the membrane through period doubling. For a fixed value of the tension parameter, the system is more chaotic for a shear thinning fluid in comparison to a shear thickening fluid.

**Acknowledgement**
PG acknowledges the financial support by NBHM (DAE, Govt. of India). AB and SC acknowledge the financial support provided by the Indian Institute of Technology Kharagpur, India [Sanction Letter no.: IIT/SRIC/ATDC/CEM/2013-14/118, dated 19.12.2013].

**Appendix A: Derivation of base velocity and pressure boundary condition**

*Base state velocity*: The base state velocity may be derived as follows: In the case of fully developed flow of a power-law fluid, the governing equation (2.2) for the flow in the elastic segment reduces to

$$0 = -\mathrm{Re}\frac{\partial \hat{p}}{\partial \hat{x}} + \frac{\partial}{\partial \hat{y}}\left(\left|\frac{\partial \hat{u}}{\partial \hat{y}}\right|^{n-1}\frac{\partial \hat{u}}{\partial \hat{y}}\right) \tag{4.1}$$

together with the boundary conditions $\hat{u}(0) = 0$ and $\hat{u}(\hat{h}) = 0$. Solving the above equation, the base velocity is obtained as:

$$\hat{u} = \int_0^{\hat{y}}\left(\mathrm{Re}\frac{\partial \hat{p}}{\partial \hat{x}}\hat{y} + C_1\right)\left|\mathrm{Re}\frac{\partial \hat{p}}{\partial \hat{x}}\hat{y} + C_1\right|^{\frac{1}{n}-1} d\hat{y} + C_2 \tag{4.2}$$

where $C_1$ and $C_2$ are the constants of integration. The relevant boundary conditions yield $C_1 = -\mathrm{Re}\frac{\partial \hat{p}}{\partial \hat{x}}\frac{\hat{h}}{2}$ and $C_2 = 0$. Subsequently, we obtain the unidirectional velocity profile as

$$\hat{u} = \frac{n}{n+1}\mathrm{Re}^{\frac{1}{n}}\frac{\partial \hat{p}}{\partial \hat{x}}\left|\frac{\partial \hat{p}}{\partial \hat{x}}\right|^{\frac{1}{n}-1}\left[\left|\left(\hat{y} - \frac{\hat{h}}{2}\right)\right|^{\frac{1}{n}+1} - \left(\frac{\hat{h}}{2}\right)^{\frac{1}{n}+1}\right] \tag{4.3}$$

Therefore, the volume flow rate at any cross section may be easily obtained as

$$\hat{q} = \int_0^{\hat{h}}\hat{u}\,d\hat{y} = -\hat{h}\frac{n}{n+1}\mathrm{Re}^{\frac{1}{n}}\frac{\partial \hat{p}}{\partial \hat{x}}\left|\frac{\partial \hat{p}}{\partial \hat{x}}\right|^{\frac{1}{n}-1}\frac{n+1}{2n+1}\left(\frac{\hat{h}}{2}\right)^{\frac{1}{n}+1} \tag{4.4}$$

With the help of equation (4.4) and eliminating the pressure gradient in equation (4.3) we obtain the velocity profile given in equation (2.6). Let us now drop the carat over the variables for convenience.

Proceeding further, to determine the pressure distribution inside the channel, we assume a fixed upstream pressure $p_u$ and a zero downstream pressure, that is at $x = -L_1$, $p = p_u$ and at $x = 1 + L_2$, $p = 0$, respectively. When $h = 1$, the flow is steady and uniform along the channel, which gives $p = p_u - \frac{2}{\text{Re}}\left(\frac{4n+2}{n}\right)^n (L_1 + x)$ with $p_u = \frac{2}{\text{Re}}\left(\frac{4n+2}{n}\right)^n (L_1 + 1 + L_2)$. From which we may consider that a non-uniform external pressure

$$p_e(x) = \frac{2}{\text{Re}}\left(\frac{4n+2}{n}\right)^n (1 + L_2 - x) \tag{4.5}$$

will act over the elastic segment $0 \le x \le 1$. In the rigid section the downstream pressure is zero and $q = q(t)$ is uniform. Therefore, in the rigid section, we may recast equation (2.7) as

$$\frac{\partial p}{\partial x} = -\frac{\partial q}{\partial t} - \frac{2}{\text{Re}}\left(\frac{4n+2}{n}\right)^n |q|^{n-1} q. \tag{4.6}$$

We may solve equation (4.6) with the boundary condition that at $x = 1 + L_2$, $p = 0$ to obtain $p = \left(\frac{\partial q}{\partial t} + \frac{2}{\text{Re}}\left(\frac{4n+2}{n}\right)^n |q|^{n-1} q\right) L_2$. This may be simplified further using equation (2.8) to obtain $\left(\frac{\partial q}{\partial t} + \frac{2}{\text{Re}}\left(\frac{4n+2}{n}\right)^n |q|^{n-1} q\right) L_2 = \frac{2}{\text{Re}}\left(\frac{4n+2}{n}\right)^n (1 + L_2 - x) - T\frac{\partial^2 h}{\partial x^2}$ which yields the pressure boundary condition at $x = 1$ as

$$T\left.\frac{\partial^2 h}{\partial x^2}\right|_{x=1} = -\left(\frac{\partial q}{\partial t} + \frac{2}{\text{Re}}\left(\frac{4n+2}{n}\right)^n \left(|q|^{n-1} q - 1\right)\right) L_2. \tag{4.7}$$

Equation (4.7) provides the required boundary condition used in equation (2.10).

*Wall Equation*: In deriving the relation between pressure and the membrane, we have followed the work of Jensen & Heil (2003) and Stewart et.al. (2009). We consider the membrane as a thin-walled elastic beam of thickness $h_0$ , density $\rho_w$ (Kirchhoff-Love model) and is in a longitudinal pre-stress position with an imposed initial tension $T_0$. In Lagrangian co-ordinate system at any time $t$, the new material position of the membrane wall is given by

$$\mathbf{R}_w(\zeta, t) = \mathbf{r}_w(\zeta) + \mathbf{d}(\zeta, t) \tag{4.8}$$

Where, $\mathbf{r}_w(\zeta) = (\zeta, H)$ is the initial position of the membrane and $\mathbf{d}(\zeta, t) = (\Delta x, \Delta y)$ is the displacement vector. For thin beam and small strains the second Piola-Kirchoff stress tensor is dominant during wall deformation and which may be represented through the linear constitutive relation $\sigma = \sigma_0 + E\gamma$, where $\sigma_0 = T_0 / h_0$ is the axial pre-stress, $E$ is the incremental Youngŝ modulus and $\gamma = \frac{\partial(\Delta x)}{\partial \zeta} + \frac{1}{2}\left\{\left(\frac{\partial(\Delta x)}{\partial \zeta}\right)^2 + \left(\frac{\partial(\Delta y)}{\partial \zeta}\right)^2\right\}$ is the nonlinear Kirchoff-Love axial extensional strain. The extension and bending due the infinitely small displacement in the membrane will make a variation in the wall strain energy. By the principal of virtual work, the balance between the wall strain energy, fluid traction on the membrane and the membrane inertia gives the equation for the membrane as

$$\int_0^L \left\{ \sigma\delta\gamma + \frac{Eh_0^2}{12}\chi\delta\chi - \left( \frac{1}{h_0}\sqrt{\frac{A_2}{A_1}}\mathbf{f} - \rho_w \frac{\partial^2 \mathbf{R}_w}{\partial t^2} \right)\cdot\delta\mathbf{R}_w \right\}\sqrt{A_1}\, d\zeta = 0 \qquad (4.9)$$

Where $\mathbf{f} = (p - p_{ext})\mathbf{n} - \boldsymbol{\tau}\cdot\mathbf{n}$ is the traction on the wall due to fluid load and imposed external pressure, $\mathbf{n} = \left( -\frac{\partial(\Delta y)}{\partial\zeta}, \left(1 + \frac{\partial(\Delta x)}{\partial\zeta}\right)\right) / \sqrt{\left(1 + \frac{\partial(\Delta x)}{\partial\zeta}\right)^2 + \left(\frac{\partial(\Delta y)}{\partial\zeta}\right)^2}$ is the unit outward normal $\chi = \det\left( \frac{\partial\mathbf{R}_w}{\partial\zeta}, \frac{\partial^2\mathbf{R}_w}{\partial\zeta^2} \right) / \left\| \frac{\partial\mathbf{R}_w}{\partial\zeta} \right\|^3$ is the wall curvature ($\|\boldsymbol{\alpha}\| = \sqrt{\boldsymbol{\alpha}\cdot\boldsymbol{\alpha}}$ denotes the norm of the vector $\boldsymbol{\alpha}$), $A_1 = \left\| \frac{\partial\mathbf{r}_w}{\partial\zeta} \right\|^2$ and $A_2 = \left\| \frac{\partial\mathbf{R}_w}{\partial\zeta} \right\|^2$. In addition to this, the fluid will satisfy the no-slip and kinematic condition $\mathbf{u} = \frac{\partial\mathbf{R}_w}{\partial t}$ at the wall. The choice of the scales $\Delta x, \zeta \sim H$, $\Delta y, h_0 \sim H$ and $t \sim L/U$ gives $\tau \sim k_p (U/H)^n$, $\chi \sim 1/H$, $A_1 \sim 1$, $A_2 \sim 1$. These will reduce the above equation (4.9) in the non-dimensional form as

$$\int_0^{\tilde L} \left\{ \frac{E}{\rho U^2}\left[ (\tilde\sigma_0 + \tilde\gamma)\delta\tilde\gamma + \frac{\tilde h_0^2}{12}\tilde\chi\delta\tilde\chi \right] - \frac{1}{\tilde h_0}\left( \sqrt{\frac{\tilde A_2}{\tilde A_1}}\tilde{\mathbf{f}}\cdot\delta\tilde{\mathbf{R}}_w - \delta^2 \frac{\rho_w \tilde h_0}{\rho}\frac{\partial^2\tilde{\mathbf{R}}_w}{\partial\tilde t^2}\cdot\delta\tilde{\mathbf{R}}_w \right) \right\}\sqrt{\tilde A_1}\, d\tilde\zeta = 0$$

(4.10)

Where '$\tilde\alpha$' denotes the non-dimensional $\alpha$ and $\delta = H/L \ll 1$. Here, $\tilde{\mathbf{f}} = (\tilde p - \tilde p_{ext})\tilde{\mathbf{n}} - \frac{1}{\mathrm{Re}}\tilde{\boldsymbol{\tau}}\cdot\tilde{\mathbf{n}}$ is the non-dimensional body force, $\tilde\gamma = \frac{\partial(\Delta\tilde x)}{\partial\tilde\zeta} + \frac{1}{2}\left\{ \left(\frac{\partial(\Delta\tilde x)}{\partial\tilde\zeta}\right)^2 + \left(\frac{\partial(\Delta\tilde y)}{\partial\tilde\zeta}\right)^2 \right\}$ is the non-dimensional extensional strain and $\mathrm{Re} = \frac{\rho U H}{k_p (U/H)^{n-1}}$ is the Reynolds number. The strain energy due to the bending stiffness is of the $O(\tilde h_0^2)$ and can be neglected for $\tilde h_0 \ll 1$. If $\delta^2 \frac{\rho_w \tilde h_0}{\rho} \ll 1$, the wall inertia is small compared to the fluid load. Now, at high $\mathrm{Re}$, the normal traction $(\tilde p - \tilde p_{ext})\tilde{\mathbf{n}}$ will become a dominating factor in the fluid load as the fluid traction is of $O(1/\mathrm{Re})$. If we assume the imposed pre-stress is much larger than the extensional strain generated by the wall displacement, then it may be assumed that $(\tilde\sigma_0 + \tilde\gamma) \approx \tilde\sigma_0$. Therefore, the above equation (4.10) can be simplified as

$$\int_0^{\tilde{L}} \left\{ \frac{E}{\rho U^2} \tilde{h}_0 \delta \tilde{T} - \frac{1}{\tilde{h}_0} \left( \sqrt{\frac{\tilde{A}_2}{\tilde{A}_1}} \left( \tilde{p} - \tilde{p}_{ext} \right) \tilde{\mathbf{n}} \cdot \delta \tilde{\mathbf{R}}_w \right) \right\} \sqrt{\tilde{A}_1} d\tilde{\xi} = 0 \tag{4.11}$$

In case of small amplitude wall deformation and long wavelength, the position of the material point can be described by $\tilde{\xi} = \tilde{x}$ and $\tilde{h} = 1 + \Delta\tilde{y}$ or equivalently, $(\Delta\tilde{x}, \Delta\tilde{y}) = (0, \tilde{h} - 1)$. Upon substituting we get $\tilde{\mathbf{R}}_w = (\tilde{x}, \tilde{h})$, $\tilde{\mathbf{n}} = \left(-\partial\tilde{h}/\partial\tilde{x}, 1\right) / \sqrt{1 + \left(\partial\tilde{h}/\partial\tilde{x}\right)^2}$, $\tilde{T} = \frac{1}{2}\left(\partial\tilde{h}/\partial\tilde{x}\right)^2$, $\tilde{A}_1 = 1$, $\tilde{A}_2 = 1 + \left(\partial\tilde{h}/\partial\tilde{x}\right)^2$. So, the wall equation (4.11) the no-slip and the kinematic condition can be written as

$$\tilde{p} = \tilde{p}_{ext} - \tilde{T} \frac{\partial^2 \tilde{h}}{\partial \tilde{x}^2} \left( 1 + \left( \frac{\partial \tilde{h}}{\partial \tilde{x}} \right)^2 \right)^{-3/2}, \tilde{u} = 0 \text{ and } \tilde{v} = \delta \frac{\partial \tilde{h}}{\partial \tilde{t}} \tag{4.12}$$

where $\tilde{T} = \tilde{h}_0 \left( E / \rho U^2 \right)$. The rescaling of the variables $x = \delta \tilde{x}$, $T = \delta^2 \tilde{T}$, $v = \tilde{v}/\delta$, $t = \tilde{t}$ and $p = \tilde{p}$ will reduce the above equation (4.12) in the form

$$p = p_{ext} - T \frac{\partial^2 h}{\partial x^2} \left( 1 + \delta^2 \left( \frac{\partial h}{\partial x} \right)^2 \right)^{-3/2}, u = 0 \text{ and } v = \frac{\partial h}{\partial t} \tag{4.13}$$

Equation (4.13) serves as a condition for the membrane motion and utilized in equation (2.8).

### Appendix B: Full numerical solution for equation (2.9) and boundary condition (2.10)

In order to obtain the time evolution of the flux and membrane displacement, we attempt to solve equation (2.9) along with boundary conditions (2.10) using a semi-implicit scheme for temporal discretization and a Chebyshev collocation method for the spatial discretization. The flow domain $0 \le x \le 1$ is transformed into the domain $-1 \le \xi \le 1$ (Gauss-Lobatto grid) by the transformation $\xi = 2x - 1$ to make it compatible with the Chebyshev collocation method. After this, the discretized equations are obtained as follows

$$\begin{gathered} \mathbf{I}\boldsymbol{\alpha}^{k+1} + 2\Delta t \mathbf{D}^{(1)} \boldsymbol{\beta}^{k+1} = \mathbf{I}\boldsymbol{\alpha}^k \\ \mathbf{I}\boldsymbol{\beta}^{k+1} - 8T\Delta t \left( \mathbf{I}\boldsymbol{\alpha}^k \right) \mathbf{D}^{(3)} \boldsymbol{\alpha}^{k+1} = \mathbf{I}\boldsymbol{\beta}^k + \frac{2}{\mathrm{Re}} \left( \frac{4n+2}{n} \right)^n \Delta t \mathbf{I}\boldsymbol{\alpha}^k + \Delta t \mathbf{f}^k \end{gathered} \tag{4.14}$$

which satisfy the following boundary conditions

$$\alpha_0^{k+1} = 1; \alpha_N^{k+1} = 1; \beta_N^{k+1} = 1;$$

$$\beta_0^{k+1} + \frac{4L_2^{-1} T \Delta t \left( \mathbf{D}^{(2)} \boldsymbol{\alpha}^k \right)_0}{1 + \frac{2}{\mathrm{Re}} \left( \frac{4n+2}{n} \right)^n \left| \beta_0^k \right|^{n-1}} = \frac{\beta_0^k + \frac{2}{\mathrm{Re}} \left( \frac{4n+2}{n} \right)^n}{1 + \frac{2}{\mathrm{Re}} \left( \frac{4n+2}{n} \right)^n \left| \beta_0^k \right|^{n-1}} \tag{4.15}$$

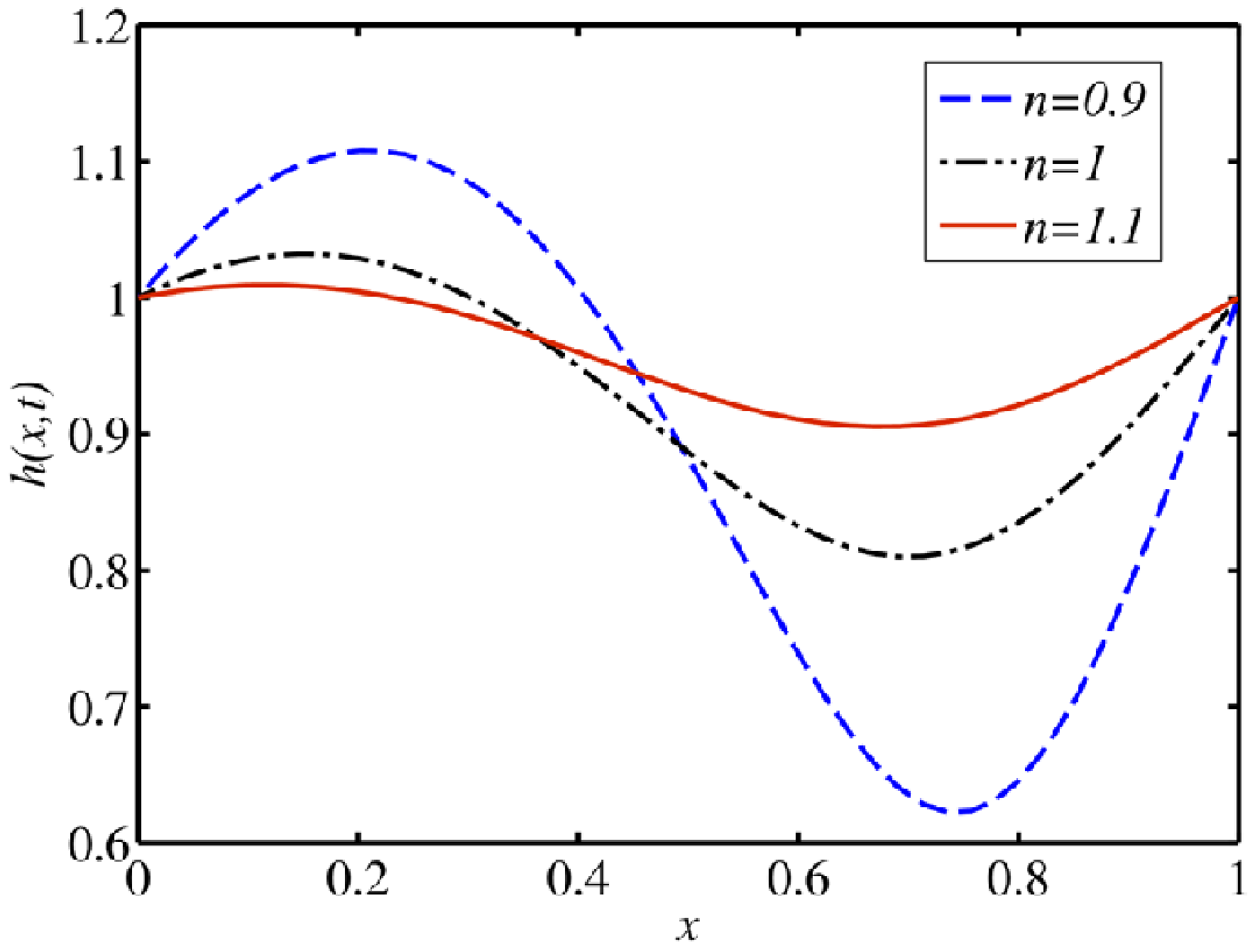

Figure B1: The variation in the membrane from its initial position $h(x,t)=1$ with respect to the spatial variable $x$ at time $t=20$, for different values of the power-law index. The other parametric values are $T=0.05$, $\mathrm{Re}^{-1}=0.01$ and $L_2=10$.

where the superscript denotes the $k^{th}$ time level and the subscript denotes the spatial coordinate with $\xi_0=-1$ and $\xi_N=1$. Where $\boldsymbol{\alpha}=(\alpha_0,\alpha_1,\alpha_2,...,\alpha_N)$, $\boldsymbol{\beta}=(\beta_0,\beta_1,\beta_2,...,\beta_N)$ are the co-efficient vectors corresponding to $h(\xi,t)$, $q(\xi,t)$, $\mathbf{I}$ is the identity matrix of required dimension, $\Delta t$ is the time interval, $\mathbf{D}^{(j)}$ is the Chebyshevʼs differentiation matrix of order $j$, and the vector $\mathbf{f}$ consist of the non-linear terms which has the components of the form

$$f_i^k=-2\left(\frac{4n+2}{3n+2}\right)\left\{2\frac{\beta_i^k}{\alpha_i^k}\left(\mathbf{D}^{(1)}\boldsymbol{\beta}^k\right)_i-\left(\frac{\beta_i^k}{\alpha_i^k}\right)^2\left(\mathbf{D}^{(1)}\boldsymbol{\alpha}^k\right)_i\right\}$$
$$-\frac{2}{\mathrm{Re}}\left(\frac{4n+2}{n}\right)^n\left\{\left|\frac{\beta_i^k}{\left(\alpha_i^k\right)^2}\right|^{n-1}\frac{\beta_i^k}{\left(\alpha_i^k\right)^2}\right\} \tag{4.16}$$

The above equations are solved by considering an initial profile for the membrane, $h(\xi,t)=1+\delta_{in}\left(1-\xi^2\right)$ ($\delta_{in}$ denotes an small perturbation parameter), and the specified initial volume flow rate $q(\xi,t)=1$. The MATLAB files describing the solutions for the eigenvalue problem will be made available on request. The shape of the membrane obtained by solving (4.14)-(4.16) is illustrated in figure B1, for different values of the power-law index and keeping other parameters fixed.

**Appendix C: Numerical solution for the eigenvalue problem**

To calculate the eigenvalues numerically, we have employed the Chebyshevôs collocation method for equation (2.11) and the boundary conditions described by equation (2.12). First, the spatial variable ó$x$ô is transformed into the ó$\xi$ô domain. With this transformation, we require to solve

$$2\frac{dQ}{d\xi}+\sigma H=0$$
$$8T\frac{d^3H}{d\xi^3}+2\frac{4n+2}{3n+2}\left(\frac{dH}{d\xi}-2\frac{dQ}{d\xi}\right)+\frac{2}{\mathrm{Re}}\left(\frac{4n+2}{n}\right)^n\left\{(2n+1)H-nQ\right\}=\sigma Q \tag{5.1}$$

subjected to

$$\text{at } \xi=-1; H=0, Q=0$$
$$\text{at } \xi=1; H=0, 4T\frac{d^2H}{\partial\xi^2}=-\left(\sigma Q+\frac{2}{\mathrm{Re}}\left(\frac{4n+2}{n}\right)^n nQ\right)L_2 \tag{5.2}$$

Considering the Chebyshev series representations for both the functions $H(\xi)$ and $Q(\xi)$ with the coefficient vectors $\mathbf{a}$ and $\mathbf{b}$ respectively, and calculating the functions at Gauss-Lobatto collocation points (i.e. $\xi_j=\cos(j\pi/N)$, $j=0,1,...,N$), the above equations can be reduced to a generalized matrix eigenvalue problem of the form

$$\mathbf{A}\begin{bmatrix}\mathbf{a}\\ \mathbf{b}\end{bmatrix}=\sigma\mathbf{B}\begin{bmatrix}\mathbf{a}\\ \mathbf{b}\end{bmatrix} \tag{5.3}$$

where $\mathbf{A}, \mathbf{B}$ are square matrices of order $2(N+1)\times 2(N+1)$. The matrices are of the form $\mathbf{A}=\begin{bmatrix}\mathbf{A}_{11} & \mathbf{A}_{12}\\ \mathbf{A}_{21} & \mathbf{A}_{22}\end{bmatrix}$ and $\mathbf{B}=\begin{bmatrix}\mathbf{0} & \mathbf{I}\\ -\mathbf{I} & \mathbf{0}\end{bmatrix}$ where the submatrices are defined as

$$\mathbf{A}_{11}=8T\mathbf{D}^{(3)}+2\frac{4n+2}{3n+2}\mathbf{D}^{(1)}+\frac{2}{\mathrm{Re}}\left(\frac{4n+2}{n}\right)^n(2n+1)\mathbf{I},$$
$$\mathbf{A}_{12}=-4\frac{4n+2}{3n+2}\mathbf{D}^{(1)}-\frac{2}{\mathrm{Re}}\left(\frac{4n+2}{n}\right)^n n\mathbf{I}, \tag{5.4}$$
$$\mathbf{A}_{21}=\mathbf{0},$$
$$\mathbf{A}_{22}=2\mathbf{D}^{(1)}$$

The boundary conditions are transform into $a_0=0, a_N=0, b_N=0$ and $4T\left(\mathbf{D}^{(2)}\mathbf{a}\right)_0+\frac{2}{\mathrm{Re}}\left(\frac{4n+2}{n}\right)^n nL_2b_0=-\sigma L_2b_0$. To incorporate the boundary conditions the modifications made into the matrices $\mathbf{A}$ and $\mathbf{B}$ are as follows:

$$\begin{gathered}
\mathbf{A}_{11}(1,1)=1,\ \mathbf{A}_{11}(N+1,N+1)=1 \\
\mathbf{A}_{22}(1,1)=\frac{2}{\mathrm{Re}}\left(\frac{4n+2}{n}\right)^{n} nL_2 \\
\mathbf{A}_{22}(1,2:N+1)=(0,0,.....,0)_{1\times N} \\
\mathbf{A}_{22}(N+1,N+1)=1 \\
\mathbf{A}_{21}(1,1:N+1)=4T\mathbf{D}^{(2)}(1,1:N+1) \\
\mathbf{B}_{22}(1,1)=-L_2
\end{gathered} \tag{5.5}$$

Where $\mathbf{A}_{ij}(l,p)$ denotes the $(l,p)$-th element of the submatrix $\mathbf{A}_{ij}$. The MATLAB files describing the solutions for the eigenvalue problem will be made available upon request.

**Appendix D: Linear stability equations**

The matrices $\mathbf{L}_D$, $\mathbf{L}_I$, $\mathbf{B}_1$ and $\mathbf{B}_2$ arising in equation (2.17) are given as

$$\begin{gathered}
\mathbf{L}_D=\begin{pmatrix} 1 & 0 & 0 & 0 \\ 0 & 1 & 0 & 0 \\ 0 & 0 & 1 & 0 \\ -\dfrac{4(2n+1)}{3n+2} & 0 & 0 & T_{k0} \end{pmatrix},\ \mathbf{L}_I=\begin{pmatrix} 0 & 0 & 0 & 0 \\ 0 & 0 & -1 & 0 \\ 0 & 0 & 0 & -1 \\ 0 & 0 & \dfrac{2(2n+1)}{3n+2} & 0 \end{pmatrix} \\
\mathbf{B}_1=\begin{pmatrix} 1 & 0 & 0 & 0 \\ 0 & 1 & 0 & 0 \\ 0 & 0 & 0 & 0 \\ 0 & 0 & 0 & 0 \end{pmatrix},\ \mathbf{B}_2=\begin{pmatrix} 0 & 0 & 0 & 0 \\ 0 & 1 & 0 & 0 \\ 0 & 0 & 0 & 0 \\ 0 & 0 & 0 & T_{k0} \end{pmatrix},
\end{gathered} \tag{6.1}$$

while the various terms in the inhomogeneous terms are defined as below

$$\begin{aligned}
&A_{10}=0; A_{20}=0; G_0=0, \\
&A_{1i}=-\sum_{j=0}^{i-1}\sigma_{i-j}H_j \\
&A_{2i}=\sum_{j=0}^{i-1}\left[\sigma_{i-j}Q_j-T_{k\,i-j}\frac{d^3H_j}{dx^3}-2R_{i-j}\left(\frac{4n+2}{n}\right)^{n}\left\{(2n+1)H_j-nQ_j\right\}\right] \\
&G_i=-\sum_{j=0}^{i-1}\left[\sigma_{i-j}Q_jL_2+T_{k\,i-j}\frac{d^2H_j}{dx^2}+2n\left(\frac{4n+2}{n}\right)^{n}L_2R_{i-j}Q_j\right]
\end{aligned} \tag{6.2}$$

The matrices appearing in the adjoint problem (2.18) are given as

$$\mathbf{L}^{*}=\begin{pmatrix} -\frac{\partial}{\partial x} & 0 & 0 & \frac{4(2n+1)}{3n+2}\frac{\partial}{\partial x} \\ 0 & -\frac{\partial}{\partial x} & 0 & 0 \\ 0 & -1 & -\frac{\partial}{\partial x} & \frac{2(2n+1)}{3n+2} \\ 0 & 0 & -1 & -T_{k0}\frac{\partial}{\partial x} \end{pmatrix},\mathbf{B}_{1}^{*}=\begin{pmatrix} 0 & 0 & 0 & 0 \\ 0 & 0 & 0 & 0 \\ 0 & 0 & 1 & 0 \\ 0 & 0 & 0 & 1 \end{pmatrix}$$

$$\mathbf{B}_{2}^{*}=\begin{pmatrix} 1 & 0 & 0 & -\frac{4(2n+1)}{3n+2} \\ 0 & 0 & 0 & 0 \\ 0 & 0 & 1 & 0 \\ 0 & 0 & 0 & 0 \end{pmatrix} \tag{6.3}$$

For $k=1$ from equations (2.17) and (2.18), one can easily find the solution for $\mathbf{\Phi}_0$ and $\mathbf{\Psi}$ as

$$\mathbf{\Phi}_0=d_0\begin{pmatrix} 0 \\ \sin(\pi x) \\ \pi\cos(\pi x) \\ -\pi^2\sin(\pi x) \end{pmatrix},\mathbf{\Psi}=d_1\begin{pmatrix} 2\left(1-\cos(\pi x)\right) \\ 1 \\ -\sin(\pi x)/\pi \\ \left(1-\cos(\pi x)\right)/6 \end{pmatrix} \tag{6.4}$$

where $d_0$ and $d_1$ are constants. Without the loss of generality, it may be assumed that $d_1=1$. Using $\mathbf{\Phi}_0$ and $\mathbf{\Psi}$ in equation (2.19), we obtain the first solvability condition as given in (2.20).

With the knowledge of the eigenfunctions, the first solvability condition for $k=2$ in $\left\langle \mathbf{\Psi}^T,\mathbf{A}_0\right\rangle = \mathbf{\Psi}^T\mathbf{L}_D\mathbf{\Phi}_0\Big|_{x=0}^{x=1}$ yields $T_{21}=0$. Therefore, considering $T=T_{20}+\varepsilon^2T_{22}$, $\mathrm{Re}^{-1}=\varepsilon R_1$, $\sigma=\varepsilon\sigma_1$ such that $T-T_{20}\approx O(\mathrm{Re}^{-2})$, the solution for $H_1(x)$ and $Q_1(x)$ at $O(\varepsilon)$ are obtained as,

$$\begin{aligned} H_1(x)&=\frac{1}{2\pi}\left[(3n+2)\left(\frac{4n+2}{n}\right)^n R_1+2\sigma_1\right]\left(\cos(2\pi x)+\pi x\sin(2\pi x)-1\right) \\ Q_1(x)&=\frac{\sigma_1}{2\pi}\left(\cos(2\pi x)-1\right) \end{aligned} \tag{6.5}$$

These functions may thus be used to obtain the next eigenfunction $\mathbf{\Phi}_1=\left[Q_1,H_1,\frac{dH_1}{dx},\frac{d^2H_1}{dx^2}\right]^T$ which is evaluated by means of equation (6.5) as

$$\mathbf{\Phi}_1 = \begin{pmatrix} \frac{\sigma_1}{2\pi}\left(\cos(2\pi x)-1\right) \\ \frac{1}{2\pi}\left[(3n+2)\left(\frac{4n+2}{n}\right)^n R_1 + 2\sigma_1\right]\left(\cos(2\pi x)+\pi x\sin(2\pi x)-1\right) \\ \frac{1}{2\pi}\left[(3n+2)\left(\frac{4n+2}{n}\right)^n R_1 + 2\sigma_1\right]\left(-\pi\sin(2\pi x)+2\pi^2 x\cos(2\pi x)\right) \\ \frac{1}{2\pi}\left[(3n+2)\left(\frac{4n+2}{n}\right)^n R_1 + 2\sigma_1\right]\left(-4\pi^3 x\sin(2\pi x)\right) \end{pmatrix} \tag{6.6}$$

We may thus employ the second solvability condition $\left\langle \mathbf{\Psi}^T, \mathbf{A}_1 \right\rangle = \mathbf{\Psi}^T \mathbf{L}_D \mathbf{\Phi}_1\Big|_{x=0}^{x=1}$ to obtain equation (2.23). From equation (2.23) one may have a quadratic equation for the eigenvalue $\sigma$ as

$$\begin{aligned} &16\pi^4\left(T_{20}-T\right)+\frac{3(17n+8)}{3n+2}\sigma^2+3(18n+10)\left(\frac{4n+2}{n}\right)^n \mathrm{Re}^{-1}\sigma \\ &+\frac{15n}{4}(3n+2)\left(\frac{4n+2}{n}\right)^{2n+1}\mathrm{Re}^{-2}=0 \end{aligned} \tag{6.7}$$

Using the solvability conditions $T_{21}=0$ and equation (2.23), the solution of equation (2.17) at $O\left(\varepsilon^2\right)$ may be obtained as

$$Q_2(x) = -\frac{d_0}{8\pi^2}\begin{bmatrix} \sigma_1\left(2\sigma_1+\alpha R_1\right)\left(2\pi x\left(\cos\left(2\pi x\right)+2\right)-3\sin\left(2\pi x\right)\right) \\ +4\pi\sigma_2\left(\cos\left(2\pi x\right)-1\right) \end{bmatrix} \tag{6.8}$$

and

$$H_2(x) = \frac{d_0\sin(\pi x)}{16\pi^2}\begin{bmatrix} 2\cos(\pi x)\left\{\left(8\pi^2x^2+\frac{43n+20}{2n+1}\right)\sigma_1^2+16\pi^2 x\sigma_2\right\} \\ +8\pi\sin(\pi x)\left\{\frac{(13n+6)x-(21n+10)}{2n+1}\sigma_1^2-4\sigma_2\right\} \\ +\alpha^2R_1^2\left\{\left(4\pi^2x^2+13\right)\cos(\pi x)+8\pi\left(2x-3\right)\sin(\pi x)\right\} \\ +16\alpha\begin{Bmatrix} \left(\frac{23n+13}{4(2n+1)}R_1\sigma_1+\pi^2x\left(R_2+xR_1\sigma_1\right)\right)\cos(\pi x)- \\ \left(R_2+R_1\sigma_1\left(\frac{(11n+6)-(7n+4)x}{2n+1}\right)\right)\pi\sin(\pi x) \end{Bmatrix} \end{bmatrix} \tag{6.9}$$

where $\alpha = \left(\frac{4n+2}{n}\right)^n(3n+2)$. The solutions for $Q_2\left(x\right)$ and $H_2\left(x\right)$ will eventually lead to the solution for $\mathbf{\Phi}_2$.

Using $\mathbf{\Phi}_2$ and $\mathbf{\Psi}$, from the third solvability condition $\left\langle \mathbf{\Psi}^T, \mathbf{A}_2 \right\rangle = \mathbf{\Psi}^T \mathbf{L}_D \mathbf{\Phi}_2 \Big|_{x=0}^{x=1}$, we can have an approximation for $T$ at the $O(\varepsilon^3)$ as given in equation (2.25). Further assuming $T = T_{20} + \varepsilon^2 T_{22}$ and $\mathrm{Re}^{-1} \approx \varepsilon^2 R_2$ such that $\sigma = \varepsilon\sigma_1 + \varepsilon^2\sigma_2$, and setting $R_1 = 0$ in (2.25) we obtain

$$\begin{aligned} T_{22} &= \frac{3}{16\pi^4}\frac{17n+8}{3n+2}\sigma_1^2 \\ R_2 &= -\frac{6(2n+1)\sigma_1^2 + (17n+8)\sigma_2}{2(3n+2)(9n+5)}\left(\frac{4n+2}{n}\right)^{-n} \end{aligned} \tag{6.10}$$

From which a second approximation for the eigenvalue is obtained as (2.26).

## Appendix E: Expansions for weakly nonlinear analysis

The functions appears in (3.2) are given in the component form as

$$g_0 = 0; f_{10} = 0; f_{20} = 0 \tag{7.1}$$

$$g_1 = -T_{k1}\frac{\partial^2 H_0}{\partial x^2} - 2n\left(\frac{4n+2}{n}\right)^n L_2 R_1 Q_0 - L_2\frac{\partial Q_0}{\partial \tau_0} \tag{7.2}$$

$$\begin{aligned} g_2 = &-\left(T_{k2}\frac{\partial^2 H_0}{\partial x^2} + T_{k1}\frac{\partial^2 H_1}{\partial x^2}\right) - L_2\left(\frac{\partial Q_1}{\partial \tau_0} + \frac{\partial Q_0}{\partial \tau_1}\right) \\ &- 2n\left(\frac{4n+2}{n}\right)^n L_2\left(R_2 Q_0 + R_1 Q_1 + \frac{n-1}{2}R_1 Q_0^2\right) \end{aligned} \tag{7.3}$$

$$\begin{aligned} g_3 = &-\left(T_{k3}\frac{\partial^2 H_0}{\partial x^2} + T_{k2}\frac{\partial^2 H_1}{\partial x^2} + T_{k1}\frac{\partial^2 H_2}{\partial x^2}\right) - L_2\left(\frac{\partial Q_2}{\partial \tau_0} + \frac{\partial Q_1}{\partial \tau_1} + \frac{\partial Q_0}{\partial \tau_2}\right) \\ &- 2n\left(\frac{4n+2}{n}\right)^n L_2\left(\begin{aligned} &R_3 Q_0 + R_2 Q_1 + (n-1)R_1 Q_0 Q_1 + \frac{1}{2}(n-1)R_2 Q_0^2 \\ &+\frac{1}{6}(n-1)(n-2)R_1 Q_0^3 \end{aligned}\right) \end{aligned} \tag{7.4}$$

$$f_{11} = -\frac{\partial H_0}{\partial \tau_0}; f_{12} = -\frac{\partial H_1}{\partial \tau_0} - \frac{\partial H_0}{\partial \tau_1}; f_{13} = -\frac{\partial H_2}{\partial \tau_0} - \frac{\partial H_1}{\partial \tau_1} - \frac{\partial H_0}{\partial \tau_2} \tag{7.5}$$

$$\begin{aligned} f_{21} = &-T_{k1}\frac{\partial^3 H_0}{\partial x^3} - (2n+1)T_{k0}H_0\frac{\partial^3 H_0}{\partial x^3} \\ &- 2\left(\frac{4n+2}{n}\right)^n\left((2n+1)R_1 H_0 - nR_1 Q_0\right) + \frac{\partial Q_0}{\partial \tau_0} \\ &-\left(\frac{4n+2}{3n+2}\right)\left\{\left((2n-1)H_0 + 2Q_0\right)\left(\frac{\partial H_0}{\partial x} - \frac{\partial Q_0}{\partial x}\right) - 2\frac{\partial Q_1}{\partial x}\right\} \end{aligned} \tag{7.6}$$

$$f_{22} = -T_{k2}\frac{\partial^3 H_0}{\partial x^3} - T_{k1}\frac{\partial^3 H_1}{\partial x^3} - \left(2n+1\right)\left(T_{k0}H_1 + nT_{k0}H_0^2 + T_{k1}H_0\right)\frac{\partial^3 H_0}{\partial x^3}$$
$$-\left(2n+1\right)T_{k0}H_0\frac{\partial^3 H_1}{\partial x^3} + \left(\frac{\partial Q_1}{\partial \tau_0} + \frac{\partial Q_0}{\partial \tau_1} + 2nH_0\frac{\partial Q_0}{\partial \tau_0}\right)$$
$$-2\left(\frac{4n+2}{n}\right)^n \begin{Bmatrix} \left(2n+1\right)\left(R_2H_0 + R_1H_1 + nR_1H_0^2\right) \\ -n\left(R_2Q_0 + R_1Q_1 + \frac{1}{2}\left(n-1\right)R_1Q_0^2\right) \end{Bmatrix}$$
$$+\left(\frac{4n+2}{3n+2}\right)\left\{Q_1\frac{\partial Q_0}{\partial x} + Q_0\frac{\partial Q_1}{\partial x} + \frac{\partial Q_2}{\partial x} + 2(2n-1)H_0\left(Q_0\frac{\partial Q_0}{\partial x} + \frac{\partial Q_1}{\partial x}\right)\right\}$$
$$-\left(\frac{4n+2}{3n+2}\right)\left\{\left(2Q_1 + Q_0^2\right)\frac{\partial H_0}{\partial x} + 2Q_0\frac{\partial H_1}{\partial x} + 2(n-1)H_0\left(2Q_0\frac{\partial H_0}{\partial x} + \frac{\partial H_1}{\partial x}\right)\right\}$$
$$-\left(\frac{4n+2}{3n+2}\right)\begin{Bmatrix} (n-1)H_0^2\left((2n-3)\frac{\partial H_0}{\partial x} - 2(2n-1)\frac{\partial Q_0}{\partial x}\right) \\ +2H_1\left((n-1)\frac{\partial H_0}{\partial x} - (2n-1)\frac{\partial Q_0}{\partial x}\right) \end{Bmatrix} \tag{7.7}$$

$$
\begin{aligned}
f_{23} = &-T_{k3}\frac{\partial^3 H_0}{\partial x^3} - T_{k2}\frac{\partial^3 H_1}{\partial x^3} - T_{k1}\frac{\partial^3 H_2}{\partial x^3} \\
&-(2n+1)T_{k0}H_0\frac{\partial^3 H_2}{\partial x^3} - (2n+1)\left(T_{k1}H_0 + nT_{k0}H_0^2 + T_{k0}H_1\right)\frac{\partial^3 H_1}{\partial x^3} \\
&-(2n+1)\begin{pmatrix} T_{k2}H_0 + T_{k1}H_1 + T_{k0}H_2 + T_{k0}H_0H_1 \\ +nT_{k1}H_0^2 + \frac{n}{3}(2n-1)T_{k0}H_0^3 \end{pmatrix}\frac{\partial^3 H_0}{\partial x^3} \\
&-2\left(\frac{4n+2}{n}\right)^n \left\{ \begin{matrix} (2n+1)\begin{pmatrix} R_3H_0 + R_2H_1 + R_1H_2 + nR_2H_0^2 \\ +\frac{1}{3}n(2n-1)R_1H_0^3 + 2nR_1H_0H_1 \end{pmatrix} \\ -n\begin{pmatrix} R_3Q_0 + R_2Q_1 + R_1Q_2 + \frac{1}{2}(n-1)R_2Q_0^2 \\ +\frac{1}{6}(n^2-3n+2)R_1Q_0^3 + (n-1)R_1Q_0Q_1 \end{pmatrix} \end{matrix} \right\} \\
&+\left(\frac{\partial Q_2}{\partial \tau_0} + \frac{\partial Q_1}{\partial \tau_1} + \frac{\partial Q_0}{\partial \tau_2} + n(2n-1)H_0^2\frac{\partial Q_0}{\partial \tau_0}\right) \\
&+2n\left(H_0\frac{\partial Q_0}{\partial \tau_1} + H_1\frac{\partial Q_0}{\partial \tau_0} + H_0\frac{\partial Q_1}{\partial \tau_0}\right) \\
&+\left(\frac{4n+2}{3n+2}\right)\left\{ \begin{array}{l}
-2(n-1)\left((2n-3)H_0H_1 + H_2 + Q_0^2 + 2H_0Q_1\right)\frac{\partial H_0}{\partial x} \\
-2\left(Q_0Q_1 + Q_2\right)\frac{\partial H_0}{\partial x} - 2\left((n-1)H_0 + Q_0\right)\frac{\partial H_2}{\partial x} \\
-\left((2n^2-5n+3)H_0^2 + 2(n-1)H_1 + Q_0^2 + 2Q_1\right)\frac{\partial H_1}{\partial x} \\
+2\left(2(n-1)(2n-1)H_0H_1 + (2n-1)\left(H_2 + H_0Q_1\right)\right)\frac{\partial Q_0}{\partial x} \\
-\frac{2}{3}(n-1)(2n-3)H_0^3\left((n-2)\frac{\partial H_0}{\partial x} - (2n-1)\frac{\partial Q_0}{\partial x}\right) \\
+2\frac{\partial Q_3}{\partial x} + 2\left((2n-1)H_0 + Q_0\right)\frac{\partial Q_2}{\partial x} + 2Q_2\frac{\partial Q_0}{\partial x} \\
-2H_1Q_0\left(2(n-1)\frac{\partial H_0}{\partial x} - (2n-1)\frac{\partial Q_0}{\partial x}\right) \\
-2(n-1)H_0^2Q_0\left((2n-3)\frac{\partial H_0}{\partial x} - (2n-1)\frac{\partial Q_0}{\partial x}\right) \\
+2\left((n-1)(2n-1)H_0^2 + (2n-1)H_1 + Q_1\right)\frac{\partial Q_1}{\partial x} \\
-2H_0Q_0\left(2(n-1)\frac{\partial H_1}{\partial x} - (2n-1)\frac{\partial Q_1}{\partial x}\right)
\end{array} \right\}
\end{aligned}
\tag{7.8}
$$

For $k=1$, the solution for $H_1$ can be obtained from the equation and boundary conditions

$$T_{10}\frac{\partial^3 H_1}{\partial x^3}+\frac{4n+2}{3n+2}\frac{\partial H_1}{\partial x}=f_{21}(x;\tau_0,\tau_1,\tau_2)$$
$$H_1(0)=0;T_{10}\frac{\partial^2 H_1(1)}{\partial x^2}=g_1(1;\tau_0,\tau_1,\tau_2) \tag{7.9}$$

as

$$H_1=\frac{1}{8\pi(2n+1)}\left\{\begin{array}{l}2(2n+1)\pi\mathrm{A}_0^2\left(\cos(2\pi x)+4\cos(\pi x)-5\right)\\ -(3n+2)\mathrm{A}_0\pi^3 T_{11}\left(2\pi\left(1-\cos(\pi x)+x\cos(\pi x)\right)-3\sin(\pi x)\right)\\ +4\left(\frac{4n+2}{n}\right)^n(3n+2)(2n+1)\mathrm{A}_0 R_1\left(2\cos(\pi x)-2+\pi x\sin(\pi x)\right)\\ +8(2n+1)\frac{\partial \mathrm{A}_0}{\partial \tau_0}\left(2\cos(\pi x)-2+\pi x\sin(\pi x)\right)\end{array}\right\} \tag{7.10}$$

Now the remaining boundary condition $H_1(1)=0$ gives the solvability condition (3.5).

For $k=2$, the solution for $Q_1$ and $H_1$ are obtained from the equations and boundary conditions of order $\varepsilon^2$

$$T_{20}\frac{\partial^3 H_1}{\partial x^3}+\frac{4n+2}{3n+2}\frac{\partial H_1}{\partial x}=f_{21}(x;\tau_0,\tau_1,\tau_2)$$
$$H_1(0)=0;H_1(1)=0;T_{20}\frac{\partial^2 H_1(1)}{\partial x^2}=g_1(1;\tau_0,\tau_1,\tau_2) \tag{7.11}$$

which yields

$$Q_1=\frac{1}{k\pi}\frac{\partial \mathrm{A}_0}{\partial \tau_0}\left(\cos(2\pi x)-1\right)$$

$$H_1=\frac{1}{\pi}\left\{\begin{array}{l}2\pi\mathrm{A}_0^2\sin^4(\pi x)+\frac{\partial \mathrm{A}_0}{\partial \tau_0}\left(\pi x\sin(2\pi x)-\sin^2(\pi x)\right)\\ -\frac{(3n+2)}{2(2n+1)}\mathrm{A}_0\pi^3 T_{21}\left(4\pi\left(1-\cos(2\pi x)+x\cos(2\pi x)\right)-3\sin(2\pi x)\right)\\ +\left(\frac{4n+2}{n}\right)^n\frac{(3n+2)}{2(2n+1)}\mathrm{A}_0 R_1\left(\cos(2\pi x)-1+\pi x\sin(2\pi x)\right)\end{array}\right\} \tag{7.12}$$

which gives the solvability condition $T_{21}=0$.

Proceeding in the similar way, the solvability condition for the equations of order $\varepsilon^3$ is given in (3.9) whereas the condition for order $\varepsilon^4$ is

$$
\begin{aligned}
32(3n+2)\pi^4 A_0 T_{23} &= 9\alpha^3(2n+1)A_0R_1^3 + \left(204n+96\right)\frac{\partial^2 A_0}{\partial\tau_0\partial\tau_1} - 12(2n+5)\frac{\partial^3 A_0}{\partial\tau_0^3} \\
&+ 2\alpha^2(2n+1)R_1\left\{15R_2A_0 + (4n+11)\pi A_0^2R_1 + 12R_1\frac{\partial A_0}{\partial\tau_0}\right\} \\
&+ \alpha(2n+1)\left\{4(2n+7)\pi R_2A_0^2 + 3(4n-3)\pi^2A_0^3R_1\right\} \\
&+ 12\alpha\left\{(9n+5)\left(R_1\frac{\partial A_0}{\partial\tau_1} + R_2\frac{\partial A_0}{\partial\tau_0}\right) - R_1\frac{\partial^2 A_0}{\partial\tau_0^2}\right\} \\
&- 16\alpha\left(n^2+10n+4\right)\pi R_1A_0\frac{\partial A_0}{\partial\tau_0} + 4\pi(n-1)A_0\frac{\partial^2 A_0}{\partial\tau_0^2} \\
&+ 4\pi^2\left\{16(3n+2)\pi^2T_{22} - 9(2n+1)A_0^2\right\}\frac{\partial A_0}{\partial\tau_0} - 2\pi(85n+44)\left(\frac{\partial A_0}{\partial\tau_0}\right)^2
\end{aligned}
\tag{7.13}
$$

Equation (7.13) gives a disturbance equation for the amplitude equation (3.14) of the form

$$
\frac{\partial^2 A_0}{\partial\tau_0\partial\tau_1} = \frac{1}{18\left(17n+8\right)^2}
\left[
\begin{aligned}
&-6\alpha\left(17n+8\right)R_2\left\{(2n+1)(2n+7)\pi A_0^2 + 3(9n+5)\frac{\partial A_0}{\partial\tau_0}\right\} \\
&-2(n-1)\pi^3A_0^2\left\{16(3n+2)\pi^2T_{22} + (6n+3)A_0^2\right\} \\
&-36(2n+1)\pi^2\left\{16(3n+2)\pi^2T_{22} - 3(11n+5)A_0^2\right\}\frac{\partial A_0}{\partial\tau_0} \\
&+3\pi(17n+8)(85n+44)\left(\frac{\partial A_0}{\partial\tau_0}\right)^2
\end{aligned}
\right]
\tag{7.14}
$$

## Appendix F: Method for heteroclinic and homoclinic orbits

The orbits of the system (3.19) for a given Hamiltonian $H$ is given by the branches

$$
Y_{\pm}^{\mathcal{H}}\left(X^{\mathcal{H}}\right) = \pm\sqrt{2}\sqrt{\left(\frac{2n+1}{17n+8}\right)\frac{\pi^2}{4}\left(X^{\mathcal{H}}\right)^4 - \frac{16}{3}\left(\frac{3n+2}{17n+8}\right)\frac{\pi^4}{2}\left(X^{\mathcal{H}}\right)^2\lambda + \mathcal{H}}
\tag{8.1}
$$

Equation (8.1) gives the branches of the periodic orbits when $0 < \mathcal{H} < \dfrac{64\pi^6(3n+2)^2\lambda^2}{9(2n+1)(17n+8)}$ and when $H \to 64\pi^6(3n+2)^2\lambda^2/9(2n+1)(17n+8)$, it gives the heteroclinic branches as

$$
Y_{\pm}^{het}\left(X_{\pm}^{het}\right) = \pm\sqrt{2}\sqrt{\left(\frac{2n+1}{17n+8}\right)\frac{\pi^2}{4}\left(X_{\pm}^{het}\right)^4 - \frac{16}{3}\left(\frac{3n+2}{17n+8}\right)\frac{\pi^4}{2}\left(X_{\pm}^{het}\right)^2\lambda + \frac{64\pi^6(3n+2)^2\lambda^2}{9(2n+1)(17n+8)}}
\tag{8.2}
$$

for $X_{\pm}^{het}$ lies between the two saddle points, i.e., $X_{\pm}^{het} \in \left[-4\pi\sqrt{\frac{\lambda}{3}\left(\frac{3n+2}{2n+1}\right)}, 4\pi\sqrt{\frac{\lambda}{3}\left(\frac{3n+2}{2n+1}\right)}\right]$. + (-) denotes the upper (lower) branches of the orbits.

The Melnikov function $\mathrm{M} = \int_{\tau} Yg\left(X,Y\right)d\tau = \int_{X} g\left(X,Y\right)dX$ for the upper (lower) heteroclinic orbits can be found as

$$\begin{aligned}\mathrm{M}^{+}\left(\mathrm{M}^{-}\right) &= \int_{-4\pi\sqrt{\lambda(3n+2)/(6n+3)}}^{4\pi\sqrt{\lambda(3n+2)/(6n+3)}} g\left(X_{+}^{het},Y_{+}^{het}\right)dX_{+}^{het} \\ &= \frac{1}{15\sqrt{3}}\frac{128\pi^4(3n+2)^{5/2}\lambda^{3/2}}{27(2n+1)^{1/2}(17n+8)^2}\left\{1376\pi^4\lambda - 15\left(\frac{4n+2}{n}\right)^n(2n+7)(17n+8)\mathrm{R}_2\right\}\end{aligned} \tag{8.3}$$

The hetereoclinic orbits are preserved under the perturbation, which implies $\mathrm{M}^{+}=0$ and gives the condition (3.21).

The Melnikov function for the periodic orbits (8.1) can be defined as

$$\mathrm{M}^{\mathcal{H}} = \int_{X_1}^{X_2}\left\{g\left(X^{\mathcal{H}},Y_{+}^{\mathcal{H}}\right) - g\left(X^{\mathcal{H}},Y_{-}^{\mathcal{H}}\right)\right\}dX^{\mathcal{H}} \tag{8.4}$$

where $X_1$ and $X_2$ $\left(X_1 < X_2\right)$ are the intersection points of the periodic orbits with the X-axis. These points can be approximated for small $H$ as

$$X_{1,2} = \pm\frac{1}{2\pi^2}\sqrt{\frac{3(17n+8)}{(3n+2)}\frac{\mathcal{H}}{2\lambda}} + O\left(\mathcal{H}^{3/2}\right) \tag{8.5}$$

This yields an approximation for the above integral as

$$\mathrm{M}^{\mathcal{H}} = -\frac{\mathcal{H}\sqrt{3(3n+2)(17n+8)}}{\sqrt{\lambda}(17n+8)^2\pi^2}\left\{\left(\frac{4n+2}{n}\right)^n(9n+5)(17n+8)R_2 - 32(2n+1)\pi^4\lambda\right\} \tag{8.6}$$

The condition $\mathrm{M}^{\mathcal{H}}=0$ as $\mathcal{H}\to 0$ gives the condition for the existence of Hopf bifurcation, as found in (2.27). The condition $\mathrm{M}^{\mathcal{H}}=0$ as $\mathcal{H}\to 64\pi^6(3n+2)^2\lambda^2/9(2n+1)(17n+8)$ gives the condition for homoclinic bifurcation which may be obtained from

$$\mathrm{M}^{\mathcal{H}}\Big|_{\mathcal{H}\to 64\pi^6(3n+2)^2\lambda^2/9(2n+1)(17n+8)} = \mathrm{M}^{+} + \mathrm{M}^{-} \tag{8.7}$$

which implies (3.22).

**Appendix G: Weakly non-linear analysis for large $L_2$ near $T_{20}$.**

For large $L_2$, substituting the expansion (3.1) and (3.24) in equations (2.9)-(2.10) and using the scaling $\left(T - T_{20}\right) \approx \varepsilon^2 T_{22}$, $\mathrm{Re}^{-1} \approx \varepsilon R_1$ and $\ell \approx \varepsilon^2 \ell_2$ for the Hopf branch, the differential equations for leading order, $O(\varepsilon)$, is given by

$$\begin{gathered}\frac{\partial Q_0}{\partial x} = 0, \\ T_{20}\frac{\partial^3 H_0}{\partial x^3} + \frac{4n+2}{3n+2}\frac{\partial H_0}{\partial x} = 0 \\ H_0\left(0\right) = 0, Q_0\left(0\right) = 0 \\ H_0\left(1\right) = 0, \frac{\partial Q_0(1)}{\partial \tau_0} + 2n\left(\frac{4n+2}{n}\right)^n R_1 Q_0(1) = 0\end{gathered} \tag{8.8}$$

Equation (8.8) have a solution of the form (3.25).
At $O(\varepsilon^2)$, the equations are

$$\begin{gathered}\frac{\partial Q_1}{\partial x}=f_{11}(x;\tau_0,\tau_1,\tau_2),\\ T_{20}\frac{\partial^3 H_1}{\partial x^3}+\frac{4n+2}{3n+2}\frac{\partial H_1}{\partial x}=f_{21}(x;\tau_0,\tau_1,\tau_2)\\ H_1(0)=0,Q_1(0)=0,H_1(1)=0\\ \ell_2 T_{20}\frac{\partial^2 H_0(1)}{\partial x^2}+\frac{\partial Q_1(1)}{\partial \tau_0}+2n\left(\frac{4n+2}{n}\right)^n R_1 Q_1(1)=0\end{gathered}\tag{8.9}$$

Where the functions $f_{11}$ and $f_{21}$ are defined in (7.5)-(7.6) of Appendix E. The solution of (8.9) requires the solvability conditions

$$-6\pi(2n+1)A_0B_0-3(2n+1)\left\{\left(\frac{4n+2}{n}\right)^n(3n+2)R_1B_0+2\frac{\partial B_0}{\partial \tau_0}\right\}=0\tag{8.10}$$

and

$$2\ell_2(2n+1)B_0-(3n+2)\left\{\left(\frac{4n+2}{n}\right)^n nR_1\frac{\partial B_0}{\partial \tau_0}+\frac{\partial^2 B_0}{\partial \tau_0^2}\right\}=0\tag{8.11}$$

The second condition (8.11) reveals that the system has an unstable mode. So, to get rid of that mode assuming $B_0=0$, the solution of the above system can be obtained as

$$\begin{gathered}Q_1=-\frac{1}{2\pi}\frac{\partial A_0}{\partial \tau_0}\left(1-\cos(2\pi x)\right)\\ H_1=\frac{A_1}{2\pi}\sin(2\pi x)+B_1\left(1-\cos(2\pi x)\right)+\frac{A_0^2}{4}\left(\cos(4\pi x)-\cos(2\pi x)\right)\\ +\left(\frac{(3n+2)}{2}\left(\frac{4n+2}{n}\right)^n R_1A_0+\frac{\partial A_0}{\partial \tau_0}\right)x\sin(2\pi x)\end{gathered}\tag{8.12}$$

where $A_1=A_1(\tau_0,\tau_1,\tau_2)$ and $B_1=B_1(\tau_0,\tau_1,\tau_2)$. Since, $\sin(2\pi x)$ is present in the solution for $H_0$, without loss of generality we can assume $A_1=0$.
At $O(\varepsilon^3)$, the set of differential equations and boundary conditions are

$$\begin{gathered}\frac{\partial Q_2}{\partial x}=f_{12}(x;\tau_0,\tau_1,\tau_2),\\ T_{20}\frac{\partial^3 H_2}{\partial x^3}+\frac{4n+2}{3n+2}\frac{\partial H_2}{\partial x}=f_{22}(x;\tau_0,\tau_1,\tau_2)\\ H_1(0)=0,Q_1(0)=0,H_1(1)=0\\ \ell_2 T_{20}\frac{\partial^2 H_1(1)}{\partial x^2}+\frac{\partial Q_2(1)}{\partial \tau_0}+\frac{\partial Q_1(1)}{\partial \tau_1}+2n\left(\frac{4n+2}{n}\right)^n R_1 Q_2(1)=0\end{gathered}\tag{8.13}$$

Where the functions $f_{12}$ and $f_{22}$ are defined in (7.5)-(7.7) of Appendix E.
In order to solve (8.13), it requires the second set of solvability conditions which are

$$\begin{aligned}0=2&\left(\frac{4n+2}{n}\right)^{n}(2n+1)(3n+2)(4n+11)\pi R_1 A_0^2+30(2n+1)\pi^2 A_0^3\\&+A_0\left\{(3n+2)\left\{3\left(\frac{4n+2}{n}\right)^{2n}\left(6n^2+7n+2\right)R_1^2-32\pi^4 T_{22}\right\}\right\}\\&+24(2n+1)\pi A_0\left\{-2\pi B_1+\frac{\partial A_0}{\partial\tau_0}\right\}-48\pi(2n+1)\frac{\partial B_1}{\partial\tau_0}+6n\frac{\partial^2 A_0}{\partial\tau_0^2}\\&+6\left\{-2\left(\frac{4n+2}{n}\right)^{n}(3n+2)R_1\left[2(2n+1)\pi B_1-(n+1)\frac{\partial A_0}{\partial\tau_0}\right]\right\}\end{aligned}\tag{8.14}$$

and

$$\begin{aligned}0=2\ell_2&\frac{(2n+1)}{(3n+2)}\left\{2\left(\frac{4n+2}{n}\right)^{n}(3n+2)R_1A_0-3\pi A_0^2+4\left(\pi B_1+\frac{\partial A_0}{\partial\tau_0}\right)\right\}\\&-2\left(\frac{4n+2}{n}\right)^{n}nR_1\left\{-\left(\frac{4n+2}{n}\right)^{n}(3n+2)R_1\frac{\partial A_0}{\partial\tau_0}+4\pi\frac{\partial B_1}{\partial\tau_0}-2\frac{\partial^2 A_0}{\partial\tau_0^2}\right\}\\&+(3n+2)\left(\frac{4n+2}{n}\right)^{n}R_1\frac{\partial^2 A_0}{\partial\tau_0^2}-4\pi\frac{\partial^2 B_1}{\partial\tau_0^2}+2\frac{\partial^3 A_0}{\partial\tau_0^3}\end{aligned}\tag{8.15}$$

Using the original parameters and defining the amplitude functions $A=\varepsilon A_0$, $B=\varepsilon^2 B_1$ and $\tau_0=\varepsilon t$ the above conditions (8.14)-(8.15) can be rewritten as

$$\begin{aligned}0=2&\left(\frac{4n+2}{n}\right)^{n}(2n+1)(3n+2)(4n+11)\pi\,\mathrm{Re}^{-1}A^2+30(2n+1)\pi^2A^3\\&+A(3n+2)\left\{3\left(\frac{4n+2}{n}\right)^{2n}\left(6n^2+7n+2\right)\mathrm{Re}^{-2}-32\pi^4T_{22}\right\}\\&+24(2n+1)\pi A\left\{-2\pi B+\frac{\partial A}{\partial t}\right\}-48\pi(2n+1)\frac{\partial B}{\partial t}+6n\frac{\partial^2 A}{\partial t^2}\\&-12\left(\frac{4n+2}{n}\right)^{n}(3n+2)\mathrm{Re}^{-1}\left\{2(2n+1)\pi B-(n+1)\frac{\partial A}{\partial t}\right\}\end{aligned}\tag{8.16}$$

and

$$\begin{aligned}0=2\ell&\frac{(2n+1)}{(3n+2)}\left\{2\left(\frac{4n+2}{n}\right)^{n}(3n+2)\mathrm{Re}^{-1}A-3\pi A^2+4\left(\pi B+\frac{\partial A}{\partial t}\right)\right\}\\&-2\left(\frac{4n+2}{n}\right)^{n}n\,\mathrm{Re}^{-1}\left\{-\left(\frac{4n+2}{n}\right)^{n}(3n+2)\mathrm{Re}^{-1}\frac{\partial A}{\partial t}+4\pi\frac{\partial B}{\partial t}-2\frac{\partial^2 A}{\partial t^2}\right\}\\&+(3n+2)\left(\frac{4n+2}{n}\right)^{n}\mathrm{Re}^{-1}\frac{\partial^2 A}{\partial t^2}-4\pi\frac{\partial^2 B}{\partial t^2}+2\frac{\partial^3 A}{\partial t^3}\end{aligned}\tag{8.17}$$

Rescaling the variables by $\tilde{\ell} = \ell / \mathrm{Re}^{-2}$, $\tilde{t} = t\,\mathrm{Re}^{-1}$, $\tilde{A} = A / \mathrm{Re}^{-1}$, $\tilde{B} = B / \mathrm{Re}^{-2}$, $\tilde{T}_{22} = T_{22} / \mathrm{Re}^{-2}$, the above two conditions can be made free from $\mathrm{Re}^{-1}$, and we can write (8.16)-(8.17) as

$$\begin{aligned} 0 = {} & 2\left(\frac{4n+2}{n}\right)^{n}(2n+1)(3n+2)(4n+11)\pi\tilde{A}^{2} + 30(2n+1)\pi^{2}\tilde{A}^{3} \\ & + \tilde{A}(3n+2)\left\{3\left(\frac{4n+2}{n}\right)^{2n}\left(6n^{2}+7n+2\right) - 32\pi^{4}\tilde{T}_{22}\right\} \\ & + 24(2n+1)\pi\tilde{A}\left\{\frac{\partial\tilde{A}}{\partial\tilde{t}} - 2\pi\tilde{B}\right\} - 48\pi(2n+1)\frac{\partial\tilde{B}}{\partial\tilde{t}} + 6n\frac{\partial^{2}\tilde{A}}{\partial\tilde{t}^{2}} \\ & - 12\left(\frac{4n+2}{n}\right)^{n}(3n+2)\left[2(2n+1)\pi\tilde{B} - (n+1)\frac{\partial\tilde{A}}{\partial\tilde{t}}\right] \end{aligned} \tag{8.18}$$

and

$$\begin{aligned} 0 = {} & 2\tilde{\ell}\frac{(2n+1)}{(3n+2)}\left\{2\left(\frac{4n+2}{n}\right)^{n}(3n+2)\tilde{A} - 3\pi\tilde{A}^{2} + 4\left(\pi\tilde{B} + \frac{\partial\tilde{A}}{\partial\tilde{t}}\right)\right\} \\ & - 2\left(\frac{4n+2}{n}\right)^{n} n\left\{-\left(\frac{4n+2}{n}\right)^{n}(3n+2)\frac{\partial\tilde{A}}{\partial\tilde{t}} + 4\pi\frac{\partial\tilde{B}}{\partial\tilde{t}} - 2\frac{\partial^{2}\tilde{A}}{\partial\tilde{t}^{2}}\right\} \\ & + (3n+2)\left(\frac{4n+2}{n}\right)^{n}\frac{\partial^{2}\tilde{A}}{\partial\tilde{t}^{2}} - 4\pi\frac{\partial^{2}\tilde{B}}{\partial\tilde{t}^{2}} + 2\frac{\partial^{3}\tilde{A}}{\partial\tilde{t}^{3}} \end{aligned} \tag{8.19}$$

Solving for $\frac{\partial^{2}\tilde{A}}{\partial\tilde{t}^{2}}$ from (8.18) and substituting in (8.19), we have a system of second order differential equation of the form (3.26) and (3.27). Choosing $\frac{\partial\tilde{A}}{\partial\tilde{t}} = \tilde{X}(\tilde{t})$ and $\frac{\partial\tilde{B}}{\partial\tilde{t}} = \tilde{Y}(\tilde{t})$, the second order system (3.26)-(3.27) with two equations can be converted into a first order system of four equations and is given by

$$\frac{\partial}{\partial\tilde{t}}\begin{bmatrix}\tilde{A}\\ \tilde{B}\\ \tilde{X}\\ \tilde{Y}\end{bmatrix} = \begin{bmatrix}0 & 0 & 1 & 0\\ 0 & 0 & 0 & 1\\ a_{11} & a_{14} & a_{16} & a_{18}\\ \beta_{11} & a_{23} & a_{24} & a_{27}\end{bmatrix}\begin{bmatrix}\tilde{A}\\ \tilde{B}\\ \tilde{X}\\ \tilde{Y}\end{bmatrix} + \begin{bmatrix}0\\ 0\\ NL_{1}\\ NL_{2}\end{bmatrix} \tag{8.20}$$

Where the non-linear terms $NL_1$, $NL_2$ are given by

$$\begin{aligned} NL_{1} &= \tilde{A}\left(a_{12}\tilde{A} + a_{13}\tilde{A}^{2} + a_{15}\tilde{B} + a_{17}\tilde{X}\right) \\ NL_{2} &= \tilde{A}\left(\beta_{12}\tilde{B} + \beta_{13}\tilde{X} + \beta_{14}\tilde{Y}\right) + \tilde{A}^{2}\left(\alpha_{11} + \alpha_{12}\tilde{B} + \alpha_{13}\tilde{X}\right) \\ & \quad + a_{21}\tilde{A}^{3} + a_{22}\tilde{A}^{4} + a_{25}\tilde{B}\tilde{X} + a_{26}\tilde{X}^{2} \end{aligned} \tag{8.21}$$

and the coefficients are given as

$$a_{11} = -\frac{3n+2}{6n}\left\{3\left(\frac{4n+2}{n}\right)^{2n}\left(6n^2+7n+2\right)-32\pi^4\tilde{T}_{22}\right\}$$

$$a_{12} = -\frac{2}{6n}\left(\frac{4n+2}{n}\right)^{n}\left(24n^3+94n^2+85n+22\right)\pi$$

$$a_{13} = -\frac{30}{6n}\left(2n+1\right)\pi^2$$

$$a_{14} = \frac{24}{6n}\left(\frac{4n+2}{n}\right)^{n}\left(6n^2+7n+2\right)\pi$$

$$a_{15} = \frac{48}{6n}\left(2n+1\right)\pi^2$$

$$a_{16} = -\frac{12}{6n}\left(\frac{4n+2}{n}\right)^{n}\left(3n^2+5n+2\right)$$

$$a_{17} = -\frac{24}{6n}\left(2n+1\right)\pi$$

$$a_{18} = \frac{48}{6n}\left(2n+1\right)\pi \tag{8.22}$$

$$a_{21} = -\frac{1}{12n}\frac{\left(2n+1\right)}{\left(7n+4\right)}\left(\frac{4n+2}{n}\right)^{n}\left(192n^3+827n^2+950n+296\right)\pi$$

$$a_{22} = -\frac{10}{n}\frac{\left(2n+1\right)^2}{\left(7n+4\right)}\pi^2$$

$$a_{23} = \frac{\left(2n+1\right)}{n\left(7n+4\right)\left(3n+2\right)}\left\{-2n^2\tilde{\ell}+\left(\frac{4n+2}{n}\right)^{2n}\left(3n+2\right)^2\left(5n^2+18n+8\right)\right\}$$

$$a_{24} = -\frac{1}{24n\left(7n+4\right)\left(3n+2\right)\pi}\left\{\begin{aligned}&48n^2\tilde{\ell}\left(2n+1\right)+96n\left(3n+2\right)^2\pi^4\tilde{T}_{22}\\&+6\left(3n+2\right)^2\left(\frac{4n+2}{n}\right)^{2n}\left(2n+1\right)\left(3n^2+18n+16\right)\end{aligned}\right\}$$

$$a_{25} = -\frac{4\left(2n+1\right)}{\left(7n+4\right)}\pi$$

$$a_{26} = \frac{2\left(6n^2+7n+2\right)}{\left(7n+4\right)\left(3n+2\right)}$$

$$a_{27} = \frac{2}{n\left(7n+4\right)}\left(\frac{4n+2}{n}\right)^{2n}\left(5n^3+34n^2+32n+8\right) \tag{8.23}$$

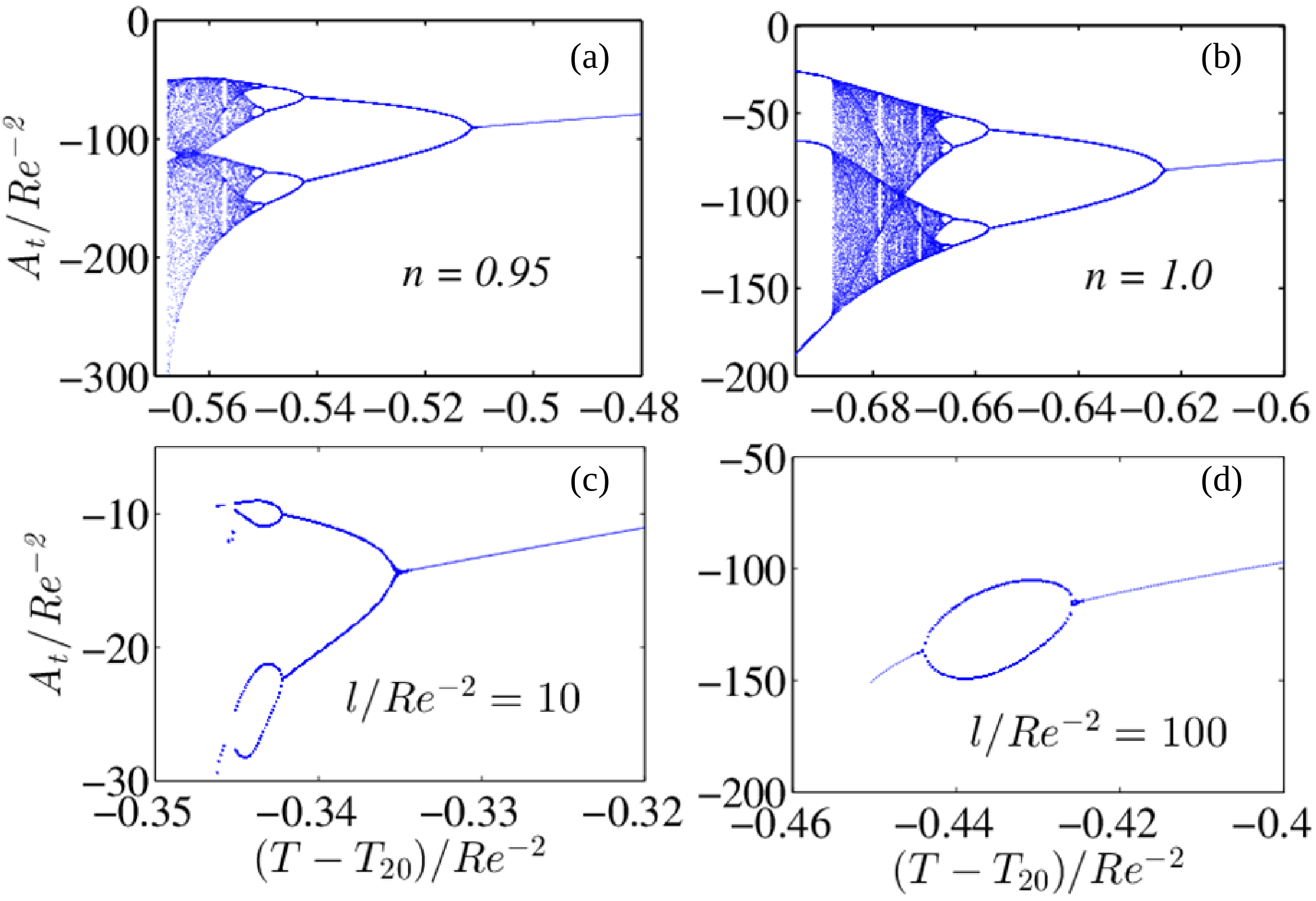

Figure G1: The Poincare section $\tilde{A}=0, \partial\tilde{A}/\partial\tilde{t}<0$ obtained from the system (8.20) is illustrated for (a) $\tilde{\ell}=100$ and shear-thinning fluid ($n$ = 0.95) (b) $\tilde{\ell}=100$ and Newtonian fluid ($n$ = 1.0) (c) $n=0.9$ , $\ell/\mathrm{Re}^{-2}=10$ (d) $n=0.9$ , $\ell/\mathrm{Re}^{-2}=100$.

$$\beta_{11}=-\frac{1}{24n(7n+4)}\left(\frac{4n+2}{n}\right)^{n}\left\{24n^2\tilde{\ell}(2n+1)+(3n+2)(5n^2+18n+8)\right\}$$
$$\left\{3\left(\frac{4n+2}{n}\right)^{2n}(6n^2+7n+2)-32\pi^4\tilde{T}_{22}\right\}$$
$$\beta_{12}=\frac{2(2n+1)}{n(7n+4)}\left(\frac{4n+2}{n}\right)^{n}(29n^2+46n+16)\pi \tag{8.24}$$
$$\beta_{13}=\frac{2(2n+1)}{3n(7n+4)}\left(\frac{4n+2}{n}\right)^{n}(-6n^3+5n^2+46n+24)$$
$$\beta_{14}=\frac{4}{n}(2n+1)\pi$$

$$\alpha_{11} = -\frac{(2n+1)}{12n(7n+4)(3n+2)}\left\{\begin{array}{l} -18n^2\tilde{\ell} - 128(3n+2)^2\pi^4\tilde{T}_{22} \\ +\left(\frac{4n+2}{n}\right)^{2n}(3n+2)^2\left(20n^3+199n^2+314n+112\right) \end{array}\right\}$$

$$\alpha_{12} = \frac{384(2n+1)}{24n(7n+4)(3n+2)}\left(6n^2+7n+2\right)\pi^2 \quad (8.25)$$

$$\alpha_{13} = -\frac{(2n+1)(17n+16)}{2n(7n+4)}\pi$$

## Appendix H: Comparison of the qualitative behaviour using integrated energy equation (following Luchini & Charru, 2010)

The momentum integral method was first proposed by Shkadov (1967) for the flow of a Newtonian fluid down an inclined plane, which is the simplest method and only involves the variables film thickness $h$, local volume flow rate $q$. Assuming a local parabolic velocity profile, the boundary layer equations are integrated to obtain a Saint-Venant type equation (which gives a good approximation for the dynamics of fluid flow when the wave length is much larger than the film thickness). Although this model fails to predict the quantitative dynamical behaviour of the stability, it is nevertheless able to give a good prediction about the qualitative behaviour. Later on, Ruyer-Quil & Manneville (1998, 2000), proposed a modified Shkadov model to address this quantitative discrepancy. Relying on the qualitative predictions, Shkadovφs model has widely been used to understand the dynamical behaviour of the membrane undergoing oscillations in presence of a fluid flow [Stewart et al. 2009, 2010; Xu et al. 2013, 2014; Xu & Jensen 2015].

In a more rigorous manner, Luchini & Charru (2010) provide a consistent equation by depth averaging the energy equation together with the mass conservation. In accordance to this approach, here we have provided the depth averaged energy equation together with a set of demonstrative numerical results to compare with the qualitative behaviour that are obtained from the momentum integral approach.

The depth average equation for the energy is given by

$$\frac{1}{2}\frac{\partial}{\partial t}\int_0^{h(x,t)} u^2 dy + \frac{1}{2}\frac{\partial}{\partial x}\int_0^{h(x,t)} u^3 dy + \int_0^{h(x,t)} u\frac{\partial p}{\partial x}dy = \frac{1}{\text{Re}}\int_0^{h(x,t)} u\frac{\partial \tau_{yx}}{\partial y}dy \quad (8.26)$$

After substituting the base state velocity profile as given in equation (2.6), it can be simplified as

$$\left(\frac{4n+2}{3n+2}\right)\left(\frac{\partial q}{\partial t} + \frac{1}{2}\frac{q}{h}\frac{\partial q}{\partial x}\right) + \frac{6(2n+1)^2}{(4n+3)(3n+2)}\left(\frac{3}{2}\frac{q}{h}\frac{\partial q}{\partial x} - \frac{q^2}{h^2}\frac{\partial h}{\partial x}\right)$$
$$= Th\frac{\partial^3 h}{\partial x^3} + \frac{2}{\text{Re}}\left(\frac{4n+2}{n}\right)^n\left\{h - \left|\frac{q}{h^2}\right|^{n-1}\frac{q}{h^2}\right\} \quad (8.27)$$

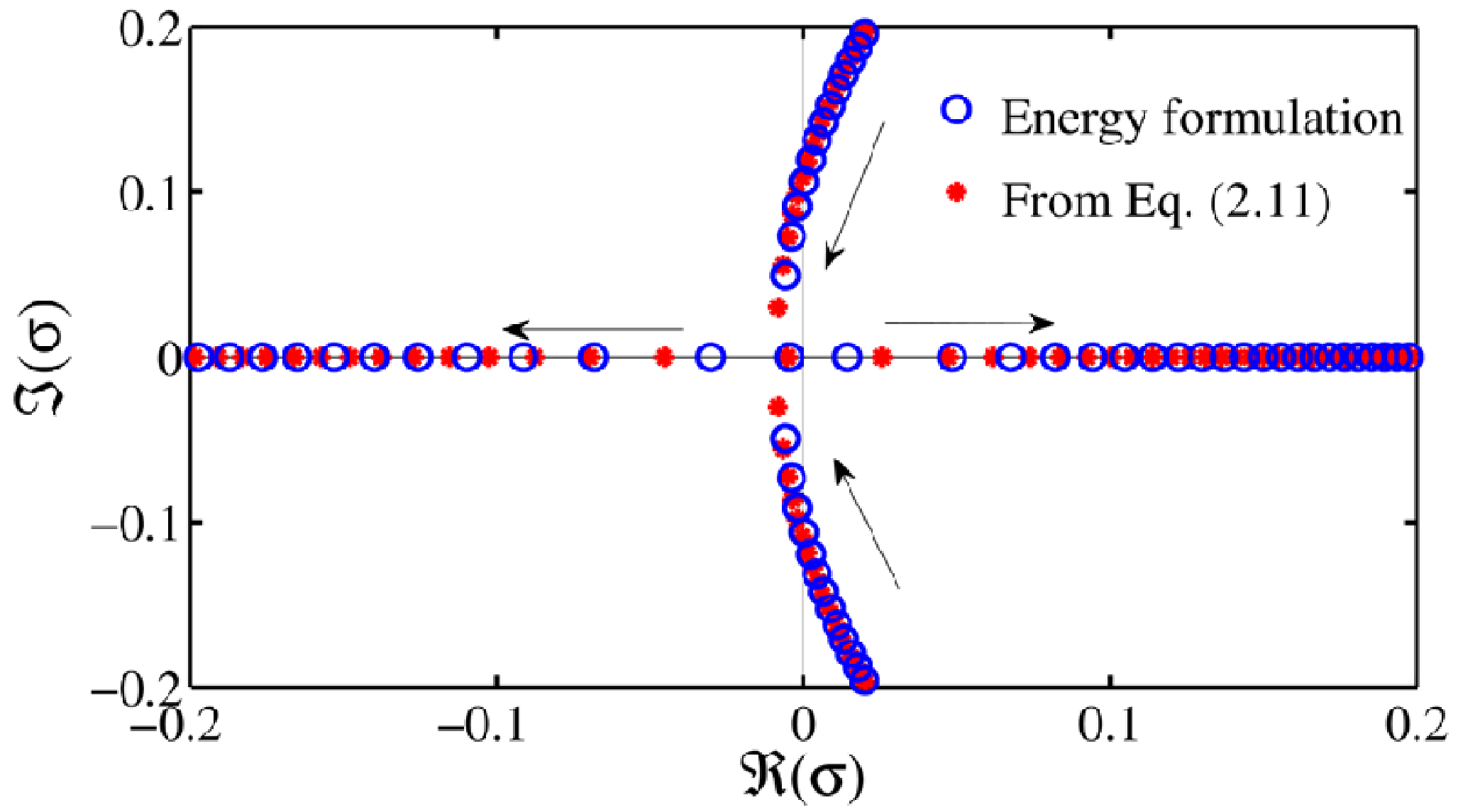

Figure H1: A comparison between the eigenvalues obtained from the linearized form of the depth-average energy equation and Eq. (2.11). The value of $\mathrm{Re}^{-1}$ = $10^{-3}$ , $n$ = 0.7 and $L_2 = 1$. Although they have qualitatively similar eigenvalue spectrum, but as of quantitative measure they are obtained for two different ranges of the tension parameter $T$. For equation (2.11), the value of $T$ ranges from 0.029 to 0.031, whereas for the equation obtained from energy, $T$ ranges from 0.036 to 0.04.

The boundary conditions at $x = 1$, are

$$h = 1 \text{ and } T\frac{\partial^2 h}{\partial x^2} = -\left(\frac{4n+2}{3n+2}\frac{\partial q}{\partial t} + \frac{2}{\mathrm{Re}}\left(\frac{4n+2}{n}\right)^n \left\{|q|^{n-1} q - 1\right\}\right)L_2 \tag{8.28}$$

These equations has same structure as that of the equation obtained in (2.9) and (2.10), the difference is in the coefficient of each term. To be more explicit, if we write this equation in a general framework as

$$\frac{\partial q}{\partial t} + a_1 \frac{q}{h}\frac{\partial q}{\partial x} + a_2 \frac{q^2}{h^2}\frac{\partial h}{\partial x} = a_3\left[Th\frac{\partial^3 h}{\partial x^3} + \frac{2}{\mathrm{Re}}\left(\frac{4n+2}{n}\right)^n \left\{h - \left|\frac{q}{h^2}\right|^{n-1}\frac{q}{h^2}\right\}\right] \tag{8.29}$$

Then for energy integral equation (8.27) we obtain the various coefficients as

$$\begin{aligned} a_1 &= \frac{1}{2} + \frac{3}{2}\left(\frac{3n+2}{4n+2}\right)\frac{6(2n+1)^2}{(4n+3)(3n+2)} \\ a_2 &= -\left(\frac{3n+2}{4n+2}\right)\frac{6(2n+1)^2}{(4n+3)(3n+2)} \\ a_3 &= \left(\frac{3n+2}{4n+2}\right) \end{aligned} \tag{8.30}$$

while for the momentum integral approach we obtain:

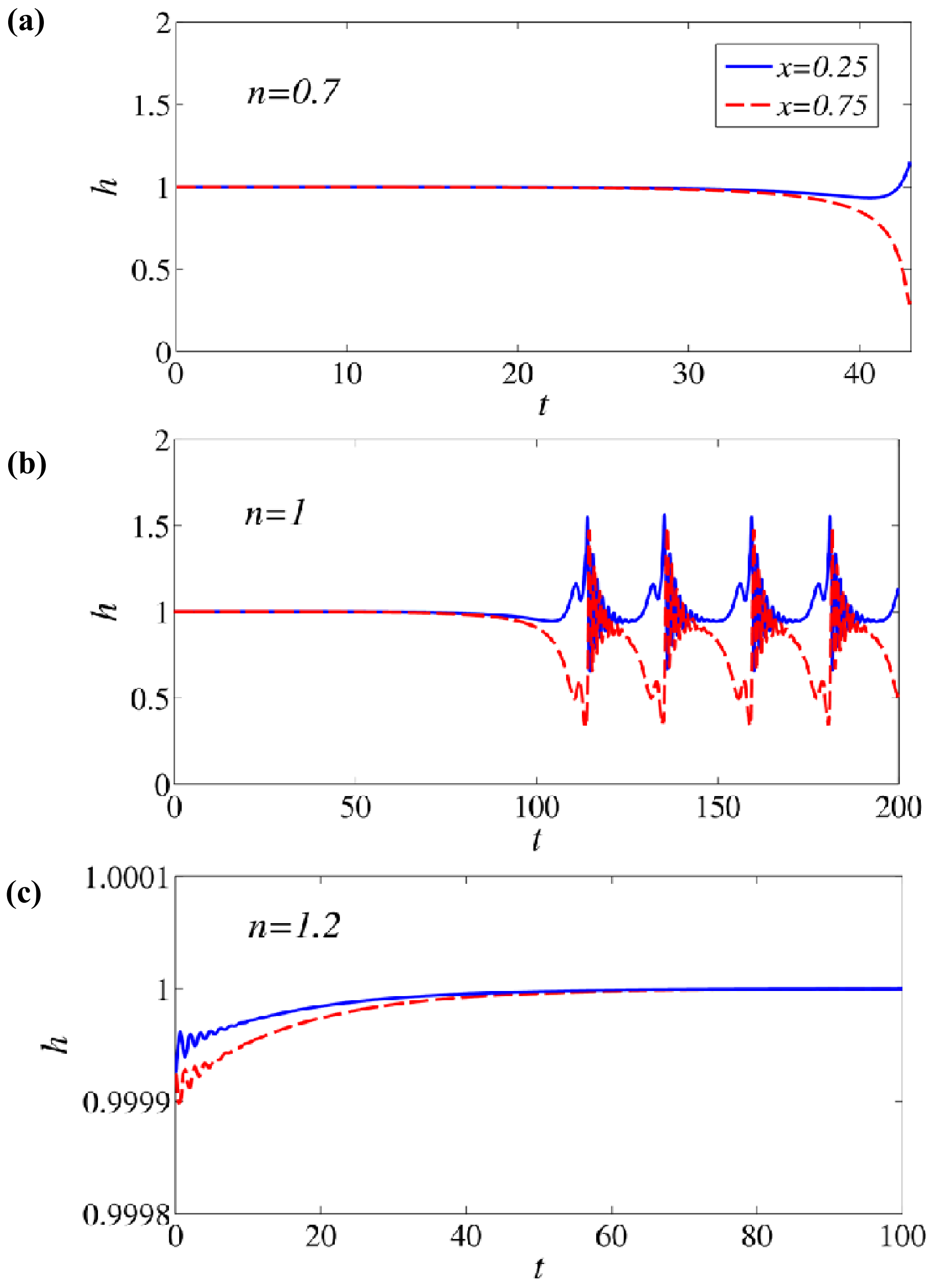

Figure H2: The temporal variation of $h(x,t)$ obtained from equation (8.27) at two different axial position, $x=0.25$ and $x=0.75$ for $T=0.08$, $\mathrm{Re}^{-1}=0.025$ and $L_2=1$. (a) $n=0.7$ (b) $n=1.0$ (c) $n=1.2$. Comparing with figure 2, we can see that the membrane dynamics appears to be qualitatively similar.

$$a_1 = 2\left(\frac{4n+2}{3n+2}\right)$$
$$a_2 = -\left(\frac{4n+2}{3n+2}\right) \qquad (8.31)$$
$$a_3 = 1$$

Now, utilizing the perturbation form for $h$ and $q$, $h = 1 + \Re\left[H(x)e^{\sigma t}\right]$ and $q = 1 + \Re\left[Q(x)e^{\sigma t}\right]$, (as described in Sec.2.3) in equation (8.29), and retaining only the linear terms, we obtain

$$\frac{dQ}{dx} + \sigma H = 0$$

$$T\frac{d^3H}{dx^3} + \frac{2}{\text{Re}}\left(\frac{4n+2}{n}\right)^n \left\{(2n+1)H - nQ\right\} - \left(\frac{4n+2}{3n+2}\right)\left\{\left(\frac{1}{2}+\frac{3}{2}c\right)\frac{dQ}{dx} - c\frac{dH}{dx}\right\} = \left(\frac{4n+2}{3n+2}\right)\sigma Q \quad (8.32)$$

where $c = \left(\frac{3n+2}{4n+2}\right)\frac{6(2n+1)^2}{(4n+3)(3n+2)}$ and the boundary conditions are of the form

$$H = 0, Q = 0; \quad \text{at } x = 0$$

$$H = 0,\ T\frac{d^2H}{\partial x^2} = -\left(\frac{4n+2}{3n+2}\sigma Q + \frac{2}{\text{Re}}\left(\frac{4n+2}{n}\right)^n nQ\right)L_2; \quad \text{at } x = 1 \quad (8.33)$$

The eigenvalues that are obtained numerically for the above equation (8.32)-(8.33) are shown in the figure H1. We have also made a comparison with the eigenvalues obtained from equation (2.11)-(2.12). In a particular range of T and $\text{Re}^{-1}$, there is a difference between the eigenvalues, but if we consider a broader range of T and $\text{Re}^{-1}$; we observe the same qualitative and quantitative nature of the eigenspectrum; only the magnitude of the real and imaginary part of a given eigenmode are different, retaining the same structure in the $(\Im(\sigma), \Re(\sigma))$ space. In addition to this we have also plotted the behaviour of $h(x,t)$ in figure H2 for some parametric values of T, $\text{Re}^{-1}$ and $n$, and have shown that the dynamical behaviour will be qualitatively similar as shown in figure 2 of the manuscript. This essentially means that the predictions obtained from the energy integral approach and the momentum integral approach are qualitatively the same.